\documentclass[a4paper,11pt,table]{article}

\pdfoutput=1 

\usepackage{jheppub} 

\usepackage[T1]{fontenc} 

\usepackage{xcolor}
\usepackage{draft} 
\usepackage{hyperref}
\usepackage{graphicx,color}
\usepackage{tcolorbox}
\usepackage{tikz-cd} 
\usepackage[all]{xy}
\usepackage{array,multirow}

\title{\huge{\fontfamily{ptm}\selectfont   Argyres-Douglas matter and S-duality: Part \Rnum{2}}}

\author[a,b]{Dan Xie}
\author[c,d]{ and Ke Ye}

\affiliation[a]{Center of Mathematical Sciences and Applications, Harvard University, Cambridge, 02138, USA}
\affiliation[b]{Jefferson Physical Laboratory, Harvard University, Cambridge, MA 02138, USA} 
\affiliation[c]{Walter Burke Institute for Theoretical Physics, California Institute of Technology, Pasadena, CA 91125, USA}
\affiliation[d]{Kavli Institute for Theoretical Physics, University of California, Santa Barbara, CA 93106, USA}

\emailAdd{dxie@cmsa.fas.harvard.edu}
\emailAdd{kye@caltech.edu}

\abstract{We study S-duality of Argyres-Douglas theories obtained by compactification of 6d (2,0) theories of $ADE$ type on a sphere with irregular punctures. The weakly coupled descriptions are given by the degeneration limit of auxiliary Riemann sphere with marked points, among which three punctured sphere represents isolated superconformal theories. We also discuss twisted irregular punctures and their S-duality.
\\
\\
\\
}

\begin{document} 
	\rightline{CALT-TH-2017-062}
	\maketitle
	
	\flushbottom

 \section{Introduction}\label{Sec: Intro}

Given a four dimensional $\mathcal{N}=2$ superconformal field theory (SCFT) with marginal deformations, it is interesting to write down its weakly coupled gauge theory descriptions. In such descriptions, gauge couplings take the role of the coordinate on the conformal manifold and the gauge theory is interpreted as conformal gauging of various strongly coupled isolated SCFTs \cite{Argyres:2007cn}. It is quite common to find more than one weakly coupled descriptions, and they are S-dual to each other as the gauge couplings are often related by $e.g.$, $\tau\propto -{1\over \tau}$. Finding all weakly coupled gauge theory descriptions is often very difficult for 
a generic strongly coupled $\mathcal{N}=2$ SCFT. 

The above questions are solved for class ${\cal S}$ theory where the Coulomb branch spectrum has integral scaling dimensions: one represents our theory by a Riemann surface $\Sigma$ with \textit{regular} singularity so that S-duality is interpreted as different degeneration limits of $\Sigma$ into three punctured sphere \cite{Gaiotto:2009we}; Once a degeneration is given, the remaining task is to identify the theory corresponding to a three punctured sphere, as well as the gauge group associated to the cylinder connecting those three punctured spheres. In class ${\cal S}$ theory framework, $\Sigma$ appears naturally as the manifold on which we compactify 6d $(2,0)$ theory. Certain $\cN=2$ SCFTs and their S-duality can be studied via geometric engineering, see \cite{DelZotto:2015rca}. 

There is a different type of $\mathcal{N}=2$ SCFT called Argyres-Douglas (AD) theories \cite{Argyres:1995jj, Xie:2012hs}. The Coulomb branch spectrum of these theories has fractional scaling dimension and they also admit marginal deformations. Again, one can engineer such AD theories by using $(2,0)$ theory on Riemann spheres $\Sigma_{g = 0}$ with \textit{irregular} singularity\footnote{We will henceforth drop the subscript $g = 0$ in what follows to denote the Riemann sphere}. Since we can not interpret the exact marginal deformations as the geometric moduli of $\Sigma$, there is no clue how  weakly coupled gauge theory descriptions can be written down in general, besides some simple cases where one can analyze the Seiberg-Witten curve directly \cite{Buican:2014hfa}.  

It came as quite a surprise that one can still interpret S-duality of $A_{N-1}$-type AD theory in terms of an auxiliary punctured Riemann surface \cite{Xie:2017vaf}. The main idea of \cite{Xie:2017vaf} is giving a map from $\Sigma$ with irregular singularities to a punctured Riemann sphere $\Sigma^{'}$, and then find  weakly coupled gauge theory as the degeneration limit of $\Sigma^{'}$ into three punctured sphere.  

The main purpose of this paper is to generalize the idea of \cite{Xie:2017vaf} to AD theories engineered using general 6d $(2,0)$ theory of type $\mfg$. The major results of this paper are
\begin{itemize}
\item We revisit the classification of irregular singularity of class $(k, b)$ in \cite{Xie:2012hs, Wang:2015mra}:
\be
\Phi \sim \frac{T_k}{z^{2+\frac{k}{b}}} + \sum_{-b \leq l < k} \frac{T_l}{z^{2+\frac{l}{b}}}
\ee
and find new irregular singularity which gives SCFT in four dimensions. Briefly, they are the configuration for which
\begin{enumerate}
\item[(\rnum{1})]  $T_k$ is regular-semisimple, whose classification was studied in \cite{Wang:2015mra}.
\item[(\rnum{2})]  The new cases are that $T_k$ is semisimple. 
\item[(\rnum{3})]  Fix a pair $(k,b)$ and type $T_k$, we can consider the degeneration of $T_k$ and the crucial constraint is that the corresponding Levi subalgebra has to be the same for $T_l$, $l > -b$. 
\end{enumerate}

\item We successfully represent our theory by an auxiliary punctured sphere from the data defining our theory from 6d (2,0) SCFT framework, and we then find weakly coupled gauge theory descriptions by studying degeneration limit of new punctured sphere. 
\end{itemize}

For instance, we find that for $\mfg = D_N$, $b = 1$ and large $k$ and all coefficient matrices regular semisimple, one typical duality frame looks like
\vspace{6pt}
\begin{adjustwidth}{0cm}{}
\begin{tikzpicture}
    \draw (16.5,0) node[anchor=west] (1) {$\cT_{N-1}$,};
    \draw (13.8,0.6) node[anchor=west] (2) {$SO(2N-2)$};
    \draw (12.4,0) node[anchor=west] (3) {$ \cT_{N-2}$};
    \draw (10.3,0.6) node[anchor=west] (4) {$ \dots \dots$};
    \draw (8.9,0) node[anchor=west] (5) {$ \cT_3$};
    \draw (7.0,0.6) node[anchor=west] (6) {$SO(6)$};
    \draw (5.6,0) node[anchor=west] (7) {$ \cT_2$};
     \draw (3.6,0.6) node[anchor=west] (8) {$SO(4)$};
     \draw (2.2,0) node[anchor=west] (9) {$\cT_1$};
    
    \path (1) edge[sedge] (2)
             (3) edge[sedge] (2)
             (3) edge[sedge] (4)
             (4) edge[sedge] (5)
             (5) edge[sedge] (6)
             (6) edge[sedge] (7)
             (7) edge[sedge] (8)
             (8) edge[sedge] (9)
    ;
\end{tikzpicture}
\end{adjustwidth}
\vspace{6pt}
where ${\cT}_i$ is given by $D_{i+1}$ theory $\pbra{\Rnum{3}_{k,1}^{[1; 2i]^{\times (k+1)}, [1^{i+1}; 0]}, [1^{2i+2}]}$. The notation we use to label the AD theories is 
\be
\pbra{\Rnum{3}_{k,b}^{ \{ \mfl_i \} },\ \ \ Q},
\ee
where $\Rnum{3}$ means type-\Rnum{3} singularity in the sense of \cite{Xie:2012hs}, and $\{ \mfl_i \}$ are Levi subalgebra for each coefficient matrix $T_i$ and $Q$ is the label for regular puncture. Each notation will be explained in the main text.

The same theory has a second duality frame, given by
\vspace{6pt}
\begin{adjustwidth}{0cm}{}
\begin{tikzpicture}
    \draw (16.5,0) node[anchor=west] (1) {${\widehat \cT}'_{N}$,};
    \draw (13.8,0.6) node[anchor=west] (2) {$SU(N)$};
    \draw (12.4,0) node[anchor=west] (3) {$ {\widehat \cT}_{N-1}$};
    \draw (10.3,0.6) node[anchor=west] (4) {$ \dots \dots$};
    \draw (8.9,0) node[anchor=west] (5) {$ {\widehat \cT}_3$};
    \draw (7.0,0.6) node[anchor=west] (6) {$SU(3)$};
    \draw (5.6,0) node[anchor=west] (7) {$ {\widehat \cT}_2$};
     \draw (3.6,0.6) node[anchor=west] (8) {$SU(2)$};
     \draw (2.2,0) node[anchor=west] (9) {${\widehat \cT}_1$};
    
    \path (1) edge[sedge] (2)
             (3) edge[sedge] (2)
             (3) edge[sedge] (4)
             (4) edge[sedge] (5)
             (5) edge[sedge] (6)
             (6) edge[sedge] (7)
             (7) edge[sedge] (8)
             (8) edge[sedge] (9)
    ;
\end{tikzpicture}
\end{adjustwidth}
\vspace{6pt}
where ${\widehat \cT}_i$, $1 \leq i \leq N-1$ is given by $\pbra{\Rnum{3}_{k,1}^{[i,1]^{\times (k+1)}, [1^{i+1}]}, [1^{i+1}]}$, and ${\widehat \cT}'_{N}$ is given by $\pbra{\Rnum{3}_{k,1}^{[N; 0]^{\times (k+1)}, [1^{2N}; 0]}, Q}$. An unexpected corollary is that the quiver with $SO(2n)$ gauge groups are dual to quivers with $SU(n)$ gauge groups, and each intermediate matter content does not have to be engineered from the same $\mfg$-type in 6d. Similar feature appears when $\mfg = E_{6,7,8}$, as will be demonstrated in this work.

The paper is organized as follows. In section \ref{Sec: PunctureClass} we briefly review regular punctures and their associated local data, and then proceed to classify (untwisted) irregular punctures for $\mfg = D_N$ and $\mfg = E_{6,7,8}$ theories. We give relevant Coulomb branch spectrum. The map from $\Sigma$ to $\Sigma'$ is described in Section \ref{Sec: mappToRiemannSphere}. Section \ref{Sec: S-duality-DN} is devoted to study the duality frames for $D_N$ theories. We consider both untwisted and twisted theories. Finally, we study S-duality frame for $E_{6,7,8}$ theories in section \ref{Sec: S-duality-E}. We conclude in section \ref{Sec: conclusion}. 
 
 \section{SCFTs from M5 branes}\label{Sec: PunctureClass}
 
 M5 brane compactifications on Riemann surface $\Sigma$ provide a large class of $\cN = 2$ superconformal theories in four dimensions. To characterize the theory, one needs to specify a Lie algebra $\mfg$ of ADE type, the genus $g$ of the Riemann surface, and the punctures on $\Sigma$. Regular punctures are the loci where the Higgs field $\Phi$ has at most simple poles; while irregular punctures are those with $\Phi$ having higher order poles. The class $\cS$ theories developed in \cite{Gaiotto:2009we} are SCFTs with $\Sigma$ of arbitrary genus and arbitrary number of regular punctures, but no irregular puncture. Later, it was realized that one may construct much larger class of theories by utilizing irregular punctures \cite{Nanopoulos:2010zb, Bonelli:2011aa, Xie:2012hs}. However, in this case the Riemann surface is highly constrained. One may use either
 \begin{enumerate}
 \item[$\bullet$] A Riemann sphere with only one irregular puncture at the north pole;
 \item[$\bullet$] A Riemann sphere with one irregular puncture at the north pole and one regular puncture at the south pole.
 \end{enumerate}
 where the genus $g = 0$ condition is to ensure the $\mC^*$ action on the Hitchin system, which guarantees $U(1)_r$ R-symmetry and superconformality. This reduces classification of theories into classification of punctures. In this section we revisit the classification and find new irregular singularity which will produce new SCFTs. 
 
 
 \subsection{Classification of punctures}
 
\subsubsection{Regular punctures}\label{subsubsec: regular puncture}

Near the regular puncture, the Higgs field takes the form
\be
\Phi \sim \frac{\Lambda}{z} + M,
\label{regularLocal}
\ee
and classification of regular puncture is essentially classification of nilpotent orbits. The puncture itself is associated with the \textit{Nahm label}, while $\Lambda$ is given by the \textit{Hitchin label}. They are related by the \textit{Spaltenstein map}. We now briefly review the classification.

\bigskip
{\bf Lie algebra} $\mfg = A_{N-1}.$ The nilpotent orbit is classified by the partition $Y=\bbra{n_1^{h_1},\ldots, n_r^{h_r}}$, where $n_i$ are column heights,  and the flavor symmetry is \cite{Gaiotto:2009we, Chacaltana:2010ks}
\begin{equation}
G_{\rm flavor}=S\pbra{\prod_{i=1}^rU(h_i)}.
\end{equation}
The spectral curve is 
\begin{equation}
\det(x-\Phi(z))=0\rightarrow x^N+\sum_{i=2}^N\phi_i(z) x^{N-i}=0.
\end{equation}
Each $\phi_i$ is the meromorphic differentials on the Riemann surface, living in the space $H^0(\Sigma, K^{\otimes i})$. The order of pole $p_i$ of the regular puncture at $\phi_i$ determines the local dimension of Coulomb branch spectrum with scaling dimension $\Delta = i$. It is given by $p_i=i-s_i$ where $s_i$ is the height of $i$-th box of the Young Tableaux $Y$; here the 
labeling is row by row starting from bottom left corner.

\bigskip
{\bf Lie algebra} $\mfg = D_{N}.$ We now review classification of regular punctures of $D_N$ algebra. For a more elaborated study, the readers may consult \cite{Chacaltana:2012zy, Chacaltana:2011ze}. 

A regular puncture of type $\mfg = D_N$ is labelled by a partition of $2N$, but not every partition is valid. It is a requirement that the even integers appear even times, which we will call a $D$\textit{-partition}. Moreover, if all the entries of the partition are even, we call it \textit{very even} D-partition. The very even partition corresponds to two nilpotent orbit, which we will label as $\cO^{\Rnum{1}}_{\bbra{\cdot}}$ and $\cO^{\Rnum{2}}_{\bbra{\cdot}}$. We again use a Young tableau with decreasing column heights to represent such a partition, and we call it a \textit{Nahm partition}. Given a Nahm partition, the residual flavor symmetry is given by
\be
G_{\rm flavor} = \prod_{h\ {\text{odd}}} \text{Spin}(n^h) \times \prod_{h\ {\text{even}}} {\rm Sp}\pbra{n^h}.
\ee

We are interested in the contribution to the Coulomb branch dimension from each puncture. When $\mfg = A_{N-1}$ case we simply take transpose and obtain a \textit{Hitchin partition} \cite{Chacaltana:2010ks}. However, for $\mfg = D_N$ the transpose does not guarantee a valid Young tableaux. Instead it must be followed by what is called \textit{D-collapse}, denoted as $(\cdot)_D$, which is described as follows: 
\begin{enumerate}
\item[(\rnum{1})] Given a partition of $2N$, take the longest even entry $n$, which occurs with odd multiplicity (if the multiplicity is greater than 1, take the last entry of that value), then picking the largest integer $m$ which is smaller than $n-1$ and then change the two entries to be $(n,m) \rightarrow (n-1, m+1)$.

\item[(\rnum{2})] Repeat the process for the next longest even integer with odd multiplicity.

\end{enumerate}
The \textit{Spaltenstein map} $\mfS$ of a given partition $d$ is given by $(d^{\mathsf{T}})_D$ and we obtain the resulting \textit{Hitchin partition} or \textit{Hitchin diagram}\footnote{Unlike \cite{Chacaltana:2011ze}, here we define the Hitchin diagram to be the one after transpose. So that when reading Young diagram one always reads column heights.}.

The Spaltenstein map is neither one-to-one nor onto; it is not an involution as the ordinary transpose either. The set of Young diagram where $\mfS$ is an involution is called \textit{special}. More generally, we have $\mfS^3 = \mfS$. 

Given a regular puncture data, one wishes to calculate its local contribution to the Coulomb branch. We begin with the special diagram. 

Using the convention in \cite{Chacaltana:2011ze}, we can construct the local singularity of Higgs field in the Hitchin system as \eqref{regularLocal} where $\Lambda$ is an $\mathfrak{so}(2N)$ nilpotent matrix associated to the Hitchin diagram and $M$ is a generic $\mathfrak{so}(2N)$ matrix. Then, the spectral curve is identified as the SW curve of the theory, which takes the form
\be
\det(x - \Phi(z)) = x^{2N} + \sum_{i = 1}^{N-1}x^{2(N-i)} \phi_{2i}(z) + {\tilde \phi}(z)^2.
\label{D_SW curve}
\ee
We call ${\tilde \phi}$ the Pfaffian. This also determines the order of poles for each coefficient $\phi_{2i}$ and ${\tilde \phi}$. We will use $p_{2i}^{\alpha}$ to label the order of poles for the former, and ${\tilde p}^{\alpha}$ to label the order of poles for the latter. The superscript $\alpha$ denotes the $\alpha$-th puncture.

The coefficient for the leading order singularity for those $\phi$'s and ${\tilde \phi}$ are not independent, but satisfy complicated relations \cite{Tachikawa:2009rb, Chacaltana:2011ze}. Note that, the Coulomb branch dimensions of $D_N$ class $\cS$ theory are not just the degrees for the differentials; in fact the Coulomb branch is the subvariety of
\be
V_{C} = \bigoplus_{k=1}^{N-1} H^0(\Sigma, K^{2k}) \oplus \bigoplus_{k=3}^{N-1}W_k \oplus H^0(\Sigma, K^{N})
\ee
where $W_k$'s are vector spaces of degree $k$. If we take $c^{(k)}_l$ to be the coefficients for the $l$-th order pole of $\phi_k$, then the relation will be either polynomial relations in $c^{(k)}_l$ or involving both $c^{(k)}_l$ and $a^{(k)}$, where $a^{(k)}$ is a basis for $W_k$. For most of the punctures, the constraints are of the form
\be
c^{(k)}_l = \dots,
\ee
while for certain very even punctures, as ${\tilde \phi}$ and $\phi_N$ may share the same order of poles, the constraints would become
\be
c^{(N)}_l \pm 2{\tilde c}_l = \dots.
\ee
For examples of these constraints, see \cite{Chacaltana:2011ze}. 

When the Nahm partition $d$ is non-special, one needs to be more careful. The pole structure of such a puncture is precisely the same as taking $d_s = \mfS^2(d)$, but some of the constraints imposed on $d_s$ should be relaxed. In order to distinguish two Nahm partitions with the same Hitchin partition, one associates with the latter a discrete group, and the map
\be
d_{{\rm Nahm}} \rightarrow \pbra{\mfS(d_{{\rm Nahm}}), \cC(d_{{\rm Nahm}})}
\ee
makes the Spaltenstein dual one-to-one. This is studied by Sommers and Achar \cite{sommers2001lusztig, achar2002local, achar2003order} and introduced in the physical context in \cite{Chacaltana:2012zy}.

Now we proceed to compute the number of dimension $k$ operators on the Coulomb branch, denoted as $d_{k}$. We have
\be
d_{2k} = (1-4k)(1-g) + \sum_{\alpha}(p_{2k}^{\alpha} - s_{2k}^{\alpha} + t_{2k}^{\alpha}),
\ee
where $g$ is the genus of Riemann surface, $s_{2k}^{\alpha}$ is the number of constraints of homogeneous degree $2k$, and $t_{2k}^{\alpha}$ is the number of $a^{(2k)}$ parameters that give the constraints $c^{(4k)}_l = \pbra{a^{(2k)}}^2$. For $d_{2k+1}$, since there are no odd degree differentials, the numbers are
\be
d_{2k+1} = \sum_{\alpha} t_{2k+1}^{\alpha},
\ee
which is independent of genus. Finally, we take special care for $d_N$. When $N$ is even, it receives contributions from both $\phi_N$ and the Pfaffian ${\tilde \phi}$. We have
\be
d_{N} = 2(1-2N)(1-g) + \sum_{\alpha}(p_{N}^{\alpha} - s_{N}^{\alpha}) + \sum_{\alpha}{\tilde p}^{\alpha}.
\ee
When $N$ is odd, it only receives contribution from the Pfaffian:
\be
d_{N} = (1-2N)(1-g) + \sum_{\alpha}{\tilde p}^{\alpha}.
\ee

\bigskip
{\bf Lie algebra} $\mfg = E_{6,7,8}.$ Unlike classical algebras, Young tableau are no longer suitable for labelling those elements in exceptional algebras. So we need to introduce some more mathematical notions. Let $\mfl$ be a Levi subalgebra, and $\cO^{\mfl}_e$ is the distinguished nilpotent orbit in $\mfl$. We have

\vspace{5pt}
{\bf Theorem} \cite{collingwood1993nilpotent}. There is one-to-one correspondence between nilpotent orbits of $\mfg$ and conjugacy classes of pairs $(\mathfrak{l}, \cO^{\mfl}_e)$ under adjoint action of $G$.

\vspace{5pt}
The theorem provides a way to label nilpotent orbits. For a given pair $(\mathfrak{l},  \cO^{\mfl}_e)$, let $X_N$ denote the Cartan type of semi-simple part of $\mathfrak{l}$. $\cO^{\mfl}_e$ in $\mfl$ gives a weighted Dynkin diagram, in which there are $i$ zero labels. Then the nilpotent orbit is labelled as $X_N(a_i)$. In case there are two orbits with same $X_N$ and $i$, we will denote one as $X_N(a_i)$ and the other as $X_N(b_i)$. Furthermore if $\mfg$ has two root lengths and one simple component of $\mathfrak{l}$ involves short roots, then we put a tilde over it. An exception of above is $E_7$, where it has one root length, but it turns out to have three pairs of nonconjugate isomorphic Levi-subalgebras. We will use a prime for one in a given pair, but a double prime for the other one. Such labels are \textit{Bala-Carter labels}.

The complete list of nilpotent orbits for $E_6$ and $E_7$ theory are given in \cite{Chacaltana:2014jba, Chacaltana:2017boe}. We will examine them in more details later in this section and in section \ref{Sec: S-duality-E}.

\subsection{Irregular puncture}\label{subsubsec: gradingKac}

\subsubsection{Grading of the Lie algebra}

We now classify irregular punctures of type $\mfg$. We adopt the Lie-algebraic techniques reviewed in the following. Recall that for an irregular puncture at $z \sim 0$, the asymptotic solution for the Higgs field $\Phi$ looks like \cite{Xie:2012hs, Wang:2015mra, Bonelli:2011aa, Nanopoulos:2010zb}
\be
\Phi \sim \frac{T_k}{z^{2+\frac{k}{b}}} + \sum_{-b \leq l < k} \frac{T_l}{z^{2+\frac{l}{b}}}, 
\label{HiggsPole}
\ee
where all $T_l$'s are semisimple elements in Lie algebra $\mfg$, and we also require that $(k, b)$ are coprime. The Higgs field shall be singled valued when $z$ circles around complex plane, $z \rightarrow z e^{2\pi i}$, which means the resulting scalar multiplication of $T_l$ comes from gauge transformation:
\be
T_l \rightarrow e^{\frac{2\pi i l}{b}}\, T_l  = \sigma\, T_l\, \sigma^{-1} 
\ee
for $\sigma$ a $G$-gauge transformation. This condition can be satisfied provided that there is a finite order automorphism (torsion automorphism) that gives grading to the Lie algebra:
\be
\mathfrak{g} = \bigoplus_{j \in \mathbb{Z}_b} \mathfrak{g}^j.
\label{Zb grading}
\ee

All such torsion automorphisms are classified in \cite{kats1969auto, reeder2010torsion, reeder2012gradings}, and they admit a convenient graphical representation called \textit{Kac diagrams}. A Kac diagram $\mD$ for $\mfg$ is an extended Dynkin diagram of $\mfg$ with labels $(s_0, s_1, \dots s_r)$ on each nodes, called \textit{Kac coordinates}, where $r$ is the rank of $\mfg$. Here $s_0$ is always set to be $2$. Let $(\alpha_1, \dots, \alpha_r)$ be simple roots, together with the highest root $-\alpha_0 = \sum_{i=1}^r a_i \alpha_i$ where $(a_1, \dots, a_r)$ are the mark. We also define the zeroth mark $a_0$ to be $1$. Then the torsion automorphism associated with $\mD$ has order $m = \sum_{i=0}^{r} a_i s_i$ and acts on an element associated with simple root $\alpha_i$ as 
\be
\sigma: \ \ \ \ \mfg_{ \alpha_i} \rightarrow \epsilon^{ s_i} \mfg_{\alpha_i}, \ \ \ \ \ i = 1,2, \dots r,
\ee
and extend to the whole algebra $\mfg$ via multiplication. Here $\epsilon$ is the $m$th primitive root of unity. It is a mathematical theorem \cite{dynkin1972semisimple} that all $s_i$ can only be $0,1$ or $2$. We call $\mD$ \textit{even} if all its Kac coordinates are even, otherwise $\mD$ is called \textit{odd}. For even diagrams, we may divide the coordinate and the order $m$ by $2$ since the odd grading never shows up in \eqref{Zb grading}. We will adopt this convention in what follows implicitly\footnote{This convention would not cause any confusion because if even diagrams are encountered, the label $s_0$ would be reduced to $1$; for odd diagrams this label remains to be $2$, so no confusion would arise.}.

There are two quantities in the grading of special physical importance. The \textit{rank} of the $G^0$ module $\mfg^j$, denoted as ${\rm rank}(G^0 | \mfg^j)$, is defined as the dimension of a maximal abelian subspace of $\mfg^j$, consisting of semisimple elements \cite{elashvili2013cyclic}. We are interested in the case where $\mfg^1$ has positive rank: $r = {\rm rank}(G^0 | \mfg^1) > 0$. Another quantity is the intersection of centralizer of semi-simple part of $\mfg^1$ with $\mfg^0$, and this will give the maximal possible flavor symmetry. 

As we get matrix $T_j$ out of $\mfg^j$, we are interested in the case where $\mfg^j$ generically contains regular semisimple element. We call such grading \textit{regular semisimple}. A natural way to generate regular semisimple grading is to use nilpotent orbits. For $\mfg = A_{N-1}$ it is given in \cite{Xie:2017vaf}. We give the details of $D_N$ and $E_{6,7,8}$ in Appendix \ref{nilpoGrading}. Note when coefficient matrices are all regular semisimple, the AD theory with only irregular singularity can be mapped to type \Rnum{2}B string probing three-fold compound Du Val (cDV) singularities \cite{Shapere:1999xr}, which we review in Appendix \ref{type2B}. We list the final results in table \ref{table:isolatedsingularitiesALEfib}.
\begin{table}[!htb]
\begin{center}
  \begin{tabular}{ |c|c|c| }
    \hline
       $\mfg$& Singularity & $b$  \\ \hline
     $A_{N-1}$ &$x_1^2+x_2^2+x_3^N+z^{k N / b}=0$&  $ b | N$ \\ \hline
 $~$& $x_1^2+x_2^2+x_3^N+x_3 z^{k (N-1) / b}=0$ & $b | (N-1)$\\ \hline
      
 $D_N$   & $x_1^2+x_2^{N-1}+x_2x_3^2+z^{k (2N-2) / b}=0$ & $b | (2N-2)$ \\     \hline
  $~$   &$x_1^2+x_2^{N-1}+x_2x_3^2+z^{k N / b} x_3=0$& $b | N$ \\     \hline

  $E_6$  & $x_1^2+x_2^3+x_3^4+z^{12k/b} =0$&$b | 12$   \\     \hline
   $~$  & $x_1^2+x_2^3+x_3^4+z^{9k / b} x_3=0$ &$b | 9$   \\     \hline
  $~$  & $x_1^2+x_2^3+x_3^4+z^{8k / b} x_2=0$  &$b |8$   \\     \hline

   $E_7$  & $x_1^2+x_2^3+x_2x_3^3+z^{18k / b}=0$& $b | 18$   \\     \hline
      $~$  & $x_1^2+x_2^3+x_2x_3^3+z^{14k / b}x_3=0$ &$b | 14$    \\     \hline

    $E_8$   & $x_1^2+x_2^3+x_3^5+z^{30k/b}=0$& $b | 30$   \\     \hline
        $~$   & $x_1^2+x_2^3+x_3^5+z^{24k/b} x_3=0$ & $b | 24$   \\     \hline
    $~$   & $x_1^2+x_2^3+x_3^5+z^{20k/b} x_2=0$ & $b | 20$ \\     \hline

  \end{tabular}
  \end{center}
  \caption{Classification of irregular singularities with regular semisimple coefficient matrices and the 3-fold singularities corresponding to them.  In the table, $b | N$ means that $b$ is a divisor of $N$.}
  \label{table:isolatedsingularitiesALEfib}
\end{table}
This is a refinement and generalization of the classification done in \cite{Wang:2015mra, Xie:2017vaf}. We emphasize here that the grading when $\mfg^j$ generically contain semisimple elements are also crucial for obtaining SCFTs; here $b$ may be more arbitrary. Such grading will be called \textit{semisimple}.

In classical Lie algebra, semisimple element $T_i$ can be represented by the matrices. In order for the spectral curve $\det(x - \Phi(z))$ to have integral power for monomials, the matrices for leading coefficient $T_k$ is highly constrained. In particular, when $\mfg = A_{N-1}$, we have
\be
T = \left( \begin{array}{cccc} a_1 \Xi & & & \\ & \ddots & & \\ & & a_r \Xi & \\ & & & 0_{(N-r b)}\end{array} \right).
\label{Adegenerate}
\ee
Here $\Xi$ is a $b \times b$ diagonal matrix with entries $\{ 1, \omega, \omega^2, \dots, \omega^{b-1} \}$ for $\omega$ a $b$-th root of unity $\exp\pbra{2\pi i / b}$. For $\mfg = D_N$, things are more subtle and $T$ depends on whether $b$ is even or odd. A representative of Cartan subalgebra is
\be
\left( \begin{array}{cc} Z & 0 \\ 0 & - Z^{\mathsf{T}} \end{array} \right),
\label{so(2N)matrix}
\ee
where $Z \in {\rm Mat}_{N\times N}(\mathbb{C})$. When $b$ is odd, we have
\be
Z = \left( \begin{array}{cccc} 0_{N-b r} & & & \\ & a_1 \Xi & & \\ & & \ddots & \\ & & & a_{r} \Xi \end{array} \right).
\ee
When $b$ is even, we define $\Xi' = \{ 1, \omega^2, \omega^4, \dots, \omega^{b-2} \}$, then $\Xi = \Xi' \cup (-\Xi')$. Then the coefficient matrix take the form
\be
Z = \left( \begin{array}{cccc} 0_{N-r b/2} & & & \\ & a_1 \Xi' & & \\ & & \ddots & \\ & & & a_{r} \Xi' \end{array} \right).
\ee
Counting of physical parameters in two cases are different, as we will see in section \ref{subsec: field data} momentarily. In particular, the allowed mass parameters are different for these two situations.

 \subsubsection{From irregular puncture to parameters in SCFT}\label{subsec: field data}
 
We have classified the allowed order of poles for Higgs field in \eqref{HiggsPole}, and write down in classical algebras the coefficient matrix $T_i$. The free parameters in $T_i$ encode exact marginal deformations and number of mass parameters.

Based on the discussion above and the coefficient matrix, we conclude that the number of  mass parameters is equal to ${\rm rank}(\mfg^0)$ and the number of exact marginal deformation is given by ${\rm rank}(G^0 | \mfg^k) - 1$ if the leading matrix is in $\mfg^k$.  we may list the maximal number of exact marginal deformations and number of mass parameters in tables \ref{table:ANpara} - \ref{table:E8para}. We focus here only in the case when $T$'s are regular semisimple, while for semisimple situation the counting is similar. 
\begin{table}
\centering
\begin{tabular}{|c|c|c|}
\hline
order of singularity $b$ & mass parameter & exact marginal deformations\\
\hline
$b | N$ & $N / b - 1$ & $N/b  -1$ \\
$b | (N-1)$ & $(N-1) / b$ & $(N-1) / b - 1$  \\ \hline
\end{tabular}
\caption{\label{table:ANpara}Summary of mass parameters and number of exact marginal deformations in $A_{N-1}$. }
\end{table}

\begin{table}
\centering
\begin{tabular}{|c|c|c|}
\hline
order of singularity $b$ & mass parameter & exact marginal deformations\\
\hline
odd, $b | N$ & $N / b$ & $N/b  -1$ \\
even, $b| N$ & $0$ & $2N / b - 1$  \\
odd, $b | (2N-2)$ & $(N-1) / b + 1$ & $ (N-1) / b - 1$ \\
even, $b | (2N-2)$ & $1$ or $0$ & $(2N-2) / b - 1$  \\ \hline
\end{tabular}
\caption{\label{table:DNpara}Summary of mass parameters and number of exact marginal deformations in $D_N$. Note when $b$ is even divisor of $2N-2$ but not a divisor of $N-1$, the number of mass parameter is zero, otherwise it is one.}
\end{table}

\begin{table}
\centering
\begin{tabular}{|c|c|c|}
\hline
order of singularity $b$ & mass parameter & exact marginal deformations\\
\hline
$12$ & $0$ & $0$ \\
$9$ & $0$ & $0$  \\
$8$ & $1$ & $0$ \\
$6$ & $0$ & $1$  \\ 
$4$ & $2$ & $1$ \\ 
$3$ & $0$ & $2$ \\
$2$ & $2$ & $3$ \\ \hline
\end{tabular}
\caption{\label{table:E6para}Summary of mass parameters and number of exact marginal deformations in $E_6$. }
\end{table}

\begin{table}
\centering
\begin{tabular}{|c|c|c|}
\hline
order of singularity $b$ & mass parameter & exact marginal deformations\\
\hline
$18$ & $0$ & $0$ \\
$14$ & $0$ & $0$  \\
$9$ & $1$ & $0$ \\
$7$ & $1$ & $0$  \\ 
$6$ & $0$ & $2$ \\ 
$3$ & $1$ & $2$ \\
$2$ & $0$ & $6$ \\ \hline
\end{tabular}
\caption{\label{table:E7para}Summary of mass parameters and number of exact marginal deformations in $E_7$. }
\end{table}

\begin{table}
\centering
\begin{tabular}{|c|c|c|}
\hline
order of singularity $b$ & mass parameter & exact marginal deformations\\
\hline
$30$ & $0$ & $0$ \\
$24$ & $0$ & $0$  \\
$20$ & $0$ & $0$ \\
$15$ & $0$ & $0$  \\ 
$12$ & $0$ & $1$ \\ 
$10$ & $0$ & $1$ \\
$8$ & $0$ & $1$ \\
$6$ & $0$ & $3$ \\ 
$5$ & $0$ & $1$ \\ 
$4$ & $0$ & $3$ \\
$3$ & $0$ & $3$ \\
$2$ & $0$ & $7$ \\ \hline
\end{tabular}
\caption{\label{table:E8para}Summary of mass parameters and number of exact marginal deformations in $E_8$. }
\end{table}

\vspace{6pt}
$\bullet$ {\bf Argyres-Douglas matter}. We call the AD theory without any marginal deformations the \textit{Argyres-Douglas matter}. They are isolated SCFTs and thus are the fundamental building blocks in S-duality. In the weakly coupled description, we should be able to decompose the theory into Argyres-Douglas matter connected by gauge groups. 

\subsubsection{Degeneration and graded Coulomb branch dimension}\label{subsubsec: irregDegeneration}

Our previous discussion focused on the case where we choose generic regular semisimple element for a given positive rank grading. More generally, we may consider $T_k$ semisimple. We first examine the singularity where $b = 1$:
\be
\Phi \sim \frac{T_{\ell}}{z^{\ell}} + \frac{T_{\ell-1}}{z^{\ell-1}} + \dots + \frac{T_{1}}{z^1},
\label{MaximalIrregSing}
\ee
with $T_{\ell} \subset \dots \subset T_2 \subset T_1$ \cite{Witten:2007td}. For this type of singularity, the \textit{local} contribution to the dimension of Coulomb branch is
\be
\dim^{\rho}_{\mathbb{C}} {\rm Coulomb} = \frac{1}{2} \sum_{i=1}^{\ell} \dim (\cO_{T_i}).
\label{CBdimension}
\ee
This formula indicates that the Coulomb branch dimensions are summation of each semisimple orbit in the irregular singularity. It is reminiscent of the regular puncture case reviewed in section \ref{subsubsec: regular puncture}, where the local contribution to Coulomb branch of each puncture is given by half-dimension of the nilpotent orbits, $\dim^{\rho}_{\mathbb{C}} {\rm Coulomb} = \frac{1}{2} \dim \mfS(\cO_\rho)$ \cite{Chacaltana:2012zy}.

To label the degenerate irregular puncture, one may specify the centralizer for each $T_{\ell}$. Given a semisimple element $x \in \mfg$, the centralizer $\mfg^x$ is called a \textit{Levi subalgebra}, denoted as $\mfl$. In general, it may be expressed by
\be
\mfl = \mfh \oplus \sum_{\Delta' \subset \Delta} \mfg_{\alpha},
\ee
where $\mfh$ is a Cartan subalgebra and $\Delta'$ is a subset of the simple root $\Delta$ of $\mfg$. We care about its semisimple part, which is the commutator $\bbra{\mfl, \mfl}$. 

The classification of the Levi subalgebra is known. For $\mfg$ of ADE type, we have
\begin{enumerate}
\item[$\bullet$] $\mfg = A_{N-1}$: $\mfl = A_{i_1} \oplus A_{i_2} \oplus \dots A_{i_k}$, with $(i_1 + 1) + \dots + (i_k+1) = N$.
\item[$\bullet$] $\mfg = D_{N}$: $\mfl = A_{i_1} \oplus A_{i_2} \oplus \dots A_{i_k} \oplus D_j$, with $(i_1 + 1) + \dots + (i_k+1) + j = N$.
\item[$\bullet$] $\mfg = E_6$: $\mfl = E_6,~D_5, ~A_5, ~A_4+A_1, ~2A_2+A_1,~D_4,~A_4, ~A_3+A_1, ~2A_2,~A_2+2A_1, ~A_3, ~A_2+A_1,~3A_1, ~A_2, ~2A_1, ~A_1,~0$.
\item[$\bullet$] $\mfg = E_7$: $E_7, ~E_6,~ D_6, ~D_5+A_1, ~A_6, ~A_5+A_1,~ A_4+A_2, ~A_3+A_2+A_1, ~D_5, ~D_4+A_1, ~A_5^{'}, ~A_5^{''}, ~A_4+A_1, ~A_3+A_2,~A_3+2A_1,~2A_2+A_1,~A_2+3A_1,~D_4,~A_4,~(A_3+A_1)^{'},~ (A_3+A_1)^{''}, ~2A_2, ~A_2+2A_1, ~4A_1, ~A_3, ~A_2+A_1, ~(3A_1)^{'}, ~(3A_1)^{''}, ~A_2, ~2A_1, ~A_1,~0$.
\item[$\bullet$] $\mfg = E_8$: $E_8, ~E_7, ~E_6+A_1, ~D_7, ~D_5+A_2, ~A_7, ~A_6+A_1, ~A_4+A_3, ~A_4+A_2+A_1, ~E_6, ~D_6, ~D_5+A_1, ~D_4+A_2, ~A_6, ~A_5+A_1,~ A_4+A_2, ~A_4+2A_1, ~2A_3, ~A_3+A_2+A_1, ~2A_2+2A_1, ~D_5, ~D_4+A_1, ~A_5, ~A_4+A_1,~ A_3+A_2, ~A_3+2A_1,~ 2A_2+A_1, ~A_2+A_1, ~D_4, ~A_4, ~A_3+A_1, ~2A_2, ~A_2+2A_1, ~4A_1, ~A_3, ~A_2+A_1, ~3A_1, ~A_2, ~2A_1, ~A_1, ~0$. 
\end{enumerate}

Fixing the Levi subalgebra for $T_i$, the corresponding dimension for the semisimple orbit is given by
\be
\dim (\cO_{T_i}) = \dim G - \dim L_i.
\ee
We emphasize here that Levi subalgebra itself completely specify the irregular puncture. However, they may share the semisimple part $\bbra{\mfl, \mfl}$. The SCFTs defined by them can be very different. Motivated by the similarity between \eqref{CBdimension} and that of regular punctures, we wish to use nilpotent orbit to label the semisimple orbit ${\cal O}_{T_i}$, so that one can calculate the graded Coulomb branch spectrum.

The correspondence lies in the theorem we introduced in section \ref{subsubsec: regular puncture}: there is a one-to-one correspondence between the nilpotent orbit $\cO^{\mfg}_{\rho}$ and the pair $(\mfl, \cO^{\mfl}_e)$. Moreover, we only consider those nilpotent orbit with principal $\cO^{\mfl}_e$. For $\mfg = A_{N-1}$, principal orbit is labelled by partition $[N]$, while for $D_N$, it is the partition $[2N-1, 1]$. Then, given a Nahm label whose $\cO^{\mfl}_e$ is principal, we take the Levi subalgebra piece $\mfl$ out of the pair $(\mfl, \cO^{\mfl}_e)$; we use the Nahm label $\rho$ as the tag such $T_i$. We conjecture that this fully characterize the coefficients $T_i$. 

To check the validity, we recall orbit induction \cite{kempken1983induced, de2009induced}. Let $\cO^{\mfl}_{\bar e}$ be an arbitrary nilpotent orbit in $\mfl$. Take a generic element $m$ in the center $\mfz$ of $\mfl$. We define
\be
{\rm Ind}^{\mfg}_{\mfl} \cO^{\mfl}_{\bar e} : = \lim_{m \rightarrow 0} {\cal O}_{m+{\bar e}},
\ee
which is a nilpotent orbit in $\mfg$. It is a theorem that the induction preserves codimension:
\be
\dim G - \dim_{\mC} {\rm Ind}^{\mfg}_{\mfl} \cO^{\mfl}_{\bar e} = \dim L - \dim_{\mC} \cO^{\mfl}_{\bar e}.
\label{inducedCodimThm}
\ee
In particular, when $\cO^{\mfl}_{\bar e}$ is zero orbit in $\mfl$, from \eqref{inducedCodimThm} we immediately conclude that
\be
\dim \cO_{T} = \dim G - \dim L = \dim_{\mC} {\rm Ind}^{\mfg}_{\mfl} \cO^{\mfl}_0,
\ee
for $T$ the semisimple orbit fixed by $L$. The Bala-Carter theory is related to orbit induction via \cite{collingwood1993nilpotent}
\be
\dim \mfS(\cO_{\rho}) = \dim_{\mC} {\rm Ind}^{\mfg}_{\mfl} \mfS(\cO^{\mfl}_{\rm principal}) = \dim_{\mC} {\rm Ind}^{\mfg}_{\mfl} \cO^{\mfl}_0 = \dim \cO_{T}.
\ee
Therefore, treating each semisimple orbit $\cO_{T}$ as a nilpotent orbit $\cO_{\rho}$, their local contribution to Coulomb branch is exactly the same.

In the $A_{N-1}$ case, Levi subalgebra contains only $A_{i}$ pieces; the distinguished nilpotent orbit in it is unique, which is $[i+1]$. Therefore, we have a one-to-one correspondence between Nahm partitions and Levi subalgebra. More specifically, a semisimple element of the form 
\be
x = \text{diag}(a_1, \dots, a_1, a_2, \dots, a_2, \dots, a_k, \dots, a_k),
\ee
where $a_i$ appears $r_i$ times, has Levi subgroup
\be
L = S\bbra{U(r_1) \times U(r_2) \times \dots \times U(r_k)}.
\label{AtypeLevi}
\ee
whose Nahm label is precisely $[r_1, r_2, \dots, r_k]$. 

For $D_N$ case, if the semisimple element we take looks like
\be
x = \text{diag}(a_1, \dots, a_1, \dots, a_k, \dots, a_k, -a_1, \dots, -a_1, \dots, -a_k, \dots, -a_k, 0,\dots, 0 ),
\ee
where $a_i$ appears $r_i$ times and $0$ appears ${\tilde r}$ times with $\sum 2r_i + {\tilde r} = 0$, the Levi subgroup is given by
\be
L = \prod_i U(r_i) \times SO({\tilde r}).
\ee
We call $L$ of type $[r_1, \dots, r_k; {\tilde r}]$. Here we see clearly the ambiguity in labelling the coefficient $T_i$ using Levi subalgebra. For instance, when $\mfg = D_4$, we have $[1;6]$ and $[4;0]$ having the same Levi subalgebra, but clearly they are different type of matrices and the SCFT associated with them have distinct symmetries and spectrum. We will examine them in more detail in section \ref{Sec: S-duality-DN}.

With Nahm labels for each $T_i$, we are now able to compute the graded Coulomb branch spectrum. For each Nahm label, we have a collection of the pole structure $\{ p^{\alpha}_{i_1}, \dots,  p^{\alpha}_{i_r} \}$ for $i_k$ the degrees of differentials. There are also constraints that reduce or modifies the moduli. Then we conjecture that, at differential of degree $k$ the number of graded moduli is given by
\be
d_k = \sum_{\alpha} \pbra{p_k^{\alpha} - s_k^{\alpha} + t_k^{\alpha}} - 2k + 1.
\ee
They come from the term $u_i$ in $(u_0+u_1z+\ldots+u_{d_k-1}z^{d_k-1})x^{h^{\vee}-k}$, with $h^{\vee}$ the dual Coxeter number. 

However, it might happen that there are constraints of the form $c^{(2k)} = \pbra{a^{(k)}}^2$ in which $k$ is not a degree for the differentials. In this case, $t_k$ should be added to the some $k' > k$ such that $d^{\rm local}_{k'} < k' - 1$.

When a regular puncture with some Nahm label is added to the south pole, one may use the same procedure to determine the contributions of each differential to the Coulomb branch moduli. We denote them as $\{ d^{(\rm reg)}_{k}\}$. Then, we simply extend the power of $z_\beta x^{2(N-k)}$ to $-d^{(\rm reg)}_{k} < \beta < d_{k}$.

\vspace{6pt}
$\bullet$ \textit{Example}: let us consider an $E_6$ irregular puncture of class $(k,1)$ where $k$ is very large. Take $T_{\ell} = \dots = T_2$ with Levi subalgebra $D_5$, and $T_1$ with Levi subalgebra $0$. We associate to $T_i$ with $i \geq 2$ Nahm label $D_5$. As a regular puncture, it has pole structure $\{1, 2,3,4,4,6 \}$ with complicated relations \cite{Chacaltana:2014jba}:
\be
& c_3^{(6)} = \frac{3}{2} c_1^{(2)} a_2^{(4)}, \ \ \ \ \ c_4^{(8)} =3\pbra{a_2^{(4)}}^2, \\[0.5em]
& c_4^{(9)} = -\frac{1}{4} c_2^{(5)} a_2^{(4)}, \ \ \ \ \ c_6^{(12)} =\frac{3}{2} \pbra{a_2^{(4)}}^3, \\[0.5em]
& c_5^{(12)} = \frac{3}{4} c_3^{(8)} a_2^{(4)}.
\ee
After subtracting it we have pole structure $\{1, 2,2,3,3,4 \}$. There is one new moduli $a^{(4)}$, and we add it to $\phi_5$. The Nahm label $0$ has pole structure $\{ 1,4,5,7,8,11 \}$. Then we have the Coulomb branch spectrum from such irregular puncture as 
\be
& \phi_2: \frac{2k}{k+1}, \dots, \frac{k+2}{k+1}, \ \ \ \ \ \phi_5: \frac{5k}{k+1}, \dots, \frac{2k+3}{k+1}, \\[0.5em]
& \phi_6: \frac{6k}{k+1}, \dots, \frac{4k+5}{k+1}, \ \ \ \ \ \phi_8: \frac{8k}{k+1}, \dots, \frac{5k+6}{k+1}, \\[0.5em]
& \phi_9: \frac{9k}{k+1}, \dots, \frac{6k+7}{k+1}, \ \ \ \ \ \phi_{12}: \frac{12k}{k+1}, \dots, \frac{8k+9}{k+1}.
\ee

\bigskip

One can carry out similar analysis for general irregular singularity of class $(k,b)$. The idea is to define a cover coordinate $\omega$ and reduce the problem to integral order of pole. Consider an irregular singularity defined by the following data $\Phi= T /  z^{2+{k\over b}}+\ldots$; we define a cover coordinate $z=\omega^b$ and the Higgs field is reduced to
\begin{equation}
\Phi= {T' \over \omega^{k+b+1}} + \ldots
\end{equation}
Here $T'$ is another semisimple element deduced from $T$, see examples in section \ref{sec: S-duality-(k,b)}. Once we go to this cover coordinate, we can use above study of degeneration of irregular singularity with integral order of pole. We emphasize here that not all degeneration are allowed due to the specific form of $T$. 

\subsubsection{Constraint from conformal invariance}

As we mentioned, not all choices of semisimple coefficient $T_i$ define SCFTs. Consider the case $b=1$, and the irregular singularity is captured by by a sequence of Levi subgroup $\mfl_{\ell}\supset \mfl_{\ell-1} \supset \ldots \supset \mfl_1$. The necessary condition is that the number of parameters in the leading order matrix $T_k$ should be no less than the number of exact marginal deformations. As will be shown later, it turns out that this condition imposes the constraint that 
\begin{equation}
\mfl_{\ell}= \mfl_{\ell-1}\ldots= \mfl_2 = \mfl,
\end{equation}
with $\mfl_1$ arbitrary. Then we have following simple counting rule of our SCFT:
\begin{itemize}
\item The maximal number of exact marginal deformation is equal to $r-r_{\mfl}-1$, where $r$ the rank of $\mfg$ and $r_{\mfl}$ the rank of 
semi-simple part of $\mfl$. The extra minus one comes from scaling of coordinates.
\item The maximal flavor symmetry is $G_{\mfl}\times U(1)^{r-r_{\mfl}}$, here $G_{\mfl}$ is the semi-simple part of $\mfl$. 
\end{itemize}

Similarly, for $b\neq 1$, the conformal invariance implies that all the coefficients except $T_1$ should have the same Levi subalgebra.  This is automatic when the grading is regular semisimple, but it is an extra restriction on general semi-simple grading.  For example, consider $A_{N-1}$ type $(2,0)$ theory with following irregular singularity whose leading order matrix takes the form:
\be
T = \left( \begin{array}{cccc} a_1 \Xi & & & \\ & \ddots & & \\ & & a_r \Xi & \\ & & & 0_{(N-r b)}\end{array} \right).
\label{Adegenerate}
\ee
When the subleading term in \eqref{HiggsPole} has integral order, the corresponding matrix can take the following general form:
\be
T^{'} = \left( \begin{array}{cccc} a'_1 {\mathbb{I}}_b & & & \\ & \ddots & & \\ & & a'_r {\mathbb{I}}_b & \\ & & & {\mathbb{K}}_{(N-r b)}\end{array} \right).
\ee
Here ${\mathbb{I}}_b$ is the identify matrix with size $b$, and ${\mathbb{K}}_{N-rb}$ is a generic diagonal matrix. However, due to the constraints, only for ${\mathbb{K}}_{N-rb}= \kappa\, {\mathbb{I}}_{N-rb}$, $T^{'}$ has the same Levi-subalgebra as $T$. This situation is missed in previous studies \cite{Xie:2017vaf}.

\subsection{SW curve and Newton polygon}

Recall that the SW curve is identified as the spectral curve $\det(x - \Phi(z))$ in the Hitchin system. For regular semisimple coefficient $T_i$ without regular puncture, we may map the curve to the mini-versal deformation of three fold singularity in type \Rnum{2}B construction. For given Lie algebra $\mfg$, we have the deformed singularity:
\begin{align}
&A_{N-1}:~x_1^2+x_2^2+ x_3^N+\phi_2(z)x_3^{N-2}+\ldots+\phi_{N-1}(z)x_3+\phi_{N}(z) =0,  \nonumber\\
&D_N:~x_1^2+x_2^{N-1}+x_2 x_3^2+\phi_2(z)x_2^{N-2}+\ldots+\phi_{2N-4}(z)x_2+\phi_{2N-2}(z)+\tilde{\phi}_N(z)x_3=0,  \nonumber\\
&E_6:~x_1^2+x_2^3+x_3^4+\phi_2(z)x_2x_3^2+\phi_5(z)x_2x_3+\phi_6(z)x_3^2+\phi_8(z)x_2+\phi_9(z)x_3+\phi_{12}(z)=0, \nonumber\\
&E_7:~x_1^2+x_2^3+x_2x_3^3+\phi_2(z)x_2^2x_3+\phi_{6}(z) x_2^2+\phi_{8}(z)x_2x_3+\phi_{10}(z)x_3^2   \nonumber\\
&~~~~~~+\phi_{12}(z)x_2+\phi_{14}(z)x_3+\phi_{18}(z)=0, \nonumber\\
& E_8:~x_1^2+x_2^3+x_3^5+\phi_2(z)x_2x_3^3+\phi_8(z)x_2x_3^2+\phi_{12}(z)x_3^3+  \nonumber\\
&~~~~~~~\phi_{14}(z)x_2x_3+\phi_{18}(z)x_3^2+\phi_{20}(z)x_2+\phi_{24}(z)x_3+\phi_{30}(z)=0,
\label{3foldforms}
\end{align}
and $\phi_i$ is the degree $i$ differential on Riemann surface. 

A useful diagrammatic approach to represent SW curve is to use Newton polygon. When irregular singularity degenerates, the spectrum is a subset of that in regular semisimple $T_i$'s, so understanding Newton polygon in regular semisimple case is enough. 

The rules for drawing and reading off scaling dimensions for Coulomb branch spectrum is explained in \cite{Xie:2012hs, Wang:2015mra}. In particular, the curve at the conformal point determines the scaling dimension for $x$ and $z$, by requiring that the SW differential $\lambda = x dz$ has scaling dimension $1$.

\bigskip

\begin{figure}[htbp]
\centering
\begin{adjustwidth}{5.0cm}{}
\includegraphics[width=5cm]{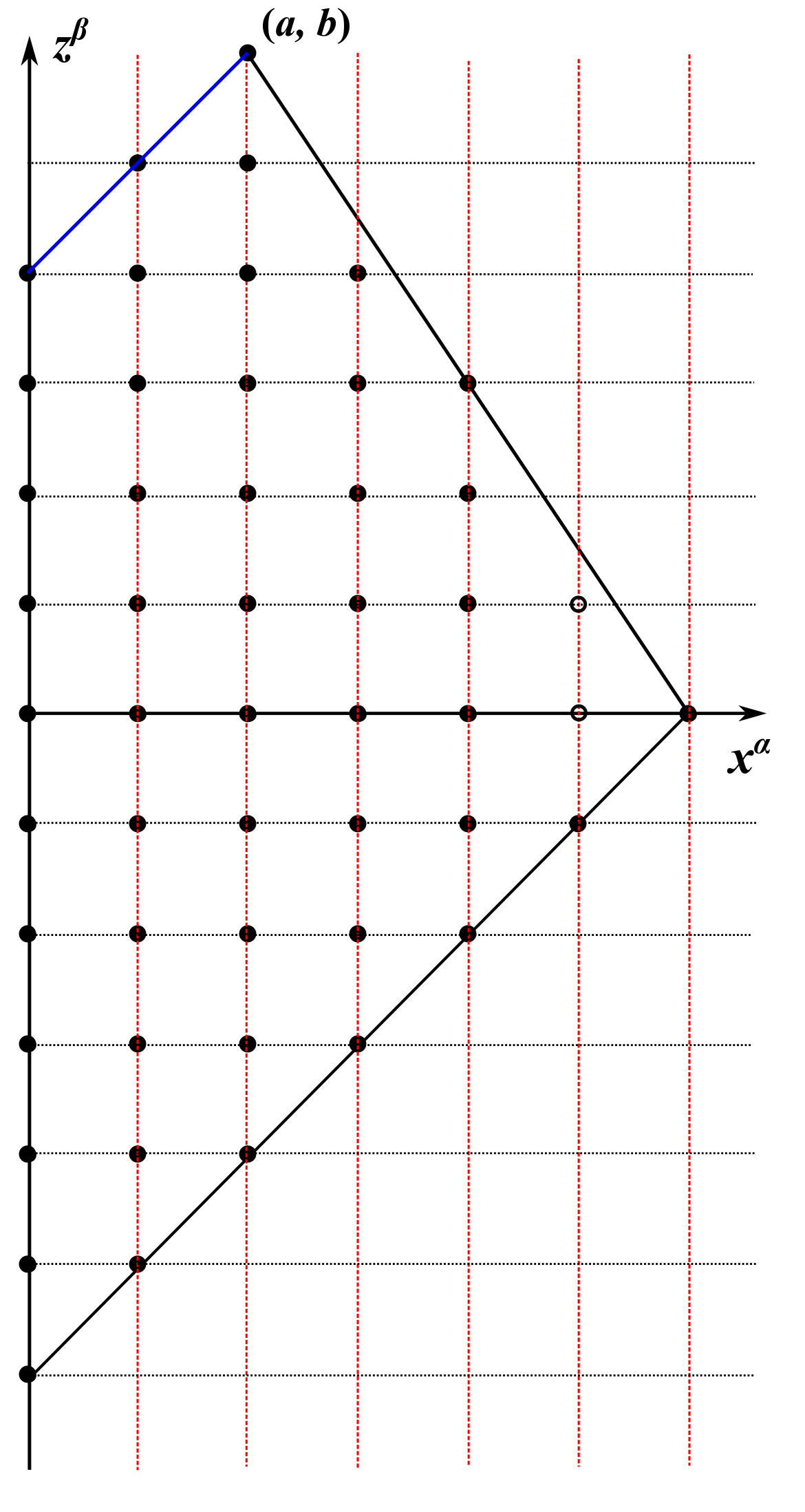}
\end{adjustwidth}
\caption[Newton polygon for $A_{N-1}$ type Argyres-Douglas theories]{An example of Newton polygon for $A_5$ theory with semisimple grading. Each black dot represents a monomial in SW curve. The white dots mean that the monomials are omitted. The letters have scaling dimension $\bbra{x} = 3/5$, $\bbra{z} = 2/5$. In general, if the vertex at the top has coordinate $(a,b)$, then we have the relation $(N-a)\bbra{x} = b \bbra{z}$ and $[x] + [z] = 1$.}
\label{ANnewton}
\end{figure}

$\bullet$ $\mfg = A_{N-1}$. The Newton polygon for regular semisimple coefficient matrices is already given in \cite{Xie:2012hs} and we do not repeat here. Here we draw the polygon when $T$ is semisimple for some semisimple grading, in the form \eqref{Adegenerate}. We give one example See figure \ref{ANnewton}.  

\bigskip

\begin{figure}[htbp]
\centering
\includegraphics[width=12cm]{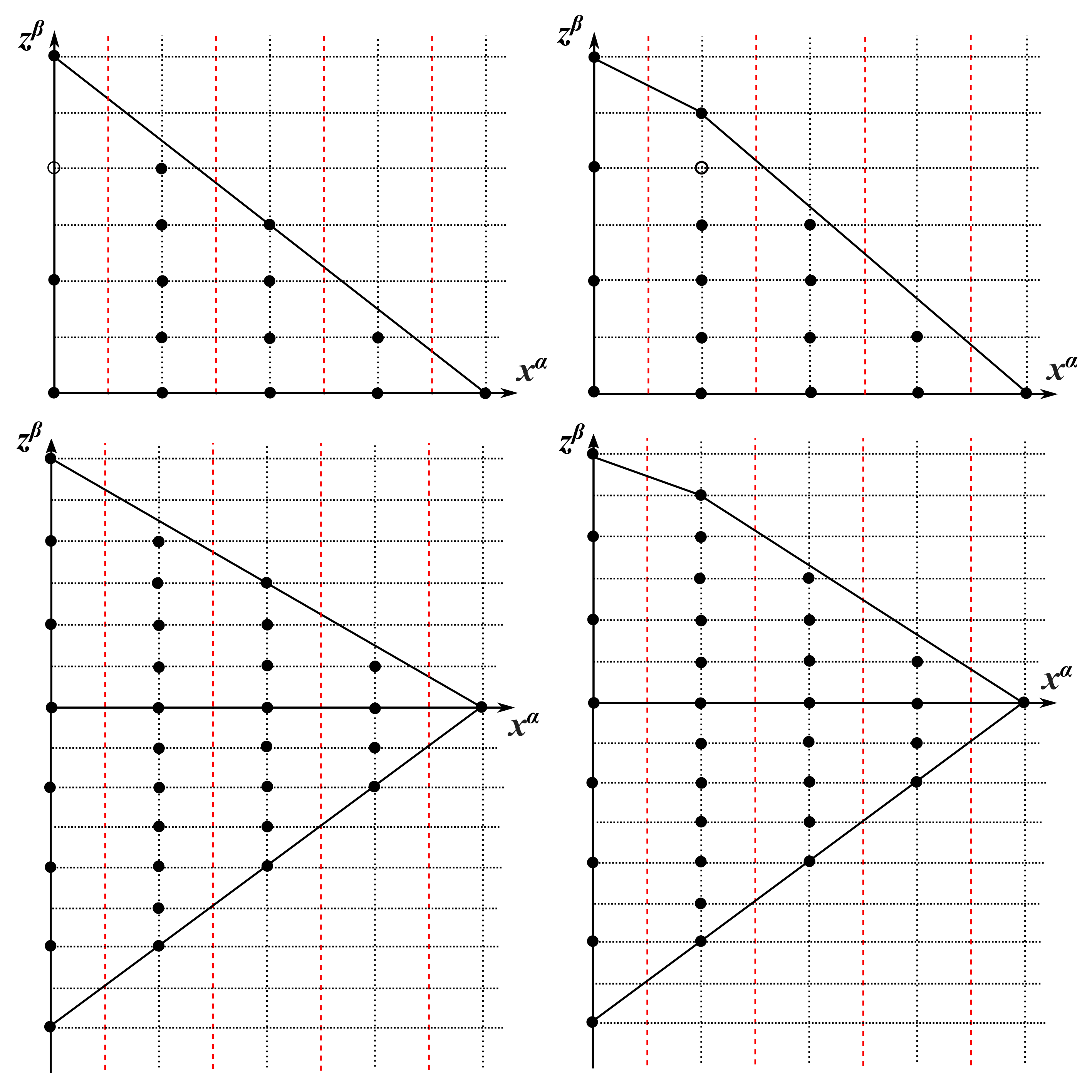}
\caption[Newton polygon for $D_N$ type Argyres-Douglas theories]{A collection of Newton polygon for examples of SCFT with $\mfg = D_N$. Each black dot represents a monomial in SW curve in the form of $x^{\alpha} z^{\beta}$; except that for the $x^0$ axis, each term represents the Pfaffian ${\tilde \phi}$, so we shall read it as $\sqrt{z^{\beta}}$. The white dots mean that the monomials are omitted. The upper left diagram gives $D_4^{(4)}[3]$ theory, while the upper right diagram gives $D_4^{(6)}[5]$. The two lower diagrams represent the same irregular puncture, but with an additional regular puncture ($e.g.$ maximal) at south pole. We denote them as $(D_4^{(4)}[3], F)$ and $(D_4^{(6)}[5], F)$ theory, respectively.}
\label{DNnewton}
\end{figure}

$\bullet$ $\mfg = D_N$. There are two types of Newton polygon, associated with Higgs field
\be
\Phi \sim \frac{T}{z^{2+\frac{k}{N}}}, \ \ \ \ \ \ \Phi \sim \frac{T}{z^{2+\frac{k}{2N-2}}},
\ee
We denote two types and their SW curves at conformal point as
\be
& D_N^{(N)}[k]: \ \ \ x^{2N} + z^{2k} = 0, \\[0.5em]
& D_N^{(2N-2)}[k]: \ \ \ x^{2N} + x^2 z^{k} = 0.
\ee
The full curve away from conformal point, and with various couplings turned on, is given by \eqref{D_SW curve}. In figure \ref{DNnewton}, we list examples of such Newton polygon. 

\bigskip

\begin{figure}[htbp]
\centering
\begin{adjustwidth}{-0.3cm}{}
\includegraphics[width=16cm]{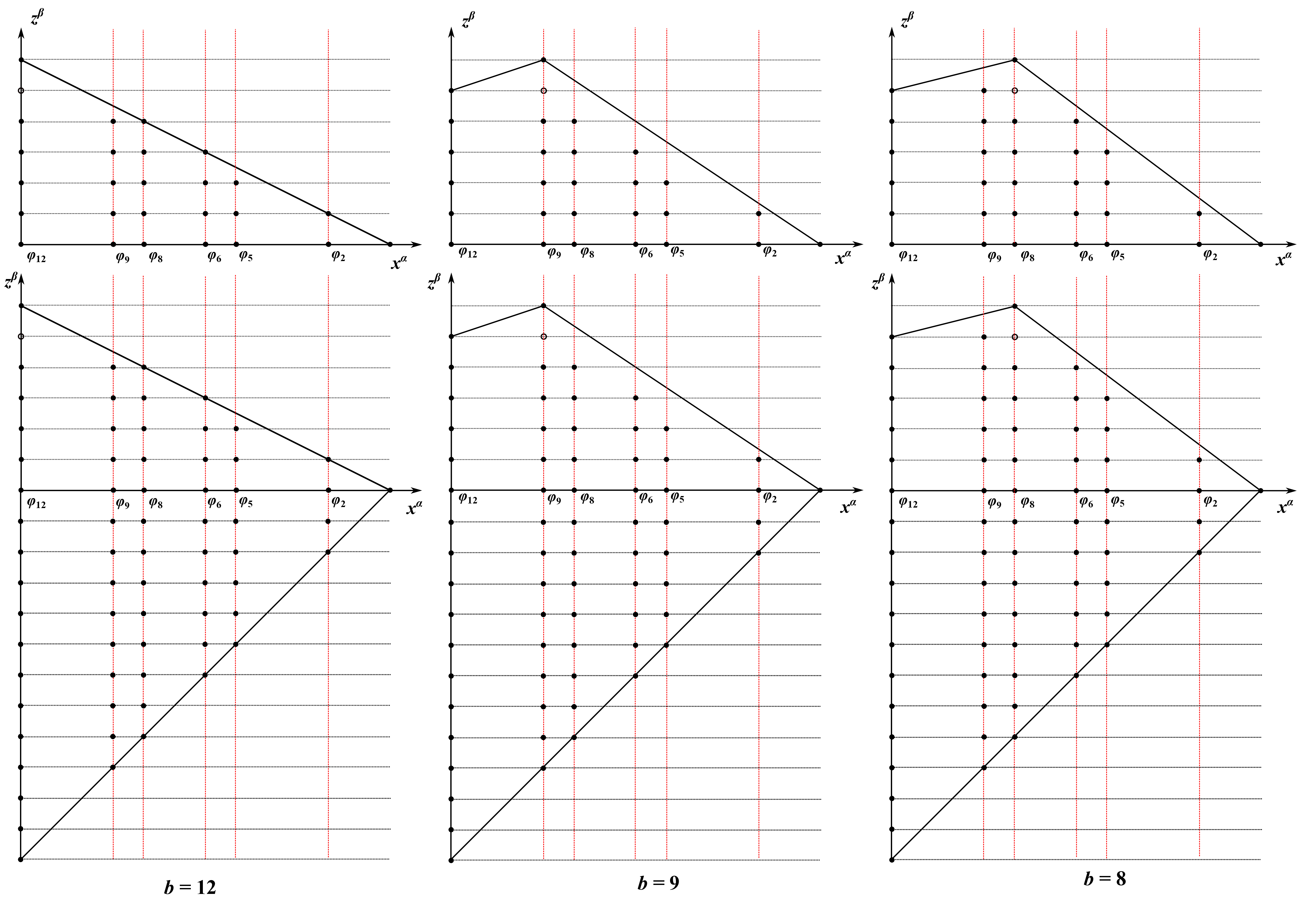}
\end{adjustwidth}
\caption[Newton polygon for $E_6$ type Argyres-Douglas theories]{A collection of Newton polygons for examples of SCFT with $\mfg = E_6$. Each black dot represents a monomial in SW curve in the three fold form. The white dots mean that the monomials are omitted. The upper left diagram gives $b= 12$, $k = 6$ theory, while the upper middle diagram gives $b = 9$, $k = 6$ theory and the upper right gives $b = 8$, $k = 6$ theory. The three lower diagrams represent the same irregular puncture, but with an additional regular puncture ($e.g.$ maximal) at south pole. }
\label{E6newton}
\end{figure}

$\bullet$ $\mfg = E_{6,7,8}$. We can consider Newton polygon from the 3-fold singularities. In this way we may draw the independent differentials unambiguously. We give the case for $E_6$ with $b = 8, 9, 12$ in figure \ref{E6newton}. The other two exceptional algebras are similar.

\section{Mapping to a punctured Riemann surface}\label{Sec: mappToRiemannSphere}

As we mentioned in section \ref{Sec: Intro}, to generate S-duality we construct an auxiliary Riemann sphere $\Sigma'$ from the initial Riemann sphere $\Sigma$ with irregular punctures. We now describe the rules. The motivation for such construction comes from 3d mirror in class $\cS$ theory \cite{Benini:2010uu, Xie:2016uqq, Nanopoulos:2010bv}. To recapitulate the idea, from 3d mirror perspective we may interpret the Gaiotto duality as splitting out the quiver theories with three quiver legs. Each quiver leg carries a corresponding flavor symmetry on the Coulomb branch and can be gauged. The 3d mirror of $A_{N-1}$ type Argyres-Douglas theories are know and they are also constructed out of quiver legs. We then regard each quiver leg as a ``marked points'' on the Riemann sphere $\Sigma'$. Unlike the class $\cS$ counterpart, now there will be more types of marked points with different rank.

Recall our setup is that the initial Riemann sphere $\Sigma$ is given by one irregular singularity of class $(k,b)$, with coefficient satisfying
\be
T_{\ell} = T_{\ell-1} = \dots = T_3 = T_2, \ \ \ \ \ T_1\ \text{arbitrary}, \ \ \ \ \ \ell = k + b + 1,
\ee
possibly with a regular puncture $Q$. We denote it as $\pbra{\Rnum{3}_{k,b}^{ \{ \mfl_i \}_{i=1}^{\ell} },\ Q}$, where $\mfl_i$ is the Levi subalgebra for the semisimple element $T_i$. We now describe the construction of $\Sigma'$.

\vspace{8pt}

$\bullet$ {\bf Lie algebra} $\mfg = A_{N-1}$. A generic matrix looks like
\begin{equation}
T_i={\rm diag} \pbra{\underbrace{a_1\Xi_b,\ldots, a_1\Xi_b}_{r_1},\ldots,\underbrace{a_s \Xi_b,\ldots, a_s \Xi_b}_{r_s},\underbrace{0,\ldots,0}_{N-(\sum r_j) b} }, \ \ \ \ \ 2 \leq i \leq \ell,
\end{equation}
The theory is represented by a sphere with one red marked point (denoted as a cross \textcolor{red}{$\times$}) representing regular singularity; one blue marked point (denoted as a square \legendbox{blue}) representing $0$'s in $T_i$, which is further associated with a Young tableaux with size $N-(\sum r_j)b$ to specify its partition in $T_1$. There are $s$ black marked points (denoted as black dots $\bullet$) with size $r_j$, $j=1,\ldots, s$ and each marked point carrying a Young tableaux of size $r_j$.  Notice that there are $s-1$ exact marginal deformations which is the same as the dimension of complex structure moduli of punctured sphere. 

There are two exceptions: if $b=1$, the blue marked point is just the same as the black marked point. If $k=1, b=1$, the red marked point is the same as the black marked point as well \cite{Xie:2017vaf}.

\vspace{8pt}

$\bullet$ {\bf Lie algebra} $\mfg = D_{N}$. We have the representative of Cartan subalgebra as \eqref{so(2N)matrix} and when $b$ is odd,
\be
Z = {\rm diag}(\underbrace{a_1 \Xi_b,\ldots, a_1 \Xi_b}_{r_1},\ldots,\underbrace{a_s \Xi_b,\ldots, a_s \Xi_b}_{r_s},\underbrace{0,\ldots,0}_{N-(\sum r_j) b}),
\ee
while when $b$ is even, 
\be
Z = {\rm diag}(\underbrace{a_1 \Xi'_{b/2},\ldots, a_1 \Xi'_{b/2}}_{r_1},\ldots,\underbrace{a_s \Xi'_{b/2},\ldots, a_s \Xi'_{b/2}}_{r_s},\underbrace{0,\ldots,0}_{N-(\sum r_j) b / 2}).
\ee
The theory is represented by a Riemann sphere with one red cross representing regular singularity, one blue puncture representing $0$'s in $T_i$; we also have a D-partition of $2\bbra{N-(\sum r_j)b}$ to specify further partition in $T_1$. Moreover, there are $s$ black marked point with size $r_j$, $j=1,\ldots, s$ and each marked point carrying a Young tableaux of size $r_j$ (no requirement on the parity of entries). 

\vspace{8pt}
$\bullet$ {\bf Lie algebra} $\mfg=E_{6,7,8}$:  Let us start with the case $b=1$, and the irregular puncture is labelled by Levi-subalgebra $L_l=\ldots=L_2=\mfl$ and a trivial Levi-subalgebra $L_1$. We note that there is at most one non-$A$ type Lie algebra for $\mfl$: $\mfl=A_{i_1}+\ldots+A_{i_k}+\mfh$; Let's define $a={\rm rank}(\mfg)-{\rm rank}(\mfh)-\sum_{j=1}^k (i_j+1)$, we have the following situations:
\begin{itemize}
\item $a\geq 0$: we have $k$ black punctures with flavor symmetry $U(i_j+1),~j=1,\ldots,k$, and $a$ more black marked point with $U(1)$ flavor symmetry; we have a blue puncture with $H$ favor symmetry ($\mfh = {\rm Lie}(H)$), and finally a red puncture representing the regular singularity.
\item $a<0$: When there is a $2A_1$ factor in $\mfl$, we regard it as $D_2$ group and use a blue puncture for it; when the rank of $\mfl$ is ${\rm rank}(\mfg)-1$, we put all $A$-type factor of $\mfl$ in a single black marked point.
\end{itemize}

The $b\neq 1$ case can be worked out similarly. 

\subsection{AD matter and S-duality}
We now discuss in more detail about the AD matter for $b=1$. Recall that the number of exact marginal deformations is equal to $r-r_{\mfl}-1$, where $r = {\rm rank}(\mfg)$, and $r_\mfl = {\rm rank}(\mfl)$. The AD matter is then given by the Levi subalgebra with rank $r-1$. We can list all the possible Levi subalgebra for AD matters in table \ref{table:AD matter}.
\begin{table}
\centering
\begin{tabular}{|c|c|}
\hline
Lie algebra $\mfg$ & Levi subalgebra associated to AD matter\\
\hline
$A_{N-1}$ & $A_{n}+A_{m},~~(n+1)+(m+1)=N$  \\
$D_N$ & $A_{n}+D_m,~~n+1+m=N$  \\
$E_6$ & $D_5,~A_5,~A_4+A_1,~2A_2+A_1$ \\
$E_7$ & $E_6,~D_6,~D_5+A_1,~A_6,~A_5+A_1,~A_4+A_2,~A_3+A_2+A_1$\\ 
$E_8$ & $E_7,~E_6+A_1, ~D_7,~D_5+A_2,~A_7,~A_6+A_1, ~A_4+A_3, ~A_4+A_2+A_1$\\ \hline
\end{tabular}
\caption{\label{table:AD matter}Possible Levi subalgebra for $T_{\ell}$ that corresponds to AD matter without exact marginal deformations.}
\end{table}

\begin{figure}[htbp]
\centering
\begin{adjustwidth}{5cm}{}
\includegraphics[width=4cm]{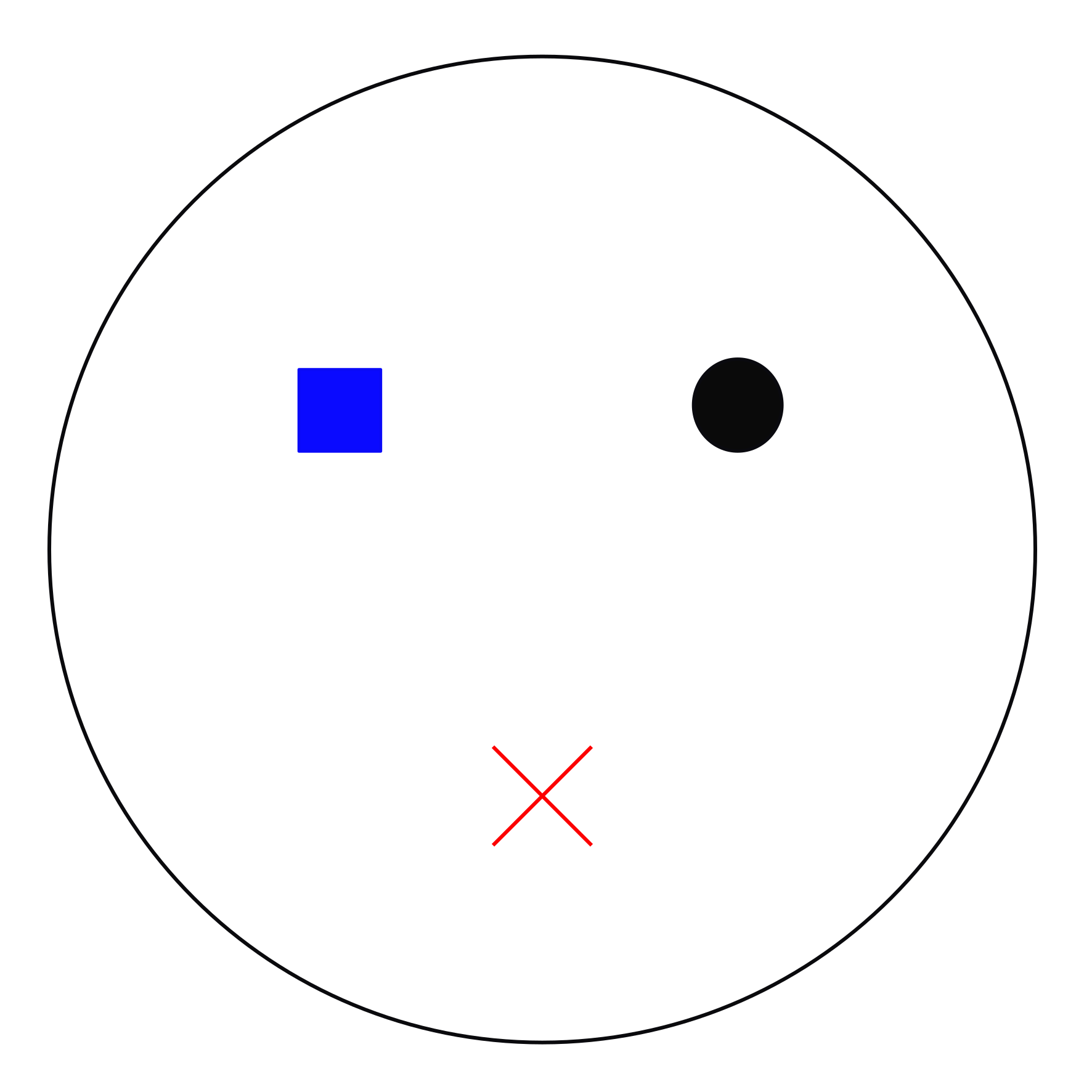}
\end{adjustwidth}
\caption[AD matter]{An example of Argyres-Douglas matter of type $\mfg$. The theory has no exact marginal deformations, and in the meantime the punctured Riemann sphere $\Sigma'$ has no complex structure moduli.}
\label{ADmatter}
\end{figure}

\vspace{8pt}
{\bf S-duality frames}. With the auxiliary Riemann sphere $\Sigma'$, we conjecture that the S-duality frame is given by different degeneration limit of $\Sigma'$; the quiver theory is given by gauge groups connecting Argyres-Douglas matter without exact marginal deformations. For AD theories of type $\mfg$, the AD matter is given by three punctured sphere $\Sigma'$: one red cross, one blue square and one black dot. The rank of black dot plus the rank of blue square should equal to the rank of the red cross.  See figure \ref{ADmatter} for an illustration. Each marked points carry a flavor symmetry. Their flavor central charge is given by \cite{Xie:2013jc, Xie:2017vaf}
\be
k_{G}^{\rm red} = h^{\vee} - \frac{b}{k+b}, \ \ \ \ k_{G}^{\rm black / blue} = h^{\vee} + \frac{b}{k+b},
\ee
where $h^{\vee}$ is the dual Coxeter number of $G$. This constraints the configuration such that one can only connect black dot and red cross, or blue square with red cross to cancel one-loop beta function.

\subsection{Central charges}\label{subsec: central charges}
The central charges $a$ and $c$ can be computed as follows \cite{Shapere:2008zf, Xie:2013jc}:
\be
2a - c = \frac{1}{4} \sum \pbra{2[u_i] - 1}, \ \ \ \ a - c = -\frac{1}{24} \dim_{\mathbb{H}} {\rm Higgs},
\label{centralCharge2}
\ee
This formula is valid for the theory admits a Lagrangian 3d mirror. We  know how to compute the Coulomb branch spectrum, and so the only remaining piece is to the dimension of Higgs branch which can be read from the mirror. 

For theories with $b=1$, the local contribution to the Higgs branch dimension with flavor symmetry $G$ for red marked point is
\begin{equation}
\dim^{\rm red}_{\mathbb{H}} {\rm Higgs} = {1\over 2}(\dim G- {\rm rank}(G)),
\end{equation}
while for blue and black marked point, we have
\begin{equation}
\dim^{\rm blue / black}_{\mathbb{H}} {\rm Higgs} = {1\over 2}(\dim G + {\rm rank}(G)).
\end{equation}
The total contribution to the Higgs branch is the summation of them, except that for $A_{N-1}$, we need to subtract one.

\section{S-duality for $D_N$ theory}\label{Sec: S-duality-DN}

\subsection{Class $(k, 1)$}\label{sec: S-duality-(k,1)}

In this section we first consider $\mfg = D_N$, and the irregular singularity we take will be 
\be
\Phi = \frac{T_{\ell}}{z^{\ell}} + \frac{T_{\ell-1}}{z^{\ell-1}} + \dots + \frac{T_1}{z} + T_{\rm reg},
\label{MaximalIrreg}
\ee
where $T_{\rm reg}$ is the regular terms. This amounts to take $k = \ell - 2$, $b = 1$\footnote{Careful readers may wonder whether $n_1 = 1$ comes from $D_N^{(N)}[k']$ or $D_N^{(2N-2)}[k']$, as their relevant matrices in section \ref{subsec: field data} are different. However, in the case $n_1 = 1$, leaving two diagonal entries to be zero has the same Levi subgroup ($SO(2)$) as that of leaving it to be diag$(a, -a)$, which is $U(1)$. So two cases actually coincide.}. We settle the questions raised in previous sections: (\rnum{1}) we show which choices of $T_i$'s give legitimate deformation for SCFT; (\rnum{2}) we illustrate how to count graded Coulomb branch spectrum and (\rnum{3}) how to obtain its S-dual theory.  In dealing with these questions, we first utilize the case $D_3 \simeq A_3$, where we already know the results \cite{Xie:2017vaf}.

\subsubsection{Coulomb branch spectrum}

Recall that in section \ref{subsubsec: irregDegeneration}, one maps each semisimple orbit $\cO_{T_i}$ to a nilpotent orbit with the same dimension. We may use the recipe of section \ref{subsubsec: regular puncture} to calculate the Coulomb branch spectrum. Let us see how this works. 

\vspace{6pt}
\textit{Example 1: non-degenerating $D_4$ theory of class $(1,1)$.} As we have $\ell = 3$, there are three regular punctures whose labels are $\bbra{1^8}$. For such a maximal puncture, the pole structure for the differential is $\{p_2, p_4, p_6; {\tilde p}\} = \{1,3,5; 3 \}$ and there are no relations. Then, the total contributions to the moduli are $\{ d_2, d_4, d_6; {\tilde d}_4 \} = \{ 0, 2, 4; 2 \}$. This is consistent with the Newton polygon of $D_4^{(4)}[4]$.

\vspace{6pt}
\textit{Example 2: degenerating $D_4$ theory of class $(1,1)$.} In this example we take $T_3$ and $T_2$ to be labelled by Levi subalgebra of type $[1, 1, 1; 2]$, while $T_1$ is still of type $[1,1,1,1; 0]$. For the former, we see that it is the same as the Levi subalgebra $[1,1,1,1;0]$. Then we are back to the previous example. This is indeed the same spectrum as indicated by Newton polygon of $D_4^{(6)}[6]$.

\vspace{6pt}
\textit{Example 3: degenerating $D_3$ theory of class $(1,1)$.} We take $T_3$ and $T_2$ to have Levi subalgebra of type $[2,1;0]$, giving a regular puncture labelled by Nahm partition $[2,2,1,1]$. In terms of Nahm partition for $A_3$, they are equivalent to $[2,1,1]$. We also take $T_1$ to be maximal. From $A_3$, the algorithm in \cite{Xie:2012hs} determines the set of Coulomb branch operators to be $\{3/2 \}$. In the language of $D_3$, the partition $[2,2,1,1]$ gives the pole structure $\{1, 2; 2 \}$, while the maximal puncture has pole structure $\{1,3;2\}$; both of them have no constraints. Then, $\{d_2, d_4; {\tilde d}_3 \} = \{ 0,0; 1\}$, giving a Coulomb branch moduli with dimension $3/2$. So we see two approaches agree.

\subsubsection{Constraints on coefficient matrices}

As we mentioned before, not every choice of $\{T_{\ell}, T_{\ell - 1}, \dots, T_1 \}$ is allowed for the SCFT to exist. Those which are allowed must have $T_{\ell} = \dots = T_2$, and $T_1$ is a further partition of them. In this section we show why this is so.

The idea of our approach is that, the total number of exact marginal deformations shall not exceed the maximum determined by the leading matrix $T_{\ell}$. We examine it on a case by case basis.

\vspace{6pt}
\textit{$D_3$}. In this case we may directly use the results of \cite{Xie:2017vaf}. Our claim holds. 

\vspace{6pt}
\textit{$D_4$}. First of all we list the correspondence between Nahm label of the regular puncture and the Levi subalgebra in table \ref{D4Induction}. The regular puncture data are taken from \cite{Chacaltana:2011ze}. There are several remarks. For very even partitions, we have two matrix representation for two nilpotent orbits; they cannot be related by Weyl group actions\footnote{The Weyl group acts on entries of $Z = {\rm diag}(a_1,a_2,\dots,a_N)$ by permuting them or simultaneously flip signs of even number of elements.}. Moreover, we also see that there are multiple coefficient matrices sharing the same Levi subalgebra; $e.g.$ $[4;0]$ and $[1;6]$. Therefore, we do need regular puncture and Nahm label to distinguish them. Finally, we need to exclude orbit which is itself distinguished in $D_4$, as their Levi subalgebra is maximal, meaning we have zero matrix. 
\begin{table}
\centering
\begin{adjustwidth}{-1.7cm}{}
\begin{tabular}{|c|c|c|c|c|c|}
\hline
Levi subalgebra & matrix $Z$ & regular puncture & pole structure & constraints & flavor symmetry \\ \hline
$\bbra{1,1,1,1;0}$ & $\pbra{\begin{array}{cccc} a & & & \\ & b & &\\ & & c &\\ & & & d \end{array}}$ & $[1^8]$ & $\{1,3,5;3\}$ & $ - $ & $- $ \\ \hline
$\bbra{2,1,1;0}$ & $\pbra{\begin{array}{cccc} a & & & \\ & a & &\\ & & b &\\ & & & c \end{array}}$ & $\bbra{2^2, 1^4}$ & $\{1,3,4;3\} $ & $ - $ & $SU(2)$ \\ \hline
$\bbra{1,1;4}$ & $\pbra{\begin{array}{cccc} 0 & & & \\ & 0 & &\\ & & b &\\ & & & c \end{array}}$ & $\bbra{3, 1^5}$ & $\{1,3,4;2\}$ & $ - $ & $SO(4)$ \\ \hline
$\bbra{2,2;0}$ & $\pbra{\begin{array}{cccc} a & & & \\ & a & &\\ & & b &\\ & & & \pm b \end{array}}$ & $\bbra{2^4}^{\Rnum{1}, \Rnum{2}}$ & $\{1,3,4;3\}$  & $ c_3^{(4)} \pm 2{\tilde c}_3 = 0$ & $SU(2) \times SU(2)$ \\ \hline
$\bbra{3,1;0}$ & $\pbra{\begin{array}{cccc} a & & & \\ & a & &\\ & & a &\\ & & & b \end{array}}$ & $\bbra{3,3,1,1}$ & $\{1,2,4;2\}$ & $ c^{(6)}_4 = (a_3)^2 $ & $SU(3)$\\  \hline
$\bbra{2; 4}$ & $\pbra{\begin{array}{cccc} a & & & \\ & a & &\\ & & 0 &\\ & & & 0 \end{array}}$ & $\bbra{3,2,2,1}^*$ & $\{1,2,4;2\}$ & $ - $  & $SU(2) \times SO(4)$ \\  \hline
$\bbra{1;6}$ & $\pbra{\begin{array}{cccc} 0 & & & \\ & 0 & &\\ & & 0 &\\ & & & a \end{array}}$ & $\bbra{5,1,1,1}$& $\{1,2,2;1\}$ & $ - $ & $SO(6)$ \\ \hline
$\bbra{4;0}$ & $\pbra{\begin{array}{cccc} a & & & \\ & a & &\\ & & a &\\ & & & \pm a \end{array}}$ & $\bbra{4,4}^{\Rnum{1}, \Rnum{2}}$& $\{1,2,3;2\}$ & \begin{tabular}[x]{@{}c@{}}$c_2^{(4)} \pm 2{\tilde c}_2 = (c_1^{(2)})^2 / 4, $ \\[0.3em] $c_3^{(6)} = \mp {\tilde c}_2 c_1^{(2)}$ \end{tabular} & $SU(4)$ \\ \hline
\end{tabular}
\end{adjustwidth}
\caption{\label{D4Induction}Association of a nilpotent orbit to a Levi subalgebra for $D_4$. Here $Z$ follows the convention in \eqref{so(2N)matrix}. The partition $[3,2,2,1]$ is non-special, and we use the * to mark it. In the last column we list the semisimple part of maximal possible flavor symmetry. The partition $[5,3]$ and $[7,1]$ are excluded; the first one is non-principal in $\mathfrak{so}(8)$ while the second gives trivial zero matrix.}
\end{table}

Now consider $\ell = 3$, and $T_3$ has the Levi subalgebra $[1,1;4]$, with one exact marginal deformation. One can further partition it into the orbit with Levi subalgebra $[2,1,1;0]$ and $[1,1,1,1;0]$. If we pick $T_2$ to be $[2,1,1;0]$, then no matter what we choose for $T_1$, there will be two dimension $2$ operators, this is a contradiction. So $T_2$ must be equal to $T_3$. 

The second example has $\ell = 3$, but $T_3$ now is associated with $[3,3,1,1]$. This puncture has a relation $c_4^{(6)} = (a^{(3)})^2$, so we remove one moduli from $\phi_6$, and add one moduli to $\phi_4$. The possible subpartitions are $[2^2,1^4]$, $[1^8]$. If $T_2 \neq T_3$ then there will be two exact marginal deformations from $\phi_4$ and ${\tilde \phi}$. This is a contradiction, so we must have $T_2 = T_3$.

As a third example, we may take $\ell = 4$ and $T_4$ corresponding to the regular punctures $[2^4]$, whose pole structure is $\{ 1,3,4; 3 \}$, with one constraints $c^{(4)}_3 \pm 2 {\tilde c}_3 = 0$. Then each of the local contribution to Coulomb moduli is $\{d_2, d_4, d_6; d_3\} = \{ 1,2,4; 3 \}$. From the matrix representation we know there is one exact marginal coupling. If we pick $T_3$ to be $[2^2, 1^4]$, then by simple calculation we see that there are two dimension $2$ operators. So we have to pick $T_3 = T_4$. Similarly, we have to pick $T_2 = T_3 = T_4$. Therefore, we again conclude that we must have $T_4 = T_3 = T_2$, while $T_1$ can be arbitrary.

\vspace{6pt}
\textit{$D_5$}. We now check the constraints for the Lie algebra $D_5$. To begin with, we list the type of Levi-subgroup and its associated regular puncture in table \ref{D5Induction}. Now we examine the constraints on coefficient matrices. We first take $\ell = 3$, and pick $T_3$ to be of the type $[3,2;0]$ whose associated regular puncture is $[3,3,2,2]$. There is a constraint $c_6^{(8)} = \pbra{c_3^{(4)}}^2 / 4$, so the local contribution to Coulomb branch is $\{d_2, d_4, d_6, d_8; d_5\} = \{1,3,4,5;3 \}$. If we take $T_2$ to be $e.g.$, $[2^4, 1^2]$, then the moduli from ${\tilde \phi}$ contribute one more exact marginal deformations other than $\phi_4$, which is a contradiction. Therefore, we again conclude that we must have $T_3 = T_2$, with arbitrary subpartition $T_1$.


\begin{table}
\centering
\begin{adjustwidth}{-2.2cm}{}
\begin{tabular}{|c|c|c|c|c|c|}
\hline
Levi subalgebra & matrix $Z$ & regular puncture & pole structure & constraints &flavor symmetry\\
\hline
$\bbra{1,1,1,1,1;0}$ & diag$(a,b,c,d,e)$ & $[1^{10}]$ & $\{1,3,5,7;4\}$  & $-$ & $-$\\ \hline
$\bbra{2,1,1,1;0}$ & diag$(a,a,b,c,d)$ & $\bbra{2^2, 1^6}$ & $\{1,3,5,6;4\}$ & $-$  & $SU(2)$ \\ \hline
$\bbra{1,1,1;4}$ & diag$(0,0,a,b,c)$ & $\bbra{3, 1^7}$ & $\{1,3,5,6;3\}$ & $-$ & $SO(4)$ \\ \hline
$\bbra{2,2,1;0}$ & diag$(a,a,b,b,c)$ & $\bbra{2^4,1^2}$ & $\{1,3,4,6;4\}$  & $-$  & $SU(2) \times SU(2)$ \\ \hline
$\bbra{3,1,1;0}$ & diag$(a,a,a,b,c)$ & $\bbra{3^2,1^4}$ & $\{1,3,4,6;3\}$ & $c_6^{(8)} = \pbra{a^{(4)}}^2$ & $SU(3)$ \\  \hline
$\bbra{2,1;4}$ & diag$(a,a,b,0,0)$ & $\bbra{3,2^2,1^3}^*$ & $\{1,3,4,6;3\}$ & $-$ & $SU(2) \times SO(4)$ \\  \hline
$\bbra{3,2;0}$ & diag$(a,a,a,b,b)$ & $\bbra{3,3,2,2}$& $\{1,3,4,6;3\}$ & $c_6^{(8)} = \pbra{c_3^{(4)}}^2 / 4$ & $SU(3) \times SU(2)$ \\ \hline
$\bbra{3;4}$ & diag$(0,0,a,a,a)$ & $\bbra{3,3,3,1}$& $\{1,2,4,5;3\}$ & $-$ & $SU(3) \times SO(4)$ \\ \hline
$\bbra{1,1;6}$ & diag$(0,0,0,a,b)$ & $\bbra{5,1^5}$& $\{1,3,4,4;2\}$ & $-$ & $SO(6)$ \\ \hline
$\bbra{4,1;0}$ & diag$(a,a,a,a,b)$ & $\bbra{4,4,1,1}$& $\{1,2,4,5;3\}$ & \begin{tabular}[x]{@{}c@{}}$c_4^{(6)} = (a^{(3)})^2, $ \\ $c_5^{(8)} = 2 a^{(3)} {\tilde c}_{3}$ \end{tabular} & $SU(4)$ \\ \hline
$\bbra{2;6}$ & diag$(0,0,0,a,a)$ & $\bbra{5,2,2,1}^*$& $\{1,2,4,4;2\}$ & $-$ & $SU(2) \times SO(6)$ \\ \hline
$\bbra{5;0}$ & diag$(a,a,a,a,a)$ & $\bbra{5,5}$& $\{1,2,3,4;2\}$ & \begin{tabular}[x]{@{}c@{}}${c'}_2^{(4)} \equiv c_2^{(4)} - (c_1^{(2)})^2 / 4, $ \\ $c_3^{(6)} = c_1^{(2)} {c'}_2^{(4)} / 2$, \\ $c_4^{(8)} = \pbra{{c'}_2^{(4)}}^2$  \end{tabular} & $SU(5)$ \\ \hline
$\bbra{1;8}$ & diag$(0,0,0,0,a)$ & $\bbra{7,1,1,1}$& $\{1,2,2,2;1\}$ & $-$ & $SO(8)$ \\ \hline
\end{tabular}
\end{adjustwidth}
\caption{\label{D5Induction}Association of a nilpotent orbit to a Levi subalgebra for $D_5$. $Z$ is the convention taken in \eqref{so(2N)matrix}. the Nahm partition $[5,3,1,1]$, $[7,3]$ and $[9,1]$ are excluded. }
\end{table}

\vspace{6pt}
Based on the above examples and analogous test for other examples, we are now ready to make a conjecture about the classification of SCFT for degenerating irregular singularities:

$\bullet$ {\bf Conjecture.} In order for the maximal irregular singularity \eqref{MaximalIrreg} of type $D$ to define a viable SCFT in four dimensions, we must have $T_{\ell} = T_{\ell - 1} = \dots = T_2\ (\ell \geq 3)$, while $T_1$ can be arbitrary subpartition of $T_i$.

We emphasize at last that when $\ell = 2$, the scaling for $x$ in SW curve is zero. Therefore, we may have arbitrary partition $T_2$ and $T_1$, so that $\cO_{T_2} \subset \cO_{T_1}$.

\subsubsection{Generating S-duality frame}\label{subsec: S-duality}

With the above ingredients in hand, we are now ready to present an algorithm that generates S-duality for various Argyres-Douglas theories of $D$ type. This may subject to various consistency checks. For example, the collection of Coulomb branch spectrum should match on both sides; the conformal anomaly coefficients (central charges) $(a, c)$ should be identical. The latter may be computed from \eqref{centralCharge2}.

\bigskip

{\bf Duality at large $k$.} For such theories with $\ell = k + 2$, if we take the Levi subalgebra of $T_{\ell} = \dots = T_2$ to be of type $\bbra{r_1, \dots r_n; {\tilde r}}$, then there are $n - 1$ exact marginal couplings. For each $r_i$, $1 \leq i \leq n$ as well as ${\tilde r}$ there is further partition of it in $T_1$: 
\be
& \bbra{r_i; 0} \rightarrow \bbra{m_1^{(i)}, \dots, m_{s_i}^{(i)}}, \ \ \ \ \sum_{j=1}^{s_i} m^{(i)}_j = r_i, \\[0.5em]
& \bbra{0; {\widetilde r}} \rightarrow \bbra{{\widetilde m}_1, \dots, {\widetilde m}_{s}; {\widetilde r}'}, \ \ \ \ \ \ 2\sum_{j=1}^s {\widetilde m}_j + {\widetilde r}'= {\widetilde r}.
\ee

The Argyres-Douglas matter is given by $Z$ in \eqref{so(2N)matrix} of the leading coefficient matrix $T_{\ell}$:
\be
Z_1 = \pbra{\begin{array}{cccccc} a & & & & &\\ & \ddots & & & & \\  & & a & & & \\ & & & 0 & &\\ & & & & \ddots & \\ & & & & & 0 \end{array}}.
\ee
They are given by a three-punctured sphere with one black dot of type $[r_1, \dots, r_m]$ with $\sum r_i = n$ for $n$ being the number of $a$'s, one blue square which is degeneration of $[0; 2N-2n]$ and one red cross. However, we note the exception when $N = 2$: in this case, since the theory is in fact given by two copies of $SU(2)$ group, so the Argyres-Douglas matter is represented differently. We will see this momentarily.

\begin{figure}[htbp]
\centering
\includegraphics[width=15cm]{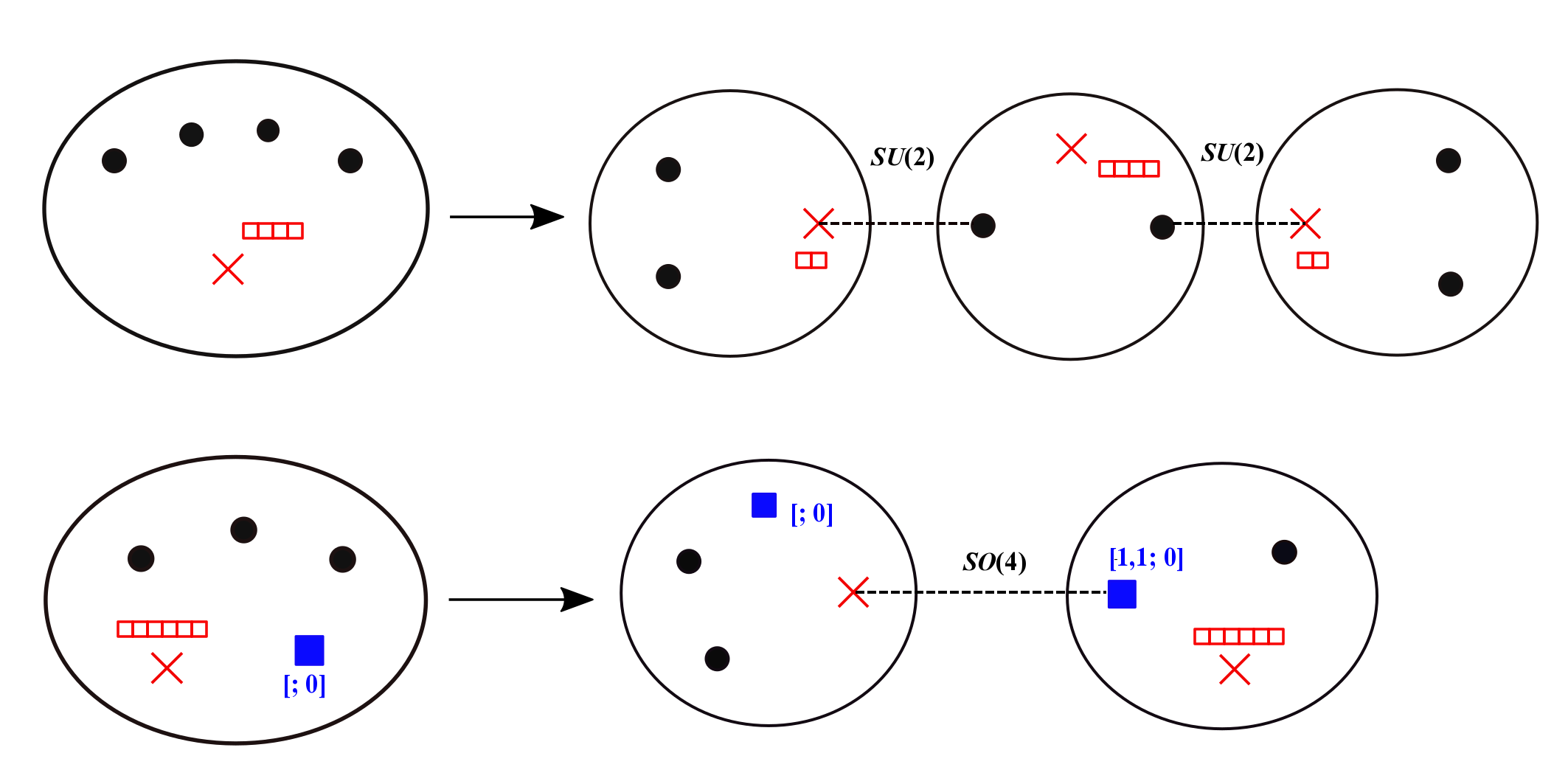}
\caption[Compare A3 and D3 S duality]{Comparison of S duality from $A_3$ (upper half) and $D_3$ (lower half) perspective. From the $A_3$ point of view, each black dot is given by $[1]$, and the new red marked point after degeneration is given by $SU(2)$ puncture $[1,1]$. The two theories on the left and right sides are $(A_1, D_{2k+2})$ theory, which is given by irregular puncture whose $T_{k+2}, \dots, T_1 = [1,1]$, and one regular puncture. The theory in the middle is $(III_{k,1}^{[2,2]^{\times (k+1)},[1,1,1,1]}, F)$ theory. Here $F$ denotes maximal puncture. From the $D_3$ point of view, two $(A_1, D_{2k+2})$ theories combine together and form a $D_2$ type theory. The theory on the right is $(III_{k,1}^{[1;4]^{\times (k+1)},[1^3; 0]}, F)$. }
\label{D3Sduality}
\end{figure}

\vspace{8pt}
\textit{Example 1: $D_3 \simeq A_3$}. This case can be analyzed from either Lie algebra perspective. Let us take $T_{\ell}$ to be regular semisimple. We also add a regular puncture labelled by a red cross. One duality frame is given in the first line of figure \ref{D3Sduality}. 

We can perform various checks for this duality. First of all, $(A_1, D_{2k+2})$ theory has Coulomb branch spectrum
\be
\Delta(\cO_i) = 2 - \frac{i}{k+1}, \ \ \ \ i = 1,2,\dots, k.
\ee
For the middle theory, for simplicity we focus on the case where the regular puncture is maximal, but replacing it with any regular puncture does not affect the result. The Coulomb branch spectrum for this theory is
\be
\Delta(\cO) = & \frac{2k+3}{k+1}, \ \frac{2k+4}{k+1}, \dots, \frac{4k+4}{k+1}, \\[0.5em]
                     & \frac{2k+3}{k+1}, \ \frac{2k+4}{k+1}, \dots, \frac{3k+3}{k+1}, \\[0.5em]
                     &\ \frac{k+2}{k+1}, \ \ \frac{k+3}{k+1}, \dots, \ \frac{2k+2}{k+1}.
\ee
We see that along with two $SU(2)$ gauge groups, the combined Coulomb branch spectrum nicely reproduces all the operators of the initial theory. Secondly, we may calculate the central charge. We know the central charges for $(A_1, D_{2k+2})$ theory are
\be
a = \frac{k}{2} + \frac{1}{12}, \ \ \ \ c =  \frac{k}{2} + \frac{1}{6}.
\ee
The central charges for the initial theory are, with the help of \eqref{centralCharge2} and three dimensional mirror, 
\be
a = 5k + \frac{55}{8}, \ \ \ \ c = 5k + \frac{58}{8}.
\ee
The central charges for the middle theory are obtained similarly:
\be
a = 4k + \frac{131}{24}, \ \ \ \ c = 4k + \frac{142}{24}.
\ee
We find that 
\be
& a_{(I_{4,4k}, F)} = 2 a^V_{SU(2)} + 2 a_{(A_1, D_{2k+2})} + a_{(III_{k,1}^{[2,2]^{\times (k+1)},[1,1,1,1]}, F)}, \\[0.5em]
& c_{(I_{4,4k}, F)} = 2 c^V_{SU(2)} + 2 c_{(A_1, D_{2k+2})} + c_{(III_{k,1}^{[2,2]^{\times (k+1)},[1,1,1,1]}, F)}.
\ee
Here $a^V$ and $c^V$ denote the contribution from vector multiplet. Finally, we may check the flavor central charge and beta functions for the gauge group. The flavor central charge for $SU(2)$ symmetry of $(A_1, D_{2k+2})$ theory is $(2k+1) / (k+1)$. The middle theory has flavor symmetry $SU(2)^2 \times SU(4)$. Each $SU(2)$ factor has flavor central charge $2 + 1/(k+1)$, so we have a total of $4$, which exactly cancels with the beta function of the gauge group.

Now we use $D_3$ perspective to analyze the S-duality. See the second line of figure \ref{D3Sduality} for illustration. It is not hard to figure out the correct puncture after degeneration of the Riemann sphere. To compare the Coulomb branch spectrum, we assume maximal regular puncture. For the theory on the left hand side, using Newton polygon we have
\be
\Delta(\cO) = & \frac{k+2}{k+1}, \ \frac{k+3}{k+1}, \ \dots, \ \frac{2k+1}{k+1}, \\[0.5em]
                     & \frac{k+2}{k+1}, \ \frac{k+3}{k+1}, \ \dots, \ \frac{2k+1}{k+1}. \\[0.5em]
\ee
We see it is nothing but the two copy of $(A_1, D_{2k+2})$ theories. For the theory on the right hand side, the spectrum is exactly the same as the $A_3$ theory $(III_{k,1}^{[2,2]^{\times (k+1)},[1,1,1,1]}, F)$. We thus conjecture that:
\be
a_{(III_{k,1}^{[1;4]^{\times (k+1)},[1^3; 0]}, F)} = 4k + \frac{131}{24}, \ \ \ \ c_{(III_{k,1}^{[1;4]^{\times (k+1)},[1^3; 0]}, F)} = 4k + \frac{142}{24}.
\label{D3centralCharge1}
\ee
This is the same as computed by the recipe in section \ref{subsec: central charges}.

\begin{figure}[htbp]
\centering
\includegraphics[width=15cm]{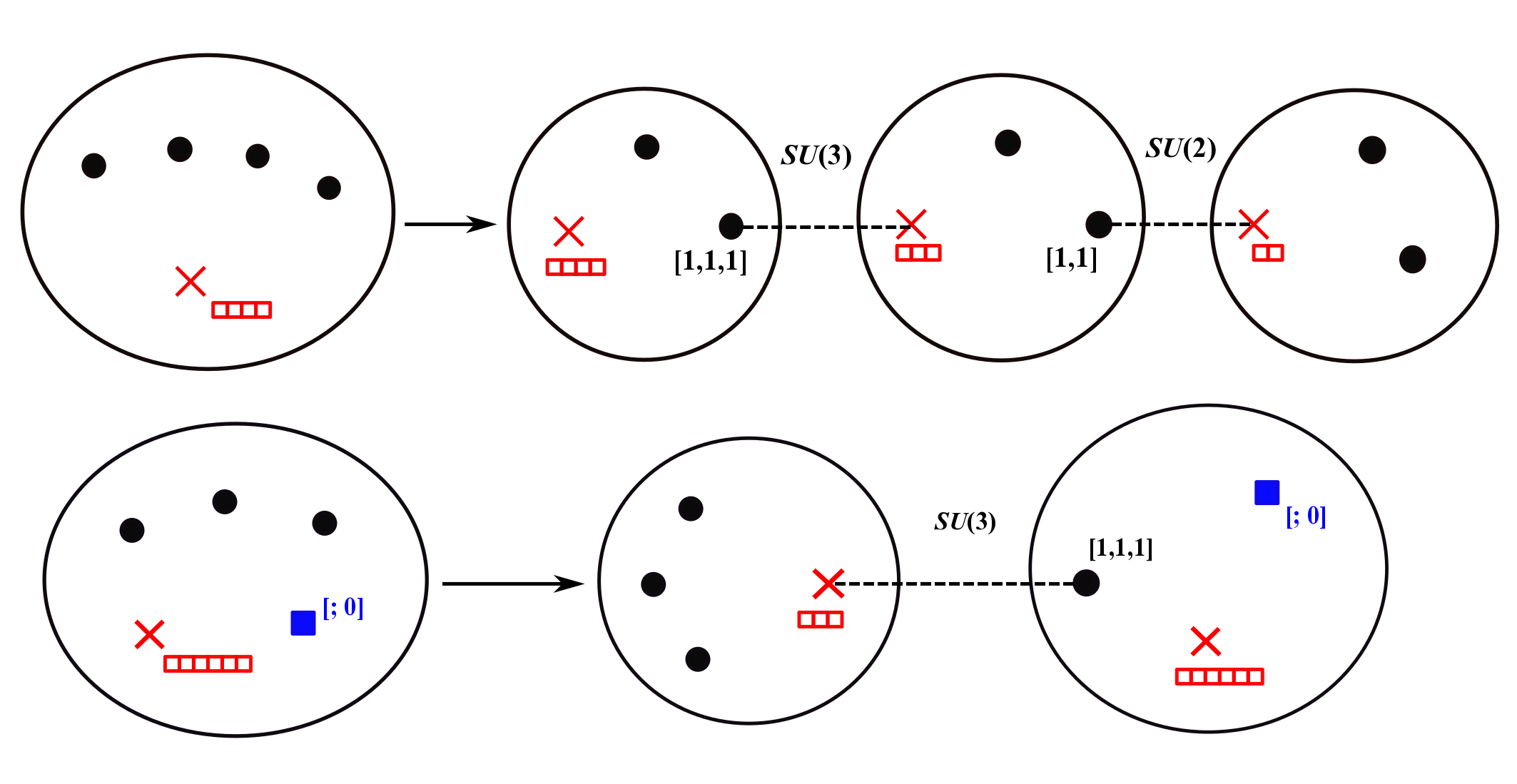}
\caption[Invisible A3 S duality]{Another S-duality frame. The upper one is from $A_3$ perspective. Here in the weakly coupled description, the rightmost theory is still $(A_1, D_{2k+2})$, the middle theory is given by $\pbra{III_{k,1}^{[2,1]^{\times (k+1)},[1,1,1]}, F}$, and the leftmost theory is given by $\pbra{III_{k,1}^{[3,1]^{\times (k+1)},[1,1,1,1]}, F}$. The lower one is from the $D_3$ perspective. The left theory without blue marked points should be understood as $A_2$ theory. The right hand theory is given by $\pbra{III_{k,1}^{[3;0]^{\times (k+1)},[1,1,1;0]}, F}$. All the computation can be done similarly by replacing full puncture $F$ to be other arbitrary regular puncture $Q$.}
\label{A3SdualityFrame2}
\end{figure}

There is another duality frame described in figure \ref{A3SdualityFrame2}. From $D_3$ perspective, we get another type of Argyres-Douglas matter and the flavor symmetry is now carried by a black dot, which is in fact $SU(3)$. It connects to the left to an $A_2$ theory with all $T_i$'s regular semisimple. This theory can further degenerate according to the rules of $A_{N-1}$ theories, and we do not picture it. We conjecture that the central charges for the theory $\pbra{III_{k,1}^{[3;0]^{\times (k+1)},[1,1,1;0]}, F}$ are
\be
a_{\pbra{III_{k,1}^{[3;0]^{\times (k+1)},[1,1,1;0]}, F}} = 3k + \frac{17}{4}, \ \ \ \ c_{\pbra{III_{k,1}^{[3;0]^{\times (k+1)},[1,1,1;0]}, F}} = 3k + \frac{19}{4}.
\ee

\begin{figure}[htbp]
\centering
\includegraphics[width=15.0cm]{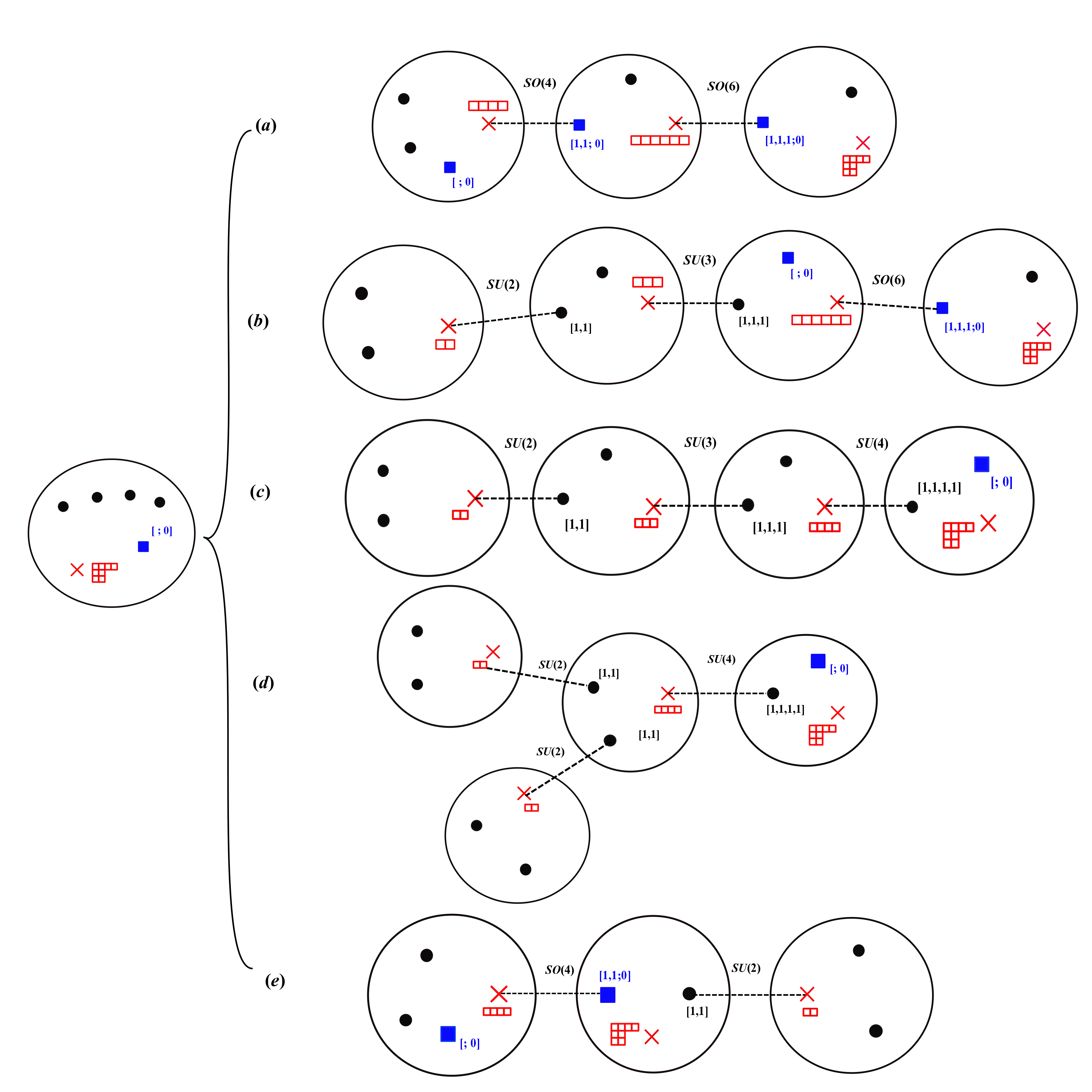}
\caption[D4 S duality]{The weakly coupled duality frame for $D_4$ theory of class $(k,1)$. For $(a)$, the leftmost theory is two copies of $(A_1, D_{2k+2})$, the middle theory is given by $\pbra{III_{k,1}^{[1;4]^{\times (k+1)},[1^3;0]}, F}$, and the rightmost theory is given by $\pbra{III_{k,1}^{[1;6]^{\times (k+1)},[1^4;0]}, Q}$ where $Q$ is any $D_4$ regular puncture. 

For $(b)$, the leftmost theory is $(A_1, D_{2k+2})$, followed by the theory $\pbra{III_{k,1}^{[2,1]^{\times (k+1)},[1^3]}, F}$. This is then followed by $\pbra{III_{k,1}^{[3,1]^{\times (k+1)},[1^4;0]}, F}$, and the rightmost theory is still $\pbra{III_{k,1}^{[1;6]^{\times (k+1)},[1^4;0]}, Q}$.

For $(c)$ and $(d)$, the rightmost theory is given by $\pbra{III_{k,1}^{[4;0]^{\times (k+1)},[1^4;0]}, Q}$. Then there are two different ways the theory $\pbra{III_{k,1}^{[1^4]^{\times (k+2)}}, F}$ can be further degenerated.

Finally for $(e)$, the leftmost theory is again two copies of $(A_1, D_{2k+2})$ theory. The middle theory is $D_4$ theory $\pbra{III_{k,1}^{[2; 4]^{\times (k+1)}, [1^4]}, F}$, and the rightmost theory is given by $(A_1, D_{2k+2})$.}
\label{D4SdualityFrame}
\end{figure}

\vspace{8pt}
\textit{Example 2: $D_4$.} Now we consider a more complicated example. Let us take a generic large $\ell > 3$ and all the coefficient matrices to be regular semisimple, $T_{\ell} = \dots = T_1 = [1^4; 0]$. There are several ways to get weakly coupled duality frame, which is described in figure \ref{D4SdualityFrame}. The regular puncture can be arbitrary. We have checked their Coulomb branch spectrum matches with the initial theory, as well as the fact that all gauge couplings are conformal.

For $(a)$ in figure \ref{D4SdualityFrame}, we can compute the central charges for the theory $\pbra{III_{k,1}^{[1;6]^{\times (k+1)},[1^4;0]}, Q}$ when $Q$ is a trivial regular puncture. Recall the initial theory may be mapped to hypersurface singularity in type \Rnum{2}B construction: 
\be
a_{\pbra{III_{k,1}^{[1^4;0]^{\times (k+2)}},\, S}} = \frac{84k^2 - 5k - 5}{6(k+1)}, \ \ \ \ c_{\pbra{III_{k,1}^{[1^4;0]^{\times (k+2)}},\, S}} = \frac{42k^2 - 2k - 2}{3(k+1)},
\ee
while we already know the central charges for $(A_1, D_{2k+2})$ and $\pbra{III_{k,1}^{[1;4]^{\times (k+1)},[1^3;0]}, F}$ theory in \eqref{D3centralCharge1}. Therefore we have
\be
a_{\pbra{III_{k,1}^{[1;6]^{\times (k+1)},[1^4;0]},\, S}} = \frac{54k^2 - 95k - 65}{6(k+1)}, \ \ \ \ c_{\pbra{III_{k,1}^{[1;6]^{\times (k+1)},[1^4;0]},\, S}} = \frac{108k^2 - 185k - 125}{12(k+1)}.
\ee
This is the same as computed from \eqref{centralCharge2}.

Notice that in $(a)$ of figure \ref{D4SdualityFrame}, the leftmost and middle theory may combine together, which is nothing but the theory $\pbra{III_{k,1}^{[1^3;0]^{\times (k+2)}},\, F}$. We can obtain another duality frame by using an $SU(3)$ gauge group. See $(b)$ of figure \ref{D4SdualityFrame}.

We can try to split another kind of Argyres-Douglas matter, and use the black dot to carry flavor symmetry. The duality frames are depicted in $(c)$ and $(d)$ in figure \ref{D4SdualityFrame}. Again, we can compute the central charges for the Argyres-Douglas matter $\pbra{III_{k,1}^{[4;0]^{\times (k+1)},[1^4;0]}, S}$:
\be
a_{\pbra{III_{k,1}^{[4;0]^{\times (k+1)},[1^4;0]},\, S}} = \frac{108k^2 - 145k - 85}{12(k+1)}, \ \ \ \ c_{\pbra{III_{k,1}^{[4;0]^{\times (k+1)},[1^4;0]},\, S}} = \frac{27k^2 - 35k - 20}{3(k+1)},
\ee
same as computed from \eqref{centralCharge2}.

By comparing the duality frames, we see a surprising fact in four dimensional quiver gauge theory. In particular, $(a)$ in figure \ref{D4SdualityFrame} has $SO(2n)$ gauge groups while $(c)$ in figure \ref{D4SdualityFrame} has $SU(n)$ gauge groups. The Argyres-Douglas matter they couple to are completely different, and our prescription says they are the same theory! 

\begin{figure}[htbp]
\centering
\begin{adjustwidth}{-0.5cm}{}
\includegraphics[width=16cm]{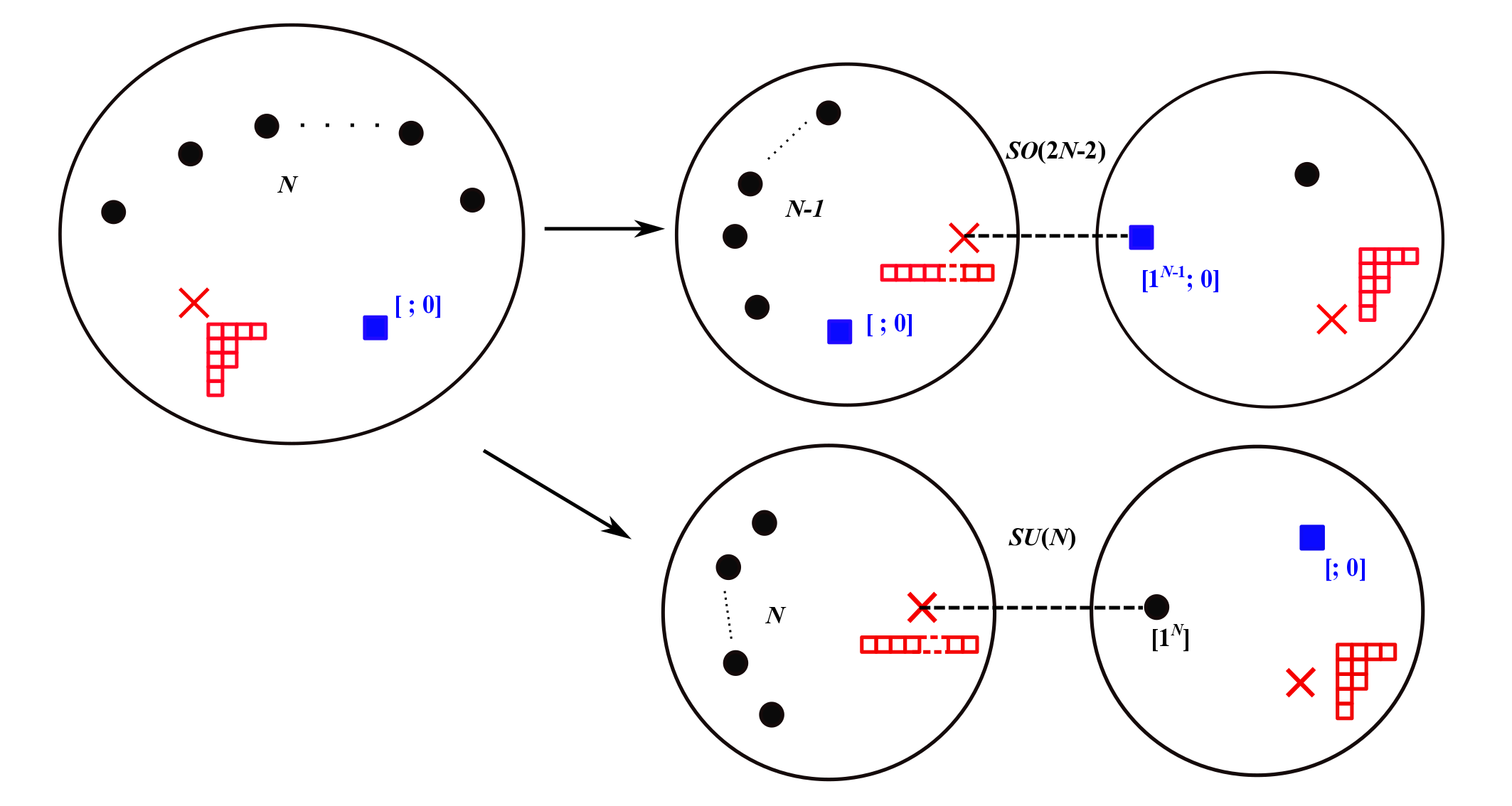}
\end{adjustwidth}
\caption[DN S duality]{The weakly coupled duality frame for $D_N$ theory of class $(k,1)$. One starts with maximal irregular puncture and a regular puncture, and recursively degenerate a sequence of Argyres-Douglas matter. The first line gives Argyres-Douglas matter $\pbra{III_{k,1}^{[1;2N-2]^{\times (k+1)},[1^{N};0]}, Q}$ and the second line gives $\pbra{III_{k,1}^{[N;0]^{\times (k+1)},[1^{N};0]}, Q}$. We get in general a quiver with $SU$ and $SO$ gauge groups.}
\label{DNSdualityFrame}
\end{figure}

\vspace{8pt}
\textit{General $D_N$.} Based on the above two examples, we may conjecture the S-duality for $D_N$ theories of class $(k,1)$ for large. The weakly coupled description can be obtained recursively, by splitting Argyres-Douglas matter one by one. See figure \ref{DNSdualityFrame} for illustration of two examples of such splitting. In the first way we get the Argyres-Douglas matter $\pbra{III_{k,1}^{[1;2N-2]^{\times (k+1)},[1^{N};0]}, Q}$, with remaining theory $\pbra{III_{k,1}^{[1^{N-1};0]^{\times (k+2)}}, F}$. The gauge group in between is $SO(2N-2)$. In the second way, we get the Argyres-Douglas matter $\pbra{III_{k,1}^{[N;0]^{\times (k+1)},[1^{N};0]}, Q}$, with remaining theory $\pbra{III_{k,1}^{[1^N]^{\times (k+1)}}, F}$. The gauge group is $SU(N)$. The central charges $(a, c)$ for special cases of regular puncture can be computed similarly. 

\bigskip 

{\bf Duality at small $k$}. We see previously that when $k$ is large enough, new punctures appearing in the degeneration limit are all full punctures. We argue here that when $k$ is small, this does not have to be so. In this section, we focus on $D_5$ theory, with coefficient matrices $T_{\ell} = \dots = T_1 = [1,\dots, 1; 0]$ and one trivial regular puncture. The auxiliary Riemann sphere is given by five black dots of type $[1]$, one trivial blue square and one trivial red cross. We will focus on the linear quiver only.

\begin{figure}[htbp]
\centering
\begin{adjustwidth}{0cm}{}
\includegraphics[width=15cm]{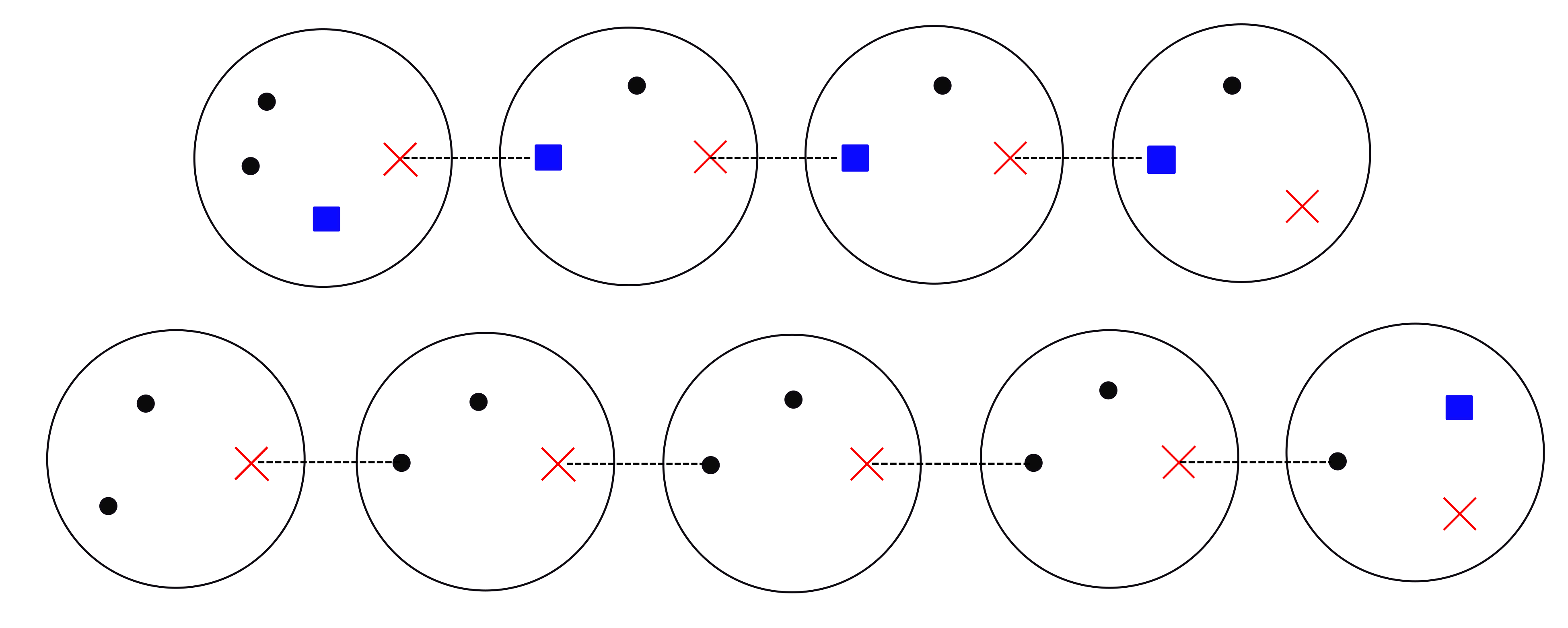}
\end{adjustwidth}
\caption[D5 Linear Quiver]{The linear quiver that we will examine for $k$ small, when $\mfg = D_5$.}
\label{D5Linear}
\end{figure}

\vspace{6pt} \textit{$D_5$ theory}. The linear quivers we consider are depicted in figure \ref{D5Linear}. 

After some lengthy calculations, we find that, for the first quiver (where red crosses are all connected with blue squares), when $k = 1$, the quiver theory is 

\vspace{6pt}
\begin{adjustwidth}{-0.5cm}{}
\begin{tikzpicture}
    \draw (18,0) node[anchor=west] (1) {$ \pbra{\Rnum{3}_{1,1}^{[1; 8]^{\times 2}, [1^2;6]}, [9,1]}$.};
    \draw (17.2,1) node[anchor=west] (2) {$SO(3)$};
    \draw (13.5,0) node[anchor=west] (3) {$ \pbra{\Rnum{3}_{1,1}^{[1; 6]^{\times 2}, [1^4;0]}, [5,1^3]}$};
     \draw (12.6,1) node[anchor=west] (4) {$SO(5)$};
     \draw (9,0) node[anchor=west] (5) {$ \pbra{\Rnum{3}_{1,1}^{[1; 4]^{\times 2}, [1^3; 0]}, [1^6]}$};
     \draw (8.5,1) node[anchor=west] (6) {$SO(4)$};
     \draw (5.5,0) node[anchor=west] (7) {$ \pbra{\Rnum{3}_{1,1}^{[1^2; 0]^{\times 3}}, [1^4]}$};
    
    \path (1) edge[sedge] (2)
             (3) edge[sedge] (2)
             (3) edge[sedge] (4)
             (4) edge[sedge] (5)
             (5) edge[sedge] (6)
             (6) edge[sedge] (7)
    ;
\end{tikzpicture}
\end{adjustwidth}
\vspace{6pt}
In particular, we have checked the central charge and confirm that the middle gauge group is indeed $SO(5)$. Moreover, its left regular puncture is superficially $[1^6]$ but only $SO(5)$ symmetry remains, similar for the right blue marked points $[; 6]$\footnote{We could imagine a similar situation of three hypermultiplets with $SO(6)$ symmetry for six half-hypermultiplets. We then only gauge five of them with $SO(5)$ gauge group. In this way, one mass parameter is frozen, so we get a total of two mass parameters.}. 

For $k = 2$, we have the quiver

\vspace{6pt}
\begin{adjustwidth}{-0.5cm}{}
\begin{tikzpicture}
    \draw (18,0) node[anchor=west] (1) {$ \pbra{\Rnum{3}_{2,1}^{[1; 8]^{\times 3}, [1^3;4]}, [9,1]}$.};
    \draw (17.2,1) node[anchor=west] (2) {$SO(5)$};
    \draw (13.5,0) node[anchor=west] (3) {$ \pbra{\Rnum{3}_{2,1}^{[1; 6]^{\times 3}, [1^4;0]}, [3,1^5]}$};
     \draw (12.5,1) node[anchor=west] (4) {$SO(6)$};
     \draw (9,0) node[anchor=west] (5) {$ \pbra{\Rnum{3}_{2,1}^{[1; 4]^{\times 3}, [1^3; 0]}, [1^6]}$};
     \draw (8.5,1) node[anchor=west] (6) {$SO(4)$};
     \draw (5.5,0) node[anchor=west] (7) {$ \pbra{\Rnum{3}_{2,1}^{[1^2; 0]^{\times 4}}, [1^4]}$};
    
    \path (1) edge[sedge] (2)
             (3) edge[sedge] (2)
             (3) edge[sedge] (4)
             (4) edge[sedge] (5)
             (5) edge[sedge] (6)
             (6) edge[sedge] (7)
    ;
\end{tikzpicture}
\end{adjustwidth}

\vspace{6pt}
For $k = 3$, we have the quiver

\begin{adjustwidth}{-0.5cm}{}
\begin{tikzpicture}
    \draw (18,0) node[anchor=west] (1) {$ \pbra{\Rnum{3}_{3,1}^{[1; 8]^{\times 4}, [1^5;0]}, [9,1]}$.};
    \draw (17.2,1) node[anchor=west] (2) {$SO(8)$};
    \draw (13.5,0) node[anchor=west] (3) {$ \pbra{\Rnum{3}_{3,1}^{[1; 6]^{\times 4}, [1^4;0]}, [1^8]}$};
     \draw (12.5,1) node[anchor=west] (4) {$SO(6)$};
     \draw (9,0) node[anchor=west] (5) {$ \pbra{\Rnum{3}_{3,1}^{[1; 4]^{\times 4}, [1^3; 0]}, [1^6]}$};
     \draw (8.5,1) node[anchor=west] (6) {$SO(4)$};
     \draw (5.5,0) node[anchor=west] (7) {$ \pbra{\Rnum{3}_{3,1}^{[1^2; 0]^{\times 5}}, [1^4]}$};
    
    \path (1) edge[sedge] (2)
             (3) edge[sedge] (2)
             (3) edge[sedge] (4)
             (4) edge[sedge] (5)
             (5) edge[sedge] (6)
             (6) edge[sedge] (7)
    ;
\end{tikzpicture}
\end{adjustwidth}

\vspace{6pt}
Finally, for $k > 3$ we reduce to the case in previous section. It is curious to see that some of the gauge group becomes smaller and smaller when $k$ decreases, due to appearance of next-to-maximal puncture. Moreover, there are theories ($i.e.$ $ \pbra{\Rnum{3}_{1,1}^{[1; 8]^{\times 2}, [1^2;6]}, [9,1]}$) whose Coulomb branch spectrum is empty. When this happens, the theory is in fact a collection of free hypermultiplets.

The same situation happens for the second type of $D_5$ quiver. When $k$ starts decreasing, the sizes of some gauge groups for the quiver theory decrease.  When $k = 1$ we get:

\vspace{6pt}
\begin{adjustwidth}{-1.7cm}{}
\begin{tikzpicture}
    \draw (19,0) node[anchor=west] (1) {$ \pbra{\Rnum{3}_{1,1}^{[5; 0]^{\times 2}, [2,2,1;0]}, [9,1]}$.};
    \draw (18.3,1) node[anchor=west] (2) {$SU(2)$};
    \draw (14.9,0) node[anchor=west] (3) {$ \pbra{\Rnum{3}_{1,1}^{[4,1]^{\times 2}, [1^5]}, [2,2,1]}$};
     \draw (14.2,1) node[anchor=west] (4) {$SU(4)$};
     \draw (11.2,0) node[anchor=west] (5) {$ \pbra{\Rnum{3}_{1,1}^{[3,1]^{\times 2}, [1^4]}, [1^4]}$};
     \draw (10.6,1) node[anchor=west] (6) {$SU(3)$};
     \draw (7.5,0) node[anchor=west] (7) {$ \pbra{\Rnum{3}_{1,1}^{[2,1]^{\times 2}, [1^3]}, [1^3]}$};
     \draw (6.9,1) node[anchor=west] (8) {$SU(2)$};
     \draw (4.4,0) node[anchor=west] (9) {$ \pbra{\Rnum{3}_{1,1}^{[1,1]^{\times 3}}, [1^2]}$};
    
    \path (1) edge[sedge] (2)
             (3) edge[sedge] (2)
             (3) edge[sedge] (4)
             (4) edge[sedge] (5)
             (5) edge[sedge] (6)
             (6) edge[sedge] (7)
             (7) edge[sedge] (8)
             (8) edge[sedge] (9)
    ;
\end{tikzpicture}
\end{adjustwidth}

\vspace{6pt}
When $k = 2$, we have the quiver 
\vspace{6pt}
\begin{adjustwidth}{-1.7cm}{}
\begin{tikzpicture}
    \draw (19,0) node[anchor=west] (1) {$ \pbra{\Rnum{3}_{2,1}^{[5; 0]^{\times 3}, [1^5;0]}, [9,1]}$.};
    \draw (18.3,1) node[anchor=west] (2) {$SU(5)$};
    \draw (15.1,0) node[anchor=west] (3) {$ \pbra{\Rnum{3}_{2,1}^{[4,1]^{\times 3}, [1^5]}, [1^5]}$};
     \draw (14.2,1) node[anchor=west] (4) {$SU(4)$};
     \draw (11.2,0) node[anchor=west] (5) {$ \pbra{\Rnum{3}_{2,1}^{[3,1]^{\times 3}, [1^4]}, [1^4]}$};
     \draw (10.6,1) node[anchor=west] (6) {$SU(3)$};
     \draw (7.5,0) node[anchor=west] (7) {$ \pbra{\Rnum{3}_{2,1}^{[2,1]^{\times 3}, [1^3]}, [1^3]}$};
     \draw (6.9,1) node[anchor=west] (8) {$SU(2)$};
     \draw (4.4,0) node[anchor=west] (9) {$ \pbra{\Rnum{3}_{2,1}^{[1,1]^{\times 4}}, [1^2]}$};
    
    \path (1) edge[sedge] (2)
             (3) edge[sedge] (2)
             (3) edge[sedge] (4)
             (4) edge[sedge] (5)
             (5) edge[sedge] (6)
             (6) edge[sedge] (7)
             (7) edge[sedge] (8)
             (8) edge[sedge] (9)
    ;
\end{tikzpicture}
\end{adjustwidth}

\vspace{6pt}
Finally when $k > 2$, all the gauge groups do not change anymore and stay as those in previous section. 

We can carry out similar analysis for all $D_N$ theory when $k$ is small. This indicates that as we vary the external data, the new punctures appearing in the degeneration limit vary as well.

\subsection{Class $(k,b)$}\label{sec: S-duality-(k,b)}

For general $b > 1$ and $(k, b)$ coprime, we need to classify which irregular punctures engineer superconformal theories, and study its duality as before. One subtlety that appears here is that, unlike $b = 1$ case in previous section, here we need to carefully distinguish between whether $b$ is an odd/even divisor of $N$/$2N-2$, as their numbers of exact marginal deformations are different. See section \ref{subsec: field data} for details. 

\subsubsection{Coulomb branch spectrum and degenerating coefficient matrices}\label{subsec: Coulomb spectrum-(k,b)}

We have mentioned in section \ref{subsubsec: irregDegeneration} how to count graded Coulomb branch dimension for general $b > 1$. We elaborate the procedure here.

\vspace{6pt} (\rnum{1}) \textit{$b$ is an odd divisor of $N$}. We may label the degenerating matrices similar to labelling the Levi subgroup: $[r_1, \dots, r_{n}; {\tilde r}]$, where $\sum 2 b r_i + {\tilde r} = 2N$, and there are $n - 1$ exact marginal deformations. To calculate the Coulomb branch spectrum, we first introduce a covering coordinate $z = w^b$, such that the pole structure becomes:
\be
\frac{T_{\ell}}{z^{2+\frac{k}{b}}} \rightarrow \frac{T'_{\ell}}{w^{k+b+1}},
\ee
and $T'_{\ell}$ is given by Levi subgroup of type $[r_1, \dots, r_1, \dots, r_n, \dots r_n; {\tilde r}]$, where $r_i$ is repeated $b$ times. Then we are back to the case $b = 1$ and we can repeat the procedure in section \ref{subsubsec: irregDegeneration}. This would give the maximal degree ${d_{2i}}$ in the monomial $w^{d_{2i}} x^{2N-2i}$ that gives Coulomb branch moduli. The monomial corresponds to the degree $2i$ differential $\phi_{2i}$, and after converting back to coordinate $z$, we have the degree of $z$ in $z^{d'_{2i}} x^{2N-2i}$ as:
\be
d'_{2i} \leq \left \lfloor \frac{d_{2i} - 2i(b-1)}{b} \right \rfloor,
\ee
and similar for the Pfaffian ${\tilde \phi}$.

\vspace{6pt} (\rnum{2}) \textit{$b$ is an even divisor of $N$}. We can label the matrix $T_{\ell}$ as $[r_1, r_2, \dots, r_n; {\tilde r}]$ such that $\sum b r_i + {\tilde r} = 2N$. Then, we take the change of variables $z = w^b$, and $T'_{\ell}$ is given by repeating each $r_i$ $(b/2)$ times, while ${\tilde r}$ is the same. This reduces to the class $(k,1)$ theories.

\vspace{6pt} (\rnum{3}) \textit{$b$ is an odd divisor of $2N-2$}. We use $[r_1, \dots, r_n; {\tilde r}]$ to label the Levi subgroup, which satisfies $2b\sum r_i + {\tilde r} = 2N-2$. To get the Coulomb branch spectrum, we again change the coordinates $z = w^b$, and the new coefficient matrix $T'_{\ell}$ is now given by Levi subgroup of type $[r_1, \dots, r_1, \dots, r_n, \dots, r_n; {\tilde r}]$, where each $r_i$ appears $b$ times. This again reduces to the class $(k,1)$ theories.

\vspace{6pt} (\rnum{4}) \textit{$b$ is an even divisor of $2N-2$}. This case is similar once we know the procedure in cases (\rnum{2}) and (\rnum{3}). We omit the details.

The above prescription also indicates the constraints on coefficient matrices in order for the resulting 4d theory is a SCFT. We conclude that $T_i$ should satisfy $T_{\ell} = \dots = T_2$, $T_1$ is arbitrary.

To see our prescription is the right one, we can check the case $D_4$. As an example, we can consider the Higgs field
\be
\Phi \sim \frac{T_{\ell}}{z^{2+\frac{1}{4}}} + \dots, \ \ \ \ \ell = 6,
\ee
and all $T_i$ to be $[1,1; 0]$. Using the above procedure, we know that at $\phi_6$ there is a nontrivial moduli whose scaling dimension is $6/5$. This is exactly the same as that given by hypersurface singularity in type \Rnum{2}B construction. Similarly, we may take $D_5$ theory:
\be
\Phi \sim \frac{T_{\ell}}{z^{2+\frac{1}{4}}} + \dots, \ \ \ \ \ell = 6,
\ee
and all $T_i$'s given by $[1,1;2]$. After changing variables we have $T'_i$ given by $[1,1,1,1;2]$, which is the same as $[1^5; 0]$. Then we have two Coulomb branch moduli with scaling dimension $\{ 6/5, 8/5 \}$, same as predicted by type \Rnum{2}B construction.

\subsubsection{Duality frames}

Now we study the S-duality for these theories. As one example, we may consider $D_4$ theory of class $(k,b) = (3,2)$, and $T_{\ell}$ is given by $[1,1,1,1;0]$. We put an extra trivial regular puncture at the south pole. This theory has Coulomb branch spectrum
\be
\Delta(\cO) = \left \{  \pbra{\frac{6}{5}}^{\times 4}, \pbra{\frac{8}{5}}^{\times 3}, (2)^{\times 3}, \pbra{\frac{12}{5}}^{\times 3}, \frac{14}{5}, \frac{16}{5}, \frac{18}{5} \right \}.
\ee
In the degeneration limit, we get three theories, described in figure \ref{D4(3,2)Sduality}. The middle theory $\pbra{\Rnum{3}_{3,2}^{[1;4]^{\times 5}, [1,1,1;0]}, [3,1,1,1]}$ gets further twisted in the sense mentioned in next subsection \ref{Sec: S-duality-twisted}, and has Coulomb branch spectrum $\{ 6/5, 8/5, 12/5, 12/5, 14/5, 16/5, 18/5 \}$. Besides it, the far left theory is two copies of $(A_1, D_5)$ theory with Coulomb branch spectrum $\{8/5, 6/5\}$ each. The far right theory is an untwisted theory, given by $\pbra{\Rnum{3}_{3,2}^{[1;6]^{\times 5}, [1,1;4]}, S}$, giving spectrum $\{12/5, 6/5 \}$. Along with the $SO(4)$ and $SU(2)$ gauge group, we see that the Coulomb branch spectrum nicely matches together. We conjecture that this is the weakly coupled description for the initial Argyres-Douglas theory.


\begin{figure}[htbp]
\centering
\begin{adjustwidth}{0cm}{}
\includegraphics[width=15cm]{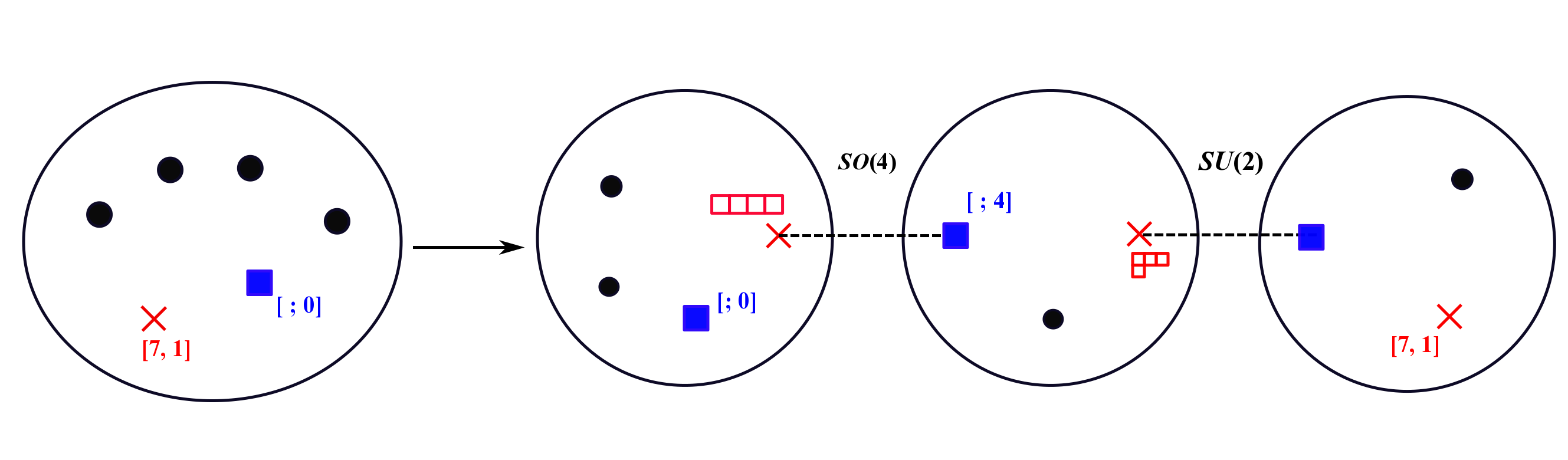}
\end{adjustwidth}
\caption[D4 S duality for (3,2) class]{S-duality for $D_4$ theory of class $(3,2)$. Here we pick the coefficient matrices to be of type $[1,1,1,1;0]$, with a trivial regular puncture (this setup can be relaxed to general $D_4$ regular punctures). In the degeneration limit, we get $SO(4) \times SU(2)$ gauge group plus three Argyes-Douglas matter. The leftmost theory is in fact two copies of $(A_1, D_5)$ theory, while the middle theory is given by twisted $D_3$ theory, given by twisting the theory $\pbra{\Rnum{3}_{3,2}^{[1;4]^{\times 5}, [1,1,1;0]}, [3,1,1,1]}$. The rightmost theory is $\pbra{\Rnum{3}_{3,2}^{[1;6]^{\times 5}, [1,1;4]}, S}$ theory.}
\label{D4(3,2)Sduality}
\end{figure}

In this example, each gauge coupling is exactly conformal as well.

As a second example, we consider $D_3$ theory of class $(3,2)$. The coefficient matrices are given by $T_{6} = \dots = T_2 = T_1 = [1,1; 2]$. We put a trivial regular puncture at the south pole. This theory has Coulomb branch spectrum
\be
\Delta(\cO) = \left \{ \frac{6}{5}, \frac{6}{5}, \frac{7}{5}, \frac{8}{5}, \frac{9}{5}, 2, \frac{12}{5} \right \},
\ee
and is represented by an auxiliary Riemann sphere with two black dots of type $[1]$, one blue square of size $2$ and one trivial red cross. See figure \ref{D3(3,2)Sduality}. After degeneration, we get two theories. We compute that the first theory is a twisting of $\pbra{\Rnum{3}^{[1;2]^{\times 5}, [1,1;0]}, [1^4]}$, having spectrum $\{ 6/5, 7/5, 8/5, 9/5 \}$. The second theory $\pbra{\Rnum{3}_{3,2}^{[1;4]^{\times 5}, [1,1,1; 0]}, S}$ has spectrum $\{12/5, 6/5 \}$. The middle gauge group is $SO(3)$, although the two sides superficially have $SO(4)$ symmetry.


\begin{figure}[htbp]
\centering
\begin{adjustwidth}{0cm}{}
\includegraphics[width=14cm]{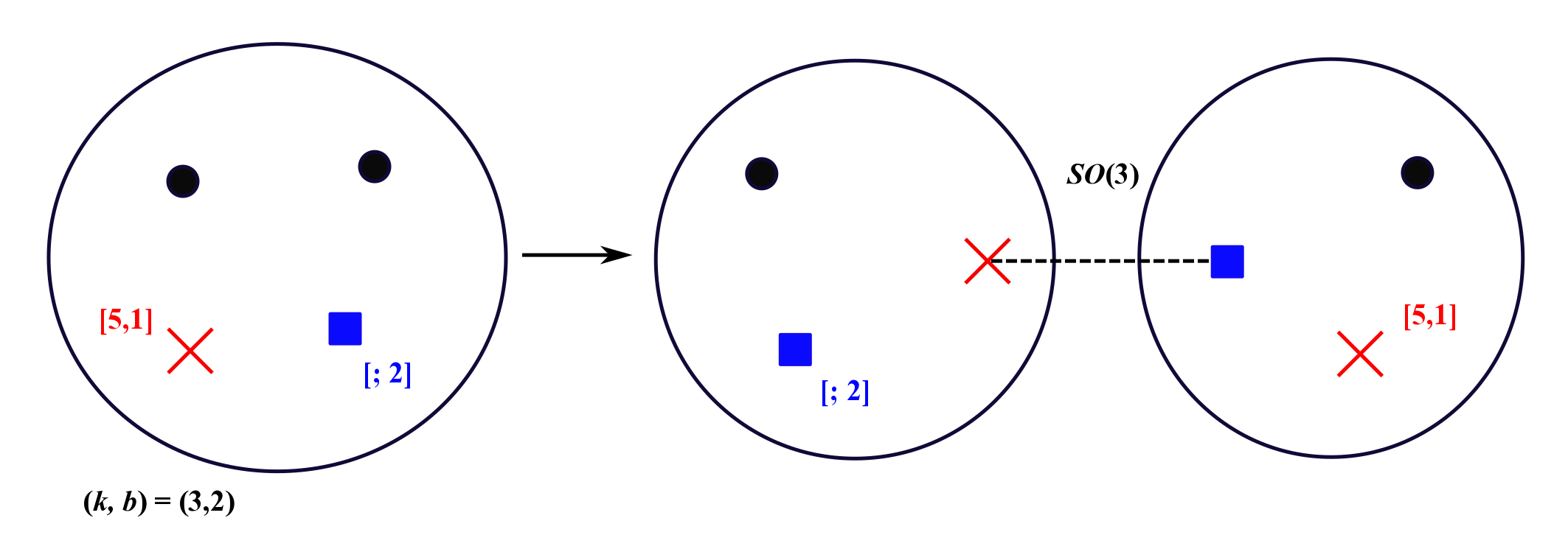}
\end{adjustwidth}
\caption[D3 S duality for (3,2) class]{S-duality for $D_3$ theory of class $(3,2)$. Here we pick the coefficient matrices to be of type $[1,1;2]$, with a trivial regular puncture (this setup can be relaxed to general $D_3$ regular punctures). }
\label{D3(3,2)Sduality}
\end{figure}

\subsection{$\mathbb{Z}_2$-twisted theory}\label{Sec: S-duality-twisted}

If the Lie algebra $\mfg$ has a nontrivial automorphism group ${\rm Out}(\mfg)$, then one may consider \textit{twisted} punctures. This means as one goes around the puncture, the Higgs field undergoes an action of nontrivial element $o \in {\rm Out}(\mfg)$:
\be
\Phi(e^{2\pi i} z) = h \bbra{o\pbra{\Phi(z)}} h^{-1},
\ee
where $h \in \mfg / \mfj^{\vee}$ with $\mfj^{\vee}$ the invariant subalgebra under ${\rm Out}(\mfg)$. Let us denote $\mfj$ the Langlands dual of $\mfj^{\vee}$.

In this section we solely consider $D_N$ theory with automorphism group $\mathbb{Z}_2$. It has invariant subalgebra $\mfj^{\vee}= B_{N-1}$ whose Langlands dual is $\mfj = C_{N-1}$. For more details of other Lie algebra $\mfg$, see \cite{Chacaltana:2012zy, Chacaltana:2012ch, Chacaltana:2013oka, Chacaltana:2014nya, Chacaltana:2015bna, Chacaltana:2016shw}. We review some background for twisted regular punctures as in \cite{Chacaltana:2013oka}, and then proceed to understand twisted irregular punctures and their S-duality. For previous study of S-duality for twisted theory, see \cite{Argyres:2007tq, Tachikawa:2010vg}.

\subsubsection{Twisted regular punctures}

Following \cite{Chacaltana:2013oka}, a regular twisted $D_N$ punctures are labelled by nilpotent orbit of $C_{N-1}$, or a $C$-\textit{partition} $d$ of $2N-2$, where all odd parts appear with even multiplicity. To fix the local Higgs field, note that $\mathbb{Z}_2$ automorphism group split the Lie algebra $\mfg$ as $\mfg = \mfj_{1} \oplus \mfj_{-1}$, with eigenvalue $\pm1$ respectively. Apparently, $\mfj_1 = B_{N-1}$. The Higgs field behaves as
\be
\Phi \sim \frac{\Lambda}{z} + \frac{\Lambda'}{z^{1/2}} + M,
\ee
where $\Lambda'$ is a generic element of $\mfj_{-1}$ and $M$ is a generic element of $\mfj_1$. $\Lambda$ is an element residing in the nilpotent orbit of $B_{N-1}$, which is given by a $B$-\textit{partition} of $2N-1$, where all even parts appear with even multiplicity. It is again related to the C-partition $d$ via the Spaltenstein map $\mfS$. To be more specific, we have $\mfS(d) = \pbra{d^{+\mathsf{T}}}_B$: 
\begin{enumerate}
\item[$\bullet$] First, ``$+$'' means one add an entry $1$ to the C-partition $d$;

\item[$\bullet$] Then, perform transpose of $d^{+}$, corresponding to the superscript $\mathsf{T}$; 

\item[$\bullet$] Finally, $(\cdot)_B$ denotes the $B$-\textit{collapse}. The procedure is the same as D-collapse in section \ref{subsubsec: regular puncture}.
\end{enumerate}
For later use we will also introduce the action $\mfS$ on a B-partition $d'$. This should give a C-partition. Concretely, we have $\mfS(d') = ({d'}^{\mathsf{T}-})_C$:
\begin{enumerate}
\item[$\bullet$] First, ``$\mathsf{T}$'' means one take transpose of $d'$;

\item[$\bullet$] Then, perform reduction of ${d'}^{\mathsf{T}}$, corresponding to subtract the last entry of ${d'}^{\mathsf{T}}$ by $1$; 

\item[$\bullet$] Finally, $(\cdot)_C$ denotes the $C$-\textit{collapse}. The procedure is the same as B- and D-collapse except that it now operates on the odd part which appears even multiplicity.
\end{enumerate}

Given a regular puncture with a C-partition, we may read off its residual flavor symmetry as
\be
G_{\rm flavor} = \prod_{h\ \text{even}} SO(n^h) \times \prod_{h\ \text{odd}} Sp\pbra{n^h}.
\ee
We may also calculate the pole structure of each differential $\phi_{2i}$ and the Pfaffian ${\tilde \phi}$ in the Seiberg-Witten curve \eqref{D_SW curve}. We denote them as $\{ p_2, p_4, \dots, p_{2N-2}; {\tilde p} \}$; in the twisted case, the pole order of the Pfaffian ${\tilde \phi}$ is always half-integer. 

As in the untwisted case, the coefficient for the leading singularity of each differential may not be independent from each other. There are constraints for $c^{(2k)}_l$, which we adopt the same notation as in section \ref{subsubsec: regular puncture}. The constraints of the form 
\be
c^{(2k)}_{l} = \pbra{a^{(k)}_{l/2}}^2
\ee
effectively remove one Coulomb branch moduli at degree $2k$ and increase one Coulomb branch moduli at degree $k$; while the constraints of the form
\be
c^{(2k)}_{l} = \dots
\ee
only removes one moduli at degree $2k$. For the algorithm of counting constraints for each differentials and complete list for the pole structures, see reference \cite{Chacaltana:2013oka}. After knowing all the pole structures and constraints on their coefficients, we can now compute the graded Coulomb branch dimensions exactly as those done in section \ref{subsubsec: regular puncture}. We can also express the local contribution to the Coulomb branch moduli as
\be
\dim^{\rho}_{\mathbb{C}}\text{Coulomb} = \frac{1}{2} \bbra{\dim_{\mathbb{C}} \mfS(\cO_{\rho}) + \dim \mfg / \mfj^{\vee}},
\ee
here $\cO_{\rho}$ is a nilpotent orbit in $C_{N-1}$ and $\mfS(\cO_{\rho})$ is a nilpotent orbit in $B_{N-1}$.

\subsubsection{Twisted irregular puncture}

Now we turn to twisted irregular puncture. We only consider the ``maximal twisted irregular singularities''. The form of the Higgs field is, in our $\mathbb{Z}_2$ twisting, 
\be
\Phi \sim \frac{T_{\ell}}{z^{\ell}} + \frac{U_{\ell}}{z^{\ell-1/2}} + \frac{T_{\ell-1}}{z^{\ell-1}} + \frac{U_{\ell-1}}{z^{\ell-3/2}} + \dots + \frac{T_1}{z} + \dots.
\label{twistedDNhiggs}
\ee
Here all the $T_{i}$'s are in the invariant subalgebra $\mathfrak{so}(2N-1)$ and all $U_i$'s are in its complement $\mfj_{-1}$. To get the Coulomb branch dimension, note that the nontrivial element $o \in \rm{Out}(\mfg)$ acts on the differentials in the SW curve as
\be
o: \ \ \ \ & \phi_{2i} \rightarrow \phi_{2i}\ \ \text{for}\ 1\leq i \leq N-1, \\[0.5em]
            & {\tilde \phi}_N \rightarrow -{\tilde \phi}_N.
\ee
Then, the Coulomb branch dimension coming from the twisted irregular singularities can be written as \cite{Wang:2015mra}:
\be
\dim^{\rho}_{\mathbb{C}}\text{Coulomb} = \frac{1}{2} \bbra{\sum_{i=1}^{\ell} \dim T_i + \sum_{i=2}^{\ell} (\dim \mfg / \mfj^{\vee} - 1) + \dim \mfg / \mfj^{\vee} }.
\ee
In the above formula, the $-1$ term in the middle summand comes from treating $U_i, 2 \leq i \leq \ell$ as parameter instead of moduli of the theory. It corresponds to the Pfaffian ${\tilde \phi}_N$ which switches sign under $o \in \rm{Out}(\mfg)$. 

As in the untwisted case, we are also interested in the degeneration of $T_i$ and the graded Coulomb branch dimension. First of all, we know that as an $\mathfrak{so}(2N-1)$ matrix, $T_i$ can be written down as
\be
\pbra{\begin{array}{ccc} 0 & u & v \\[0.5em] -v^{\mathsf{T}}& Z_1 & Z_2 \\[0.5em] -u^{\mathsf{T}}& Z_3 & -Z_1 \end{array}},
\ee
with $Z_{1,2,3}$ $(N-1) \times (N-1)$ matrices, and $Z_{2,3}$ are skew symmetric; while $u$, $v$ are row vectors of size $N-1$. After appropriate diagonalization, only $Z_1$ is nonvanishing. So a Levi subalgebra can be labelled by $[r_1, \dots, r_n; {\tilde r}+1]$, with ${\tilde r} + 1$ always an odd number. The associated Levi subgroup is
\be
L = \prod_i U(r_i) \times SO({\tilde r}+1).
\ee

Now we state our proposal for whether a given twisted irregular puncture defines a SCFT in four dimensions. Similar to untwisted case, we require that $T_{\ell} = T_{\ell - 1} = \dots = T_2$ and $T_1$ can be further arbitrary partition of $T_{i \geq 2}$. When all the $T_i$'s are regular semisimple, we can draw Newton polygon for these theories. They are the same as untwisted case, except that the monomials living in the Pfaffian ${\tilde \phi}_N$ get shift down one half unit \cite{Wang:2015mra}. 

\vspace{6pt}
\textit{Example: $D_4$ maximal twisted irregular puncture with $\ell = 3$.} We consider all $T_i$ to be regular semisimple $\mathfrak{so}(7)$ element $[1,1,1; 1]$, plus a trivial twisted regular puncture. From Newton polygon, we know the spectrum for this theory is $\{ 2, 3/2, 3, 5/2, 2, 3/2, 7/4, 5/4 \}$. 

\subsubsection{S-duality for twisted $D_N$ theory of class $(k,1)$}

Having all the necessary techniques at hand, we are now ready to apply the algorithm previously developed and generate S-duality frame. We state our rules as follows for theory of class $(k,1)$ with $k = \ell - 2$. 

\begin{enumerate}
\item[$\bullet$] Given coefficient matrices $T_{\ell} = \dots = T_2 = [r_1, \dots, r_n; {\tilde r}+1]$, and $T_1$ being further partition of $T_i$, we represent the theory on an auxiliary Riemann sphere with $n$ black dots with size $r_i, 1 \leq i \leq n$, a blue square with size ${\tilde r}$, and a red cross representing the regular puncture, labelled by a C-partition of $2N-2$. 

\item[$\bullet$] Different S-duality frames are given by different degeneration limit of the auxiliary Riemann sphere. 

\item[$\bullet$] Finally, one needs to figure out the newly appeared punctures. The gauge group can only connect a red cross and a blue square (${Sp}$ gauge group). This is different from untwisted case we considered before.  
\end{enumerate}

Let us proceed to examine examples. We first give a comprehensive discussion of $D_4$ theory. 

\begin{figure}[htbp]
\centering
\begin{adjustwidth}{0cm}{}
\includegraphics[width=14.6cm]{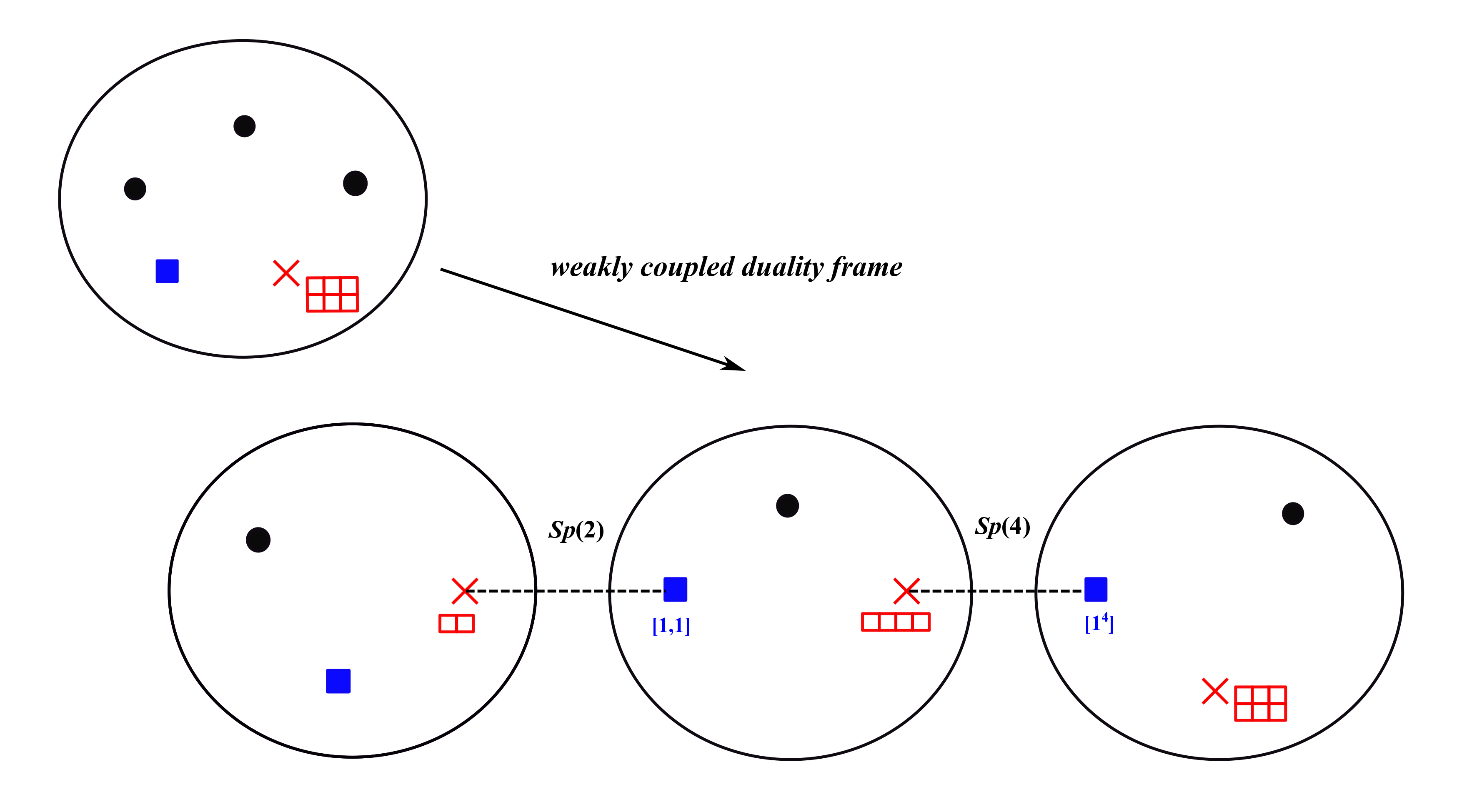}
\end{adjustwidth}
\caption[D4 S duality for twisted punctures]{S-duality for twisted $D_4$ theory of class $(k,1)$ with large $k$.  Each Argyres-Douglas matter is connected with $Sp$ gauge group. Assembling the black dot and the blue square we can read off the data for the irregular puncture and thus identify the theory.}
\label{D4TwistedSduality}
\end{figure}

\vspace{6pt}
{\bf Duality at large $k$}. We have initially three black dots of type $[1]$, a trivial blue square and an arbitrary red cross representing a regular puncture. This theory has a part of Coulomb branch spectrum coming from irregular puncture:
\be
\Delta(\cO) & = \frac{k+2}{k+1}, \dots, \frac{2k}{k+1}, \\[0.5em]
                  & = \frac{k+2}{k+1}, \dots, \frac{4k}{k+1}, \\[0.5em] 
                  & = \frac{k+2}{k+1}, \dots, \frac{6k}{k+1}, \\[0.5em] 
                  & = \frac{k+3/2}{k+1}, \dots, \frac{4k - 1/2}{k+1}.
\ee
The S-duality frame for this theory is given in figure \ref{D4TwistedSduality}. 

The duality frame in figure \ref{D4TwistedSduality} tells us the Coulomb branch spectrum of each piece. The leftmost theory $\pbra{\Rnum{3}_{k,1}^{[1;1]^{\times (k+2)}}, F}$ has the spectrum
\be
\Delta_1(\cO) & = \frac{k+2}{k+1}, \dots, \frac{2k+1}{k+1}, \\[0.5em]
                  & = \frac{k+3/2}{k+1}, \dots, \frac{2k+3/2}{k+1}.  
\ee
The rightmost theory is given by $\pbra{\Rnum{3}_{k,1}^{[1; 5]^{\times (k+1)}, [1,1,1;1]}, Q}$ whose spectrum comes from the irregular part is
\be
\Delta_2(\cO) & = \frac{k+2}{k+1}, \dots, \frac{2k}{k+1}, \\[0.5em]
                  & = \frac{2k+3}{k+1}, \dots, \frac{4k}{k+1}, \\[0.5em] 
                  & = \frac{4k+5}{k+1}, \dots, \frac{6k}{k+1}, \\[0.5em] 
                  & = \frac{3k+7/2}{k+1}, \dots, \frac{4k - 1/2}{k+1}. 
\ee
Finally, the middle theory is $\pbra{\Rnum{3}_{k,1}^{[1; 3]^{\times (k+1)}, [1,1;1]}, F}$. It contributes to the Coulomb branch spectrum coming from the irregular puncture
\be
\Delta_3(\cO) & = \frac{k+2}{k+1}, \dots, \frac{2k+1}{k+1}, \\[0.5em]
                  & = \frac{2k+3}{k+1}, \dots, \frac{4k+3}{k+1}, \\[0.5em] 
                  & = \frac{2k+5/2}{k+1}, \dots, \frac{3k + 5/2}{k+1}. 
\ee
These three pieces nicely assemble together and form the total spectrum of original theory. We thus have $Sp(2) \times Sp(4)$ gauge groups.

\vspace{6pt}
{\bf Duality at small $k$}. Similar to the untwisted case, we expect that some of the gauge group would be smaller. We now focus on a trivial twisted regular puncture in figure \ref{D4TwistedSduality}. Analysis for other twisted regular punctures are analogous. 

We find that for $k = 1$, 

\vspace{6pt}
\begin{adjustwidth}{0cm}{}
\begin{tikzpicture}
    \draw (13,0) node[anchor=west] (1) {$ \pbra{\Rnum{3}_{1,1}^{[1; 5]^{\times 2}, [1,1;3]}, [6]}$.};
    \draw (11.5,1) node[anchor=west] (2) {$Sp(2)$};
    \draw (7.1,0) node[anchor=west] (3) {$ \pbra{\Rnum{3}_{1,1}^{[1;3]^{\times 2}, [1,1;1]}, [2,1,1]}$};
     \draw (5.9,1) node[anchor=west] (4) {$Sp(2)$};
     \draw (2.2,0) node[anchor=west] (5) {$ \pbra{\Rnum{3}_{1,1}^{[1; 1]^{\times 3}}, [1,1]}$};
    
    \path (1) edge[sedge] (2)
             (3) edge[sedge] (2)
             (3) edge[sedge] (4)
             (4) edge[sedge] (5)
    ;
\end{tikzpicture}
\end{adjustwidth}
When $k \geq 2$, the second $Sp(2)$ gauge group becomes $Sp(4)$ and we reduce to the large $k$ calculations. 

\begin{figure}[htbp]
\centering
\begin{adjustwidth}{0cm}{}
\includegraphics[width=14.6cm]{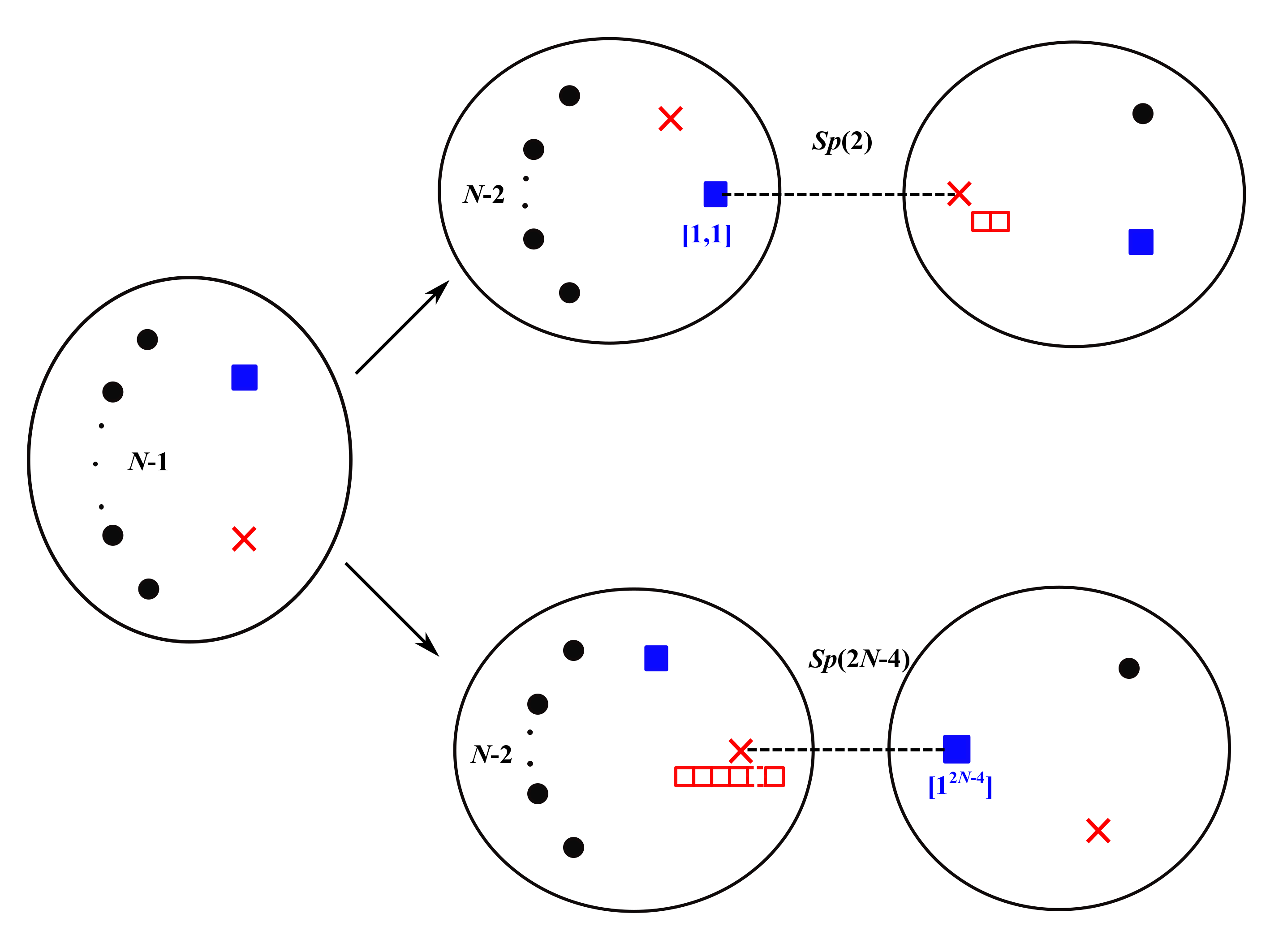}
\end{adjustwidth}
\caption[DN S duality for twisted punctures]{S-duality for twisted $D_N$ theory of class $(k,1)$ with large $k$. Here we present the duality frame recursively by splitting the Argyres-Douglas matter. In the first line we split a theory $\pbra{\Rnum{3}_{k,1}^{[1;1]^{\times (k+2)}}, F}$ with $F$ a full $D_2$ twisted puncture; in the second line we split a theory $\pbra{\Rnum{3}_{k,1}^{[1;2N-3]^{\times (k+1)}, [1^{N-1};1]}, Q}$ with original regular puncture $Q$.}
\label{DNTwistedSduality}
\end{figure}

\vspace{6pt}
\textit{S-duality of $D_N$ theory}. When $k$ is large, the intermediate gauge group in the degeneration limit does not depend on which twisted regular puncture one puts, and they are all full punctures. To obtain the duality frames, we can again follow the recursive procedure by splitting the Argyres-Douglas matter one by one. See the example of such splitting in figure \ref{DNTwistedSduality}. Again, due to twisting things become more constraining, and all matter should have a blue square on its auxiliary Riemann sphere. 

When $k$ is small, some of the intermediate puncture would be smaller. One needs to figure out those punctures carefully. We leave the details to interested readers. 

\section{Comments on S-duality for $E$-type theories}\label{Sec: S-duality-E}

Finally, we turn to the duality frames for $\mfg = \mathfrak{e}_{6,7,8}$. We focus on the Lie algebra $\mathfrak{e}_6$ while state our conjecture for $\mathfrak{e}_7$ and $\mathfrak{e}_8$ case. 

A complete list of all the relevant data for regular punctures can be found in \cite{Chacaltana:2014jba, Chacaltana:2017boe, Chacaltana:2015bna}.  We will use some of their results here for studying irregular puncture.

\subsection{Irregular puncture and S-duality for $E_6$ theory}

We focus on the irregular singularity \eqref{MaximalIrreg}. The first task is to characterize the degeneration of coefficient matrices. Those matrices $T_i$, $1 \leq i \leq \ell$ shall be represented by a Levi subalgebra $\mfl$. See section \ref{subsubsec: irregDegeneration} for the list of conjugacy classes. For each Levi subalgebra $\mfl$, we associate a nilpotent orbit with Nahm label. Since we are already using Bala-Carter's notation, we can directly read of $\mfl$. See table \ref{table:E6 orbit}. Here we exclude Bala-Carter label of the form $E_6(\cdot)$, as it gives maximal Levi subalgebra so the irregular puncture is trivial.

\begin{table}
\centering
\begin{adjustwidth}{4.0cm}{}
\begin{tabular}{|c|c|}
\hline
Levi subalgebra $\mfl$ & Nahm Bala-Carter label \\ [0.1em] \hline
$0$ & $0$\\
$A_1$ & $A_1$\\
$2A_1$ & $2A_1$\\
$3A_1$ & $(3A_1)^*$\\
$A_2$ & $A_2$\\
$A_2 + A_1$ & $A_2+A_1$\\
$2A_2$ & $2A_2$\\
$A_3$ & $A_3$\\
$2A_2 + A_1$ & $(2A_2 + A_1)^*$\\
$A_2 + 2A_1$ & $A_2 + 2A_1$\\
$A_3 + A_1$ & $(A_3+A_1)^*$\\
$D_4$ & $D_4$ \\
$A_4$ & $A_4$ \\
$A_4 + A_1$ & $A_4+A_1$ \\
$A_5$ & $(A_5)^*$\\
$D_5$ & $D_5$ \\ \hline
\end{tabular}
\end{adjustwidth}
\caption{\label{table:E6 orbit}The correspondence between Nahm label and the Levi subalgebra. The Levi subalgebra $E_6$ is omitted as it does not give any irregular puncture. We use $*$ to denote the non-special nilpotent orbit. The pole structure and constraints can be found in \cite{Chacaltana:2014jba}. Again, we exclude those with non-principal orbit in the Levi subalgebra.}
\end{table}

We are now ready to count the Coulomb branch spectrum for a given $E_6$ irregular puncture of class $(k,1)$, were $\ell = k + 2$. We use the SW curve from type \Rnum{2}B construction, whose isolated singularity has the form\footnote{As we consider $(k,1)$ theory, there is no distinction between whether it comes from $b = 8, 9$ or $12$. We can simply pick anyone of them.}
\be
x_1^2 + x_2^3 + x_3^4 + z^{12k} = 0,
\ee
whose deformation looks like
\be
x^2_1 + x^3_2 + x^4_3 + \phi_2(z) x_2 x^2_3 + \phi_5(z) x_2 x_3 + \phi_6(z) x^2_3 + \phi_8 (z) x_2 + \phi_9 (z) x_3 + \phi_{12}(z) = 0,
\label{E_6 deform sing}
\ee
where at the singularity $\phi_{12} = z^{12k}$. The Coulomb branch spectrum is encoded in these Casimirs. For example, when $k$ = 1 and regular semisimple coefficients, we know the scaling dimensions for each letter are
\be
\bbra{x_1} = 3, \ \ [x_2] = 2, \ \ [x_3] = \frac{3}{2}, \ \ [z] = \frac{1}{2}.
\ee
By enumerating the quotient algebra generator of this hypersurface singularity we know that the number of moduli for each differential is $\{ d_2, d_5, d_6, d_8, d_9, d_{12} \} = \{ 0, 3, 4, 6, 7, 10\}$. This is consistent with adding pole structures and subtract global contribution of three maximal $E_6$ regular punctures. 

\subsubsection{S-duality for $E_6$ theory} 

We now study the S-duality for $E_6$ theory of class $(k,1)$, with coefficient all regular semisimple. From the $D_N$ S-duality, we know that the Levi subalgebra directly relates to the flavor symmetry. If we take the coefficient matrix to be regular semisimple, then our initial theory is given by a sphere with six black dots, one trivial blue square and one red cross (which is an arbitrary $E_6$ regular puncture.

We only consider large $k$ situation. In type \Rnum{2}B construction \eqref{E_6 deform sing}, the scaling dimensions for each letter are
\be
\bbra{x_1} = \frac{6k}{k+1}, \ \ [x_2] =  \frac{4k}{k+1}, \ \ [x_3] =  \frac{3k}{k+1}, \ \ [z] = \frac{1}{k+1}.
\ee
So we have the spectrum of initial theory coming from irregular puncture as:
\be
& \phi_2: \frac{2k}{k+1}, \dots, \frac{k+2}{k+1}, \ \ \ \ \ \phi_5: \frac{5k}{k+1}, \dots, \frac{k+2}{k+1}, \\[0.5em]
& \phi_6: \frac{6k}{k+1}, \dots, \frac{k+2}{k+1}, \ \ \ \ \ \phi_8: \frac{8k}{k+1}, \dots, \frac{k+2}{k+1}, \\[0.5em]
& \phi_9: \frac{9k}{k+1}, \dots, \frac{k+2}{k+1}, \ \ \ \ \ \phi_{12}: \frac{12k}{k+1}, \dots, \frac{k+2}{k+1}.
\ee
There are several ways to split Argyres-Douglas matter. For example, we may pop out two black dots and one trivial blue square. We get the duality frame
\vspace{6pt}
\begin{adjustwidth}{3cm}{}
\begin{tikzpicture}
    \draw (7.1,0) node[anchor=west] (1) {$ \pbra{\Rnum{3}_{k,1}^{[1,1;0]^{\times k+2}}, [1^4]}$,};
     \draw (5.9,1) node[anchor=west] (2) {$SO(4)$};
     \draw (2.2,0) node[anchor=west] (3) {$ \pbra{\Rnum{3}_{k,1}^{(2A_1)^{\times (k+1)}, 0}, Q_{E_6}}$};
    
    \path (1) edge[sedge] (2)
             (3) edge[sedge] (2)
    ;
\end{tikzpicture}
\end{adjustwidth}
and here the right hand side theory is two copies of $(A_1, D_{2k+2})$ theory. This duality frame persists to $k = 1$. We have checked that the central charge matches.

The second way is to pop out a trivial black dot and the $E_6$ regular puncture. This results in $D_5$ gauge group:

\vspace{6pt}
\begin{adjustwidth}{3cm}{}
\begin{tikzpicture}
    \draw (7.1,0) node[anchor=west] (1) {$ \pbra{\Rnum{3}_{k,1}^{(D_5)^{\times (k+1)}, 0}, Q}$,};
     \draw (5.9,1) node[anchor=west] (2) {$SO(10)$};
     \draw (2.2,0) node[anchor=west] (3) {$ \pbra{\Rnum{3}_{k,1}^{[1^5; 0]^{\times (k+2)}}, [1^{10}]}$};
    
    \path (1) edge[sedge] (2)
             (3) edge[sedge] (2)
    ;
\end{tikzpicture}
\end{adjustwidth}
\vspace{6pt}
where the theory $\pbra{\Rnum{3}_{k,1}^{[1^5; 0]^{\times (k+2)}}, [1^{10}]}$ can be further degenerate according to $D_N$ type rules. The spectrum counting is explained in the example in section \ref{subsubsec: irregDegeneration}. We see it correctly reproduces $SO(10)$ flavor symmetry. We have also checked that the central charge matches.

Another way is to give $SU(6)$ gauge group in the degeneration limit, by poping out a trivial blue puncture and red cross. 

\vspace{6pt}
\begin{adjustwidth}{3cm}{}
\begin{tikzpicture}
    \draw (7.1,0) node[anchor=west] (1) {$ \pbra{\Rnum{3}_{k,1}^{(A_5)^{\times (k+1)}, 0}, Q_{E_6}}$,};
     \draw (5.9,1) node[anchor=west] (2) {$SU(6)$};
     \draw (2.2,0) node[anchor=west] (3) {$ \pbra{\Rnum{3}_{k,1}^{[1^6]^{\times (k+2)}}, [1^{6}]}$};
    
    \path (1) edge[sedge] (2)
             (3) edge[sedge] (2)
    ;
\end{tikzpicture}
\end{adjustwidth}
\vspace{6pt}
We find that the central charges match as well.

    

\subsection{$E_7$ and $E_8$ theory}

Finally, we turn to $E_7$ and $E_8$ Argyres-Douglas theories. Tinkertoys for $E_7$ theories have been worked out in \cite{Chacaltana:2017boe}. Similar ideas go through and we will outline the steps here. The key ingredient is to use type \Rnum{2}B construction to count the moduli. For $E_7$ theory, the deformed singularity has the form
\be
x_1^2  & + x_2^3 + x_2 x_3^3 + \phi_2(z) x_2^2 x_3 + \phi_6(z) x_2^2 + \phi_8(z) x_2 x_3\\[0.5em]
 & + \phi_{10}(z)x_3^2 + \phi_{12}(z)x_2 + \phi_{14}(z) x_3 + \phi_{18}(z) = 0, 
\ee
where $\{ \phi_2, \phi_6, \phi_8, \phi_{10}, \phi_{12}, \phi_{14}, \phi_{18} \}$ are independent differentials. For $E_8$ theory, the deformed hypersurface singularity has the form:
\be
x_1^2  & + x_2^3 + x_3^5 + \phi_2(z) x_2 x^3_3 + \phi_8(z) x_2 x^2_3 + \phi_{12}(z) x^3_3\\[0.5em]
 & + \phi_{14}(z)x_2 x_3 + \phi_{18}(z)x^2_3 + \phi_{20}(z) x_2 + \phi_{24}(z)x_3 + \phi_{30}(z) = 0, 
\ee
where $\{ \phi_2, \phi_8, \phi_{12}, \phi_{14}, \phi_{18}, \phi_{20}, \phi_{24}, \phi_{30} \}$ are independent differentials. 

The regular puncture for these two exceptional algebras are again given the Bala-Carter label. One can read off the Levi subalgebra similar as before. This then provides the way of counting Coulomb branch spectrum. The duality frame can then be inferred by comparing the spectrum in the degeneration limit, and checked with central charge computation \eqref{centralCharge2}. 

For example, we have in $\mfe_7$ theory one duality frame which looks like

\vspace{6pt}
\begin{adjustwidth}{1.8cm}{}
\begin{tikzpicture}
    \draw (9.1,0) node[anchor=west] (1) {$ \pbra{\Rnum{3}_{k,1}^{(E_6)^{\times (k+1)}, 0}, Q}$,};
     \draw (6.4,1) node[anchor=west] (2) {$E_6$};
     \draw (1.4,0) node[anchor=west] (3) {$ \pbra{\Rnum{3}_{k,1}^{(0)^{\times (k+2)}}, F_{\mfe_6}}$};
    
    \path (1) edge[sedge] (2)
             (3) edge[sedge] (2)
    ;
\end{tikzpicture}
\end{adjustwidth}
\vspace{6pt}
where $F_{\mfe_6}$ is the full $E_6$ regular puncture. Another duality frame is
\vspace{6pt}
\begin{adjustwidth}{1.8cm}{}
\begin{tikzpicture}
    \draw (9.1,0) node[anchor=west] (1) {$ \pbra{\Rnum{3}_{k,1}^{(A_6)^{\times (k+1)}, 0}, Q}$.};
     \draw (6.4,1) node[anchor=west] (2) {$SU(7)$};
     \draw (1.4,0) node[anchor=west] (3) {$ \pbra{\Rnum{3}_{k,1}^{[1^7]^{\times (k+2)}}, [1^7]}$};
    
    \path (1) edge[sedge] (2)
             (3) edge[sedge] (2)
    ;
\end{tikzpicture}
\end{adjustwidth}

For $\mfe_8$ theory, we have the duality frames
\vspace{6pt}
\begin{adjustwidth}{1.8cm}{}
\begin{tikzpicture}
    \draw (9.1,0) node[anchor=west] (1) {$ \pbra{\Rnum{3}_{k,1}^{(E_7)^{\times (k+1)}, 0}, Q}$,};
     \draw (6.4,1) node[anchor=west] (2) {$E_7$};
     \draw (1.4,0) node[anchor=west] (3) {$ \pbra{\Rnum{3}_{k,1}^{(0)^{\times (k+2)}}, F_{\mfe_7}}$};
    
    \path (1) edge[sedge] (2)
             (3) edge[sedge] (2)
    ;
\end{tikzpicture}
\end{adjustwidth}
and
\vspace{6pt}
\begin{adjustwidth}{1.8cm}{}
\begin{tikzpicture}
    \draw (9.1,0) node[anchor=west] (1) {$ \pbra{\Rnum{3}_{k,1}^{(A_7)^{\times (k+1)}, 0}, Q}$,};
     \draw (6.4,1) node[anchor=west] (2) {$SU(8)$};
     \draw (1.4,0) node[anchor=west] (3) {$ \pbra{\Rnum{3}_{k,1}^{(0)^{\times (k+2)}}, [1^8]}$};
    
    \path (1) edge[sedge] (2)
             (3) edge[sedge] (2)
    ;
\end{tikzpicture}
\end{adjustwidth}

\vspace{6pt}
We have checked that the central charges and the Coulomb branch spectrum matches. The left hand theory of each duality frames can be further degenerated according to known rules for lower rank ADE Lie algebras, and we do not picture them anymore. Here we see the interesting duality appears again: the quivers with $E_N$ type gauge group is dual to quivers with $A_{N-1}$ type quivers. 

\section{Conclusion and discussion}\label{Sec: conclusion}

In this paper, we classified the Argyres-Douglas theory of $D_N$ and $E_{6,7,8}$ type based on classification of irregular punctures in the Hitchin system. We developed a systematic way of counting graded dimension. Generalizing the construction in \cite{Xie:2017vaf}, we also obtained duality frames for these AD theories, and find a novel duality between quivers with $SO / E_N$ gauge groups and quivers with $SU$ gauge groups. 

An interesting question to ask is whether one can understand the duality from geometry. In other words, whether one can engineer these quiver theories in string theory, and the duality is interpreted as operations on the geometry side. A related question would be whether such exotic duality exist in three dimensions. In $A_{N-1}$-type AD theories, we can perform dimensional reduction and mirror symmetry to get a Lagrangian theory, which is in general a quiver with $SU$ gauge groups \cite{boalch2008irregular}. One expects that such mirror theory also exists for $D_N$ and $E$-counterpart. Then, the three dimensional mirror of the above duality would be a natural construction.

S-duality in four dimensional superconformal theories sometimes facilitate the calculation of partition functions \cite{Gadde:2010te}. It will be interesting to see the duality frames obtained for AD theories can give partition function of some of them. Partition functions of certain $A_{N-1}$ type AD theories were recently computed in \cite{Cordova:2015nma, Song:2015wta, Buican:2015ina, Buican:2017uka}. In particular, the Schur index encodes two dimensional chiral algebra \cite{Beem:2013sza, Beem:2017ooy} while Coulomb branch index gives geometric quantization of Hitchin moduli space \cite{Gukov:2016lki, Fredrickson:2017yka, Fredrickson:2017jcf} and new four manifold invariants \cite{Gukov:2017zao}. As we mentioned in section \ref{subsubsec: regular puncture}, there are more fundamental invariants arise for $D_N$ Hitchin system, so one may wonder its Hitchin fibration structure, as well as its fixed point under $U(1)$ action.

In our construction, we have obtained many AD theories whose coefficient matrices in the Higgs field degenerate. Then one can try to study their chiral algebra, characters and representations. A useful approach is taken in \cite{Buican:2017fiq}. Study of the associated chiral algebra would have further implication on the dynamics of the theory, for instance chiral ring structure, symmetries and the presence of a decoupled free sector. Furthermore, one may explore if there are corresponding $\cN = 1$ Lagrangian theories that flows to $D_N$ type AD theory, following the construction in \cite{Maruyoshi:2016tqk, Maruyoshi:2016aim, Agarwal:2016pjo, Agarwal:2017roi}.

Our study of S-duality may have many implication for the general investigation of conformal manifold for four dimensional $\cN = 2$ superconformal theories. In particular, with those duality frames, one can ask if they exhaust all the possible frames, what is the group action on the conformal manifold and how the cusps look like. There are progress in computing S-duality group from homological algebra point of view \cite{Caorsi:2016ebt, Caorsi:2017bnp}. We hope to better understand these structures in future publications.

\acknowledgments{The authors wish to thank Sergei Gukov and Yifan Wang for discussions and comment. K.Y. would like to thank 2017 Simons Summer Workshop where part of the work is done. K.Y. is supported by DOE Grant DE-SC0011632, the Walter Burke Institute for Theoretical Physics, and Graduate Fellowship at Kavli Institute for Theoretical Physics. D.X. is supported by Center for Mathematical Sciences and Applications at Harvard University, and in part by the Fundamental Laws Initiative of the Center for the Fundamental Laws of Nature, Harvard University.
}

\appendix

\section{Type \Rnum{2}B construction for AD theories}\label{type2B}

Consider type \Rnum{2}B string theory on isolated hypersurface singularity in $\mC^4$:
\be
W(x_1, x_2, x_3, x_4) = 0, \ \ \ \ \ \ \ \ W\pbra{\lambda^{q_i} x_i} = \lambda W(x_i),
\ee
where the condition of isolation at $x_i = 0$ means $dW = 0$ if and only if $x_i = 0$. The quasi-homogeneity in above formula plus the constraint $\sum q_i > 1$ guarantees that the theory has $U(1)_r$ symmetry, $i.e$ it is superconformal. 

The Coulomb branch of resulting four dimensional $\cN = 2$ SCFT is encoded in the mini-versal deformation of the singularity:
\be
F(x_i, \lambda_a) = W(x_i) + \sum_{a = 1}^{\mu} \lambda_a \phi_a,
\ee
where $\{\phi_a \}$ are a monomial basis of the quotient algebra
\be
\cA_W = \mC[x_1, x_2, x_3, x_4] \left / \left \langle  \frac{\partial W}{\partial x_1},  \frac{\partial W}{\partial x_2},  \frac{\partial W}{\partial x_3},  \frac{\partial W}{\partial x_4} \right \rangle \right. .
\ee
The dimension  $\mu$ of the algebra as a vector space is the \textit{Minor number}, given by
\be
\mu = \prod_{i=1}^4 \pbra{\frac{1}{q^i} - 1}.
\ee
The mini-versal deformation can be identified with the SW curve of the theory.

BPS particles in the SCFT can be thought of as D3 brane wrapping special Lagrangian cycles in the deformed geometry. The  integration of the holomorphic three form,
\be
\Omega = \frac{dx_1 \wedge d x_2 \wedge d x_3 \wedge d x_4}{dF}
\ee
on the three cycles give the BPS mass of the theory. Thus, we require that $\Omega$ should have mass dimension $1$. This determines the scaling dimension of the parameter $\lambda_a$:
\be
\bbra{\lambda_a} = \alpha \pbra{1- \bbra{\phi_a}}, 
\ee
where $\alpha = 1 / \pbra{\sum q_i - 1}$. 

The central charges of the theory is given by \cite{Shapere:2008zf}:
\be
a = \frac{R(A)}{4} + \frac{R(B)}{6} + \frac{5r}{24} + \frac{h}{24}, \ \ \ \ c = \frac{R(B)}{3} + \frac{r}{6} + \frac{h}{12}.
\label{WcentralCharge}
\ee
Here $R(A)$ is given by summation of Coulomb branch spectrum:
\be
R(A) = \sum_{[u_i] > 1} \pbra{\bbra{u_i} - 1},
\ee
and $r$, $h$ are number of free vector multiplets and hypermultiplets of the theory at generic point of the Coulomb branch. In our cases, $r$ equals the rank of Coulomb branch and $h$ is zero. Finally, we have \cite{Xie:2015rpa}
\be
R(B)  = \frac{\mu \alpha}{4}.
\ee

\section{Grading of Lie algebra from nilpotent orbit}\label{nilpoGrading}

A natural way of generating torsion automorphism is to use nilpotent orbit in $\mfg$. Let $e$ be a nilpotent element, which may be included in an $\mathfrak{sl}_2$ triple $\{ e, h, f \}$ such that $\bbra{e, f} = h$, $\bbra{h, e} = 2e$, $\bbra{h, f} = -2f$. With respect to the adjoint action ${\rm ad}\, h$, $\mfg$ decompose into eigenspaces:
\be
\mfg = \bigoplus_{i = -d}^{d} \mfg_i,
\label{nilgrading}
\ee
where $d$ is called the depth. Proper re-assembling of $\mfg_i$ gives \eqref{Zb grading}, hence fixes a torsion automorphism $\sigma_e$ of order $m$. We call the nilpotent element $e$ even (odd) if the corresponding Kac diagram $\mD_e$ is even (odd). In fact $\mD_e$ is identical to the weighted Dynkin diagram ${\widehat \mD}_e$ \cite{collingwood1993nilpotent}. Moreover, we have the relation $m = d+2$ and $\mfg^2 = \mfg_2 + \mfg_{-d}$. 

A \textit{cyclic} element of the semisimple Lie algebra $\mfg$ associated with nilpotent element $e$ is the one of the form $e +F$, for $F \in \mfg_{-d}$. We say $e$ is of \textit{nilpotent} (resp. \textit{semisimple} or \textit{regular semisimple}) \textit{type} if any cyclic element associated with $e$ is nilpotent (resp. any \textit{generic} cyclic element associated with $e$ is semisimple or regular semisimple). Otherwise, $e$ is called mixed type \cite{elashvili2013cyclic}. A theorem of \cite{elashvili2013cyclic} is that $e$ is of nilpotent type if and only if the depth $d$ is odd. We see that $T_2$ precisely corresponds to the cyclic element. In order to get regular semisimple coefficient matrices, it is clear that one needs $e$ of regular semisimple type. In fact, except for $\mfg = A_{N-1}$ case, all nilpotent elements of regular semi-simple type generate even Kac diagram $\mD_e$\footnote{By this we mean that the nilpotents with partition $\bbra{n, n, \dots, n, 1}$ for $\mfg = A_{N-1}$, though of regular semisimple type, are not even.}.  

However, nilpotents $e$ of regular semisimple type do not exhaust all the torsion automorphism we are interested in. To complete the list, we examine the problem from another point of view.  When a cyclic element $e+F$ is regular semisimple, its centralizer $\mfh'$ is a Cartan subalgebra. $\sigma_e$ leaves $\mfh'$ invariant, thus induces a regular element $w_e$ in the Weyl group. When $e$ gives even $\mD_e$, $w_e$ and $\sigma_e$ have the same order, called the \textit{regular number} of $w_e$. Regular element and its regular number are classified in \cite{springer1974regular}, and nilpotents of regular semisimple type do not cover all of them.

The remaining regular numbers, fortunately, are all divisors of those of $\sigma_e$. Hence, we can obtain the Kac diagrams from taking appropriate power of some $\sigma_e$. Their Kac coordinates are determined from the following algorithm \cite{reeder2010torsion, reeder2012gradings}. Suppose we start with automorphism $\sigma_e$ of order $m$ and Kac coordinates $(s_0, s_1, \dots, s_r)$ and we wish to construct automorphism of order $n < m$ by taking $\sigma^{m / n}$. We first replace the label $s_0$ by
\be
s_0 \rightarrow n - \sum_{i = 1}^N a_i s_i.
\label{KacAlgorithm1}
\ee
Now $s_0$ will be necessarily negative. After that, we pick one negative label $s_j$ at each time for $j = 0, 1, \dots, N$, and change the label into $(s'_0, s'_1, \dots, s'_r)$ such that
\be
s'_i = s_i - \langle \alpha_i, \alpha_j^{\vee} \rangle s_j, \ \ \ \ i = 0, 1, \dots, r, 
\label{KacAlgorithm2}
\ee
where $\alpha^{\vee}$ is the coroot. One repeats the procedure until finally all $(s_0, \dots, s_r)$ are positive. This gives the Kac diagram that corresponds to the automorphism with order $n$. The Kac diagram obtained is unambiguous, independent of which element $e$ we start with. 

We now use nilpotent elements to obtain the grading. For $\mfg = A_{N-1}$, this is done in \cite{Xie:2017vaf}. We mainly examine the classification when $\mfg = D_{N}$ and $E_{6,7,8}$. 

\bigskip

$\bullet$ {\bf The Lie algebra} $\mfg = D_N$. Nilpotent element $e$ is of semi-simple type if and only if
\begin{enumerate}
\item[(\rnum{1})] The embedding is $[n_1, \dots, n_1, 1, \dots, 1 ]$ where $n_1$ has even multiplicity;

\item[(\rnum{2})] $[2m+1, 2m-1, 1, \dots, 1 ]$ with $m \geq 1$;

\item[(\rnum{3})] $[n_1, 1, \dots, 1]$ for $n_1 \geq 5$.
\end{enumerate}
In particular, $e$ is of regular semi-simple type if and only if in (\rnum{1}) $n_1$ is odd and $1$ occurs at most twice; in (\rnum{2}) $p \leq 4$; in (\rnum{3}) $p \leq 2$. In each case we can compute $b = d + 2$ where $d$ is the depth. They are (\rnum{1}) $d = 2n_1 - 2$; (\rnum{2}) $d = 2n_1 - 4 = 4m - 2$; (\rnum{3}) $d = 2n_1 - 4$ \cite{elashvili2013cyclic}. As is known, these nilpotent elements are all even. Next we examine each case of regular semi-simple type in more detail.

\vspace{5pt}
\textit{Nilpotent embedding of case (\rnum{1})}. When the partition is $[n_1, n_1, \dots, n_1]$, we see $n_1$ must be a divisor of $N$. Therefore we have the Higgs field
\be
\Phi \sim \frac{T}{z^{2 + \frac{k}{n_1}}}
\label{HiggsCase1}
\ee
with $(k, n_1) = 1$. Note that when $N$ is even, the partition $[N, N]$ is not allowed. This case will be recovered in case (\rnum{2}).

When the partition is $[n_1, \dots, n_1, 1]$, then we know $n_1$ divides $2N-1$. But $n_1$ must have even multiplicity, so this case is excluded.

When the partition is $[n_1, \dots, n_1, 1, 1]$, then $n_1$, being an odd number, must divide $N-1$. Then we get \eqref{HiggsCase1} as well (but the matrix $T$ is different). 

\vspace{5pt}
\textit{Nilpotent embedding of case (\rnum{2})}. There can only be no $1$ or two $1$'s in the Young tableaux. For the former, we have $4m = 2N$. So this case exists only when $N$ is even number. The Higgs field is
\be
\Phi \sim \frac{T}{z^{2 + \frac{k}{N}}}
\label{HiggsCase2}
\ee
with $(k, N) = 1$. For the latter, we have $4m = 2N - 2$ (which means $N-1$ must be even), and the Higgs field is
\be
\Phi \sim \frac{T}{z^{2 + \frac{k}{N-1}}}
\label{HiggsCase3}
\ee
for $(k, N-1) = 1$. 

\vspace{5pt}
\textit{Nilpotent embedding of case (\rnum{3})}. When $p = 1$, we have the partition $[2N ]$. This violates the rule for D-partition.

When $p = 2$ we have $n_1 = 2N-1$, so the order of $\epsilon$ is $4N - 4$. We get the Higgs field
\be
\Phi \sim \frac{T}{z^{2 + \frac{k}{2N-2}}}.
\label{HiggsCase4}
\ee

\vspace{5pt} 
In summary, with classification of nilpotent orbit of regular semi-simple type, for $N$ odd, we have recovered $b = N$ and all its divisors $b = n_1$ (no even divisors). For $N$ even, we can recover $b =N$ as well and all its odd divisor. But we \textit{could not recover its even divisors} using the above technique. Similarly, we have recovered $b = 2N-2$ and $b = N-1$ as well as all odd divisors of $N-1$, but we missed all the even divisors of $2N-2$ except $N-1$ itself. 

The recovery of the missing cases can be achieved with the prescription introduced around \eqref{KacAlgorithm1} and \eqref{KacAlgorithm2}. We give some examples in appendix \ref{generateKD}. Here we only mention that such procedure is unambiguous, $i.e.$ the resulting Kac diagram is the same regardless of which parent torsion automorphism we use\footnote{More specifically, they should descend from the \textit{same} ``parent''. For instance, fix $D_N$, if $n_1$ and $n_2$ are \textit{both} divisors of $N$ and $n_1 | n_2$, then the torsion automorphism of $\sigma_1$ of order $n_1$ is the same whether we start with $\sigma_{\bbra{2m+1, 2m-1}}$ by taking $N/n_1$-th power, or with $\sigma_2$ of order $n_2$ by taking $n_2/n_1$-th power. See appendix \ref{generateKD} for more detail.}.

\bigskip

$\bullet$ {\bf The Lie algebra} $\mfg = E_{6,7,8}$. As in the previous case, we would like to first find all nilpotent elements of regular semi-simple type. They are listed in table \ref{table:E6} - table \ref{table:E8}, along with their order and the singular Higgs field behavior. One can also use the pole data to read off the 3-fold singularity.

\begin{table}
\centering
\begin{tabular}{|c|c|c|c|}
\hline
nilpotent orbit & depth & order & Higgs field \\
\hline
$D_4(a_1)$ & $6$ & $4$ & $\Phi \sim T / z^{2+\frac{k}{4}}$\\[0.1em]
$E_6(a_3)$ & $10$ & $6$ & $\Phi \sim T / z^{2+\frac{k}{6}}$ \\[0.1em]
$D_5$ & $14$ & $8$ &  $\Phi \sim T / z^{2+\frac{k}{8}}$\\[0.1em]
$E_6(a_1)$ & $16$ & $9$  &  $\Phi \sim T / z^{2+\frac{k}{9}}$\\[0.1em]
$E_6$ & $22$ & $12$ &  $\Phi \sim T / z^{2+\frac{k}{12}}$ \\ \hline
\end{tabular}
\caption{\label{table:E6}Summary of nilpotent elements of regular semi-simple type in $E_6$.}
\end{table}

\begin{table}
\centering
\begin{tabular}{|c|c|c|c|}
\hline
nilpotent orbit & depth & order & Higgs field \\
\hline
$E_7(a_5)$ & $10$ & $6$ & $\Phi \sim T / z^{2+\frac{k}{6}}$\\[0.1em]
$A_6$ & $12$ & $7$ & $\Phi \sim T / z^{2+\frac{k}{7}}$ \\[0.1em]
$E_6(a_1)$ & $16$ & $9$ &  $\Phi \sim T / z^{2+\frac{k}{9}}$\\[0.1em]
$E_7(a_1)$ & $26$ & $14$  &  $\Phi \sim T / z^{2+\frac{k}{14}}$\\[0.1em]
$E_7$ & $34$ & $18$ &  $\Phi \sim T / z^{2+\frac{k}{18}}$ \\ \hline
\end{tabular}
\caption{\label{table:E7}Summary of nilpotent elements of regular semi-simple type in $E_7$.}
\end{table}

\begin{table}
\centering
\begin{tabular}{|c|c|c|c|}
\hline
nilpotent orbit & depth & order & Higgs field \\
\hline
$E_8(a_7)$ & $10$ & $6$ & $\Phi \sim T / z^{2+\frac{k}{6}}$\\[0.1em]
$E_8(a_6)$ & $18$ & $10$ & $\Phi \sim T / z^{2+\frac{k}{10}}$ \\[0.1em]
$E_8(a_5)$ & $22$ & $12$ &  $\Phi \sim T / z^{2+\frac{k}{12}}$\\[0.1em]
$E_8(a_4)$ & $28$ & $15$  &  $\Phi \sim T / z^{2+\frac{k}{15}}$\\[0.1em]
$E_8(a_2)$ & $38$ & $20$ &  $\Phi \sim T / z^{2+\frac{k}{20}}$ \\[0.1em]
$E_8(a_1)$ & $46$ & $24$  &  $\Phi \sim T / z^{2+\frac{k}{24}}$\\[0.1em]
$E_8$ & $58$ & $30$  &  $\Phi \sim T / z^{2+\frac{k}{30}}$\\  \hline
\end{tabular}
\caption{\label{table:E8}Summary of nilpotent elements of regular semi-simple type in $E_8$.}
\end{table}

Again, the above classification does not exhaust the possibility of the order of poles. We expect that we should be able to get all divisors for the denominator. We still can use the same algorithm to generate them; and they are unambiguous. We recover the missing Kac diagram in appendix \ref{generateKD}.

\section{Recover missing Kac diagrams}\label{generateKD}

Here we shall give examples of how to generate those Kac diagrams of torsion automorphisms that are missing from considering nilpotent embedding, as in appendix \ref{nilpoGrading}. To begin with, we first explain in $\mfg = D_N$ case how to write down the weighted Dynkin diagrams for automorphisms of the form $\sigma_e$. For a thorough mathematical treatment, the readers may consult \cite{collingwood1993nilpotent}.

Assume that $e$ is represented by a Young tableau $Y = \bbra{n_1, n_2, \dots, n_p}$, and $n_1 + \dots + n_p = 2N$. Moreover we assume $Y$ is not very even\footnote{For weighted Dynkin diagrams of very even element, see \cite{collingwood1993nilpotent}.}, which is what we concern. For each $n_i$ we get a sequence $\{n_i-1, n_i -3, \dots, -n_i + 3, -n_i + 1 \}$. Combining the sequences for all $i$, we may arrange them in a decreasing order and the first $N$ elements are apparently non-negative, and we denote them as $\{h_1, h_2, \dots, h_N \}$. Now the Kac coordinate on the Dynkin diagram of $D_N$ is given as follows: 

\begin{tikzpicture}
\centering
    \draw (-1,0) node[anchor=east]  {$\sigma_Y:$};

    \node[dnode,label=below:$h_1-h_2$] (1) at (0,0) {};
    \node[dnode,label=below:$h_2 - h_3$] (2) at (2,0) {};
    \node[dnode,label=below:$$] (3) at (4,0) {};
    \node[dnode,label=below:$h_{N-2}-h_{N-1}$] (4) at (6,0) {};
    \node[dnode,label=above:$h_{N-1}-h_N$] (5) at (6,1.5) {};
    \node[dnode,label=below right:$h_{N-1}+h_N$] (6) at (8,0.0) {};

    \path (1) edge[sedge] (2)
          (2) edge[sedge,dashed] (3)
          (3) edge[sedge] (4)
          (4) edge[sedge] (5)
              edge[sedge] (6)
          ;
\end{tikzpicture}
\vspace{3pt}

\noindent Then, we add the highest root $\alpha_0$ and make it an extended Dynkin diagram, and put the label $s_0 = 2$ for it. If in addition the Kac diagram is even, by our convention we divide each label by $2$.

Now we present examples showing the unambiguity of generating Kac diagrams. We take $N=12$. The order $12$ torsion automorphism is obtained by the nilpotent element with partition $[13, 11]$, so its affine weighted Dynkin diagram is

\begin{tikzpicture}
\centering
    \draw (-1,0) node[anchor=east]  {$\sigma_{[13,11]}:$};

    \node[dnode,label=below:$1$] (1) at (0,0) {};
    \node[dnode,label=below:$0$] (2) at (1,0) {};
    \node[dnode,label=above:$1$] (0) at (1,1) {};
    \node[dnode,label=below:$1$] (3) at (2,0) {};
    \node[dnode,label=below:$0$] (4) at (3,0) {};
    \node[dnode,label=below:$1$] (5) at (4,0) {};
    \node[dnode,label=below:$0$] (6) at (5,0) {};
    \node[dnode,label=below:$1$] (7) at (6,0) {};
    \node[dnode,label=below:$0$] (8) at (7,0) {};
    \node[dnode,label=below:$1$] (9) at (8,0) {};
    \node[dnode,label=below:$0$] (10) at (9,0) {};
    \node[dnode,label=above:$1$] (11) at (9,1) {};
    \node[dnode,label=below:$1$] (12) at (10,0) {};

    \path (1) edge[sedge] (2)
          (2) edge[sedge] (3)
          (3) edge[sedge] (4)
          (4) edge[sedge] (5)
           (5) edge[sedge] (6)
           (6) edge[sedge] (7)
           (7) edge[sedge] (8)
           (8) edge[sedge] (9)
            (9) edge[sedge] (10)
            (10) edge[sedge] (11)
             (10) edge[sedge] (12)
             (0) edge[sedge, dashed] (2)  
             ;
\end{tikzpicture}

\vspace{3pt}
\noindent where we used dashed line to indicate the affine root. We may use the algorithm from \eqref{KacAlgorithm1} and \eqref{KacAlgorithm2} to generate an order $6$ torsion automorphism. It is given by:

\begin{tikzpicture}
\centering
    \draw (-1,0) node[anchor=east]  {$\sigma_{\{ 6 \}}:$};

    \node[dnode,label=below:$0$] (1) at (0,0) {};
    \node[dnode,label=below:$1$] (2) at (1,0) {};
    \node[dnode,label=above:$0$] (0) at (1,1) {};
    \node[dnode,label=below:$0$] (3) at (2,0) {};
    \node[dnode,label=below:$0$] (4) at (3,0) {};
    \node[dnode,label=below:$0$] (5) at (4,0) {};
    \node[dnode,label=below:$1$] (6) at (5,0) {};
    \node[dnode,label=below:$0$] (7) at (6,0) {};
    \node[dnode,label=below:$0$] (8) at (7,0) {};
    \node[dnode,label=below:$0$] (9) at (8,0) {};
    \node[dnode,label=below:$1$] (10) at (9,0) {};
    \node[dnode,label=above:$0$] (11) at (9,1) {};
    \node[dnode,label=below:$0$] (12) at (10,0) {};

    \path (1) edge[sedge] (2)
          (2) edge[sedge] (3)
          (3) edge[sedge] (4)
          (4) edge[sedge] (5)
           (5) edge[sedge] (6)
           (6) edge[sedge] (7)
           (7) edge[sedge] (8)
           (8) edge[sedge] (9)
            (9) edge[sedge] (10)
            (10) edge[sedge] (11)
             (10) edge[sedge] (12)
             (0) edge[sedge, dashed] (2)  
             ;
\end{tikzpicture} 
\vspace{3pt}

\noindent Since this diagram does not come from any nilpotent element $e$, we just use a subscript $\{6\}$ to indicate its order. With this diagram, we can further generate an order $3$ nilpotent element by taking a twice power of $\sigma_{\{6\}}$. The same algorithm gives a Kac diagram:

\begin{tikzpicture}
\centering
    \draw (-1,0) node[anchor=east]  {$\sigma_{\{ 3 \}}:$};

    \node[dnode,label=below:$0$] (1) at (0,0) {};
    \node[dnode,label=below:$0$] (2) at (1,0) {};
    \node[dnode,label=above:$1$] (0) at (1,1) {};
    \node[dnode,label=below:$0$] (3) at (2,0) {};
    \node[dnode,label=below:$0$] (4) at (3,0) {};
    \node[dnode,label=below:$0$] (5) at (4,0) {};
    \node[dnode,label=below:$0$] (6) at (5,0) {};
    \node[dnode,label=below:$0$] (7) at (6,0) {};
    \node[dnode,label=below:$1$] (8) at (7,0) {};
    \node[dnode,label=below:$0$] (9) at (8,0) {};
    \node[dnode,label=below:$0$] (10) at (9,0) {};
    \node[dnode,label=above:$0$] (11) at (9,1) {};
    \node[dnode,label=below:$0$] (12) at (10,0) {};

    \path (1) edge[sedge] (2)
          (2) edge[sedge] (3)
          (3) edge[sedge] (4)
          (4) edge[sedge] (5)
           (5) edge[sedge] (6)
           (6) edge[sedge] (7)
           (7) edge[sedge] (8)
           (8) edge[sedge] (9)
            (9) edge[sedge] (10)
            (10) edge[sedge] (11)
             (10) edge[sedge] (12)
             (0) edge[sedge, dashed] (2)  
             ;
\end{tikzpicture} 

\vspace{3pt}

\noindent This Kac diagram is precisely the same as the affine weighted Dynkin diagram of the nilpotent element $\bbra{3^8}$. So we see there is no ambiguity.

As a second example, we take $N = 9$. The same argument as above shows that the Kac diagram for order $8$ torsion automorphism constructed from nilpotent element of partition $[9, 7, 1, 1]$, is exactly identical to the one obtained by square of the torsion automorphism from the element $[17, 1]$. 

For $\mfg = E_{6,7,8}$ case, the Kac diagrams for nilpotent elements of regular semisimple type are given in \cite{elashvili2013cyclic}. With the same procedure, we can recover missing Kac diagrams as follows.

For $\mfg = E_6$, we missed order $2$ and order $3$ element, their Kac diagrams are, respectively:

\begin{tikzpicture} 
\hspace*{-0.05cm}

    \draw (-2,0.5) node[anchor=east]  {$\sigma^{E_6}_{\{2\}}: $};
    \draw (3.5,0.5) node[anchor=east]  {$,$};
    \draw (5.5,0.5) node[anchor=east]  {$\sigma^{E_6}_{\{3\}}: $};

    \node[dnode,label=right:$0$] (0) at (1,2) {};
    \node[dnode,label=below:$0$] (1) at (-1,0) {};
    \node[dnode,label=right:$1$] (2) at (1,1) {};
    \node[dnode,label=below:$0$] (3) at (0,0) {};
    \node[dnode,label=below:$0$] (4) at (1,0) {};
    \node[dnode,label=below:$0$] (5) at (2,0) {};
    \node[dnode,label=below:$0$] (6) at (3,0) {};
    
    \node[dnode,label=right:$0$] (00) at (8.5,2) {};
    \node[dnode,label=below:$0$] (11) at (6.5,0) {};
    \node[dnode,label=right:$0$] (22) at (8.5,1) {};
    \node[dnode,label=below:$0$] (33) at (7.5,0) {};
    \node[dnode,label=below:$1$] (44) at (8.5,0) {};
    \node[dnode,label=below:$0$] (55) at (9.5,0) {};
    \node[dnode,label=below:$0$] (66) at (10.5,0) {};

    \path (0) edge[dashed, sedge] (2)
          (1) edge[sedge] (3)
          (3) edge[sedge] (4)
          (4) edge[sedge] (5)
 		      edge[sedge] (2)
          (5) edge[sedge] (6);
          
    \path (00) edge[dashed, sedge] (22)
          (11) edge[sedge] (33)
          (33) edge[sedge] (44)
          (44) edge[sedge] (55)
 		      edge[sedge] (22)
          (55) edge[sedge] (66);
          
\end{tikzpicture} 

For $\mfg = E_7$ we also missed the order $2$ and order $3$ torsion automorphisms. There Kac diagram can also be obtained:

\begin{tikzpicture} 
    \draw (-2,0.5) node[anchor=east]  {$\sigma^{E_7}_{\{2\}}:$};

   \node[dnode,label=below:$0$] (0) at (-1,0) {};
    \node[dnode,label=below:$0$] (1) at (0,0) {};
    \node[dnode,label=above:$1$] (2) at (2,1) {};
    \node[dnode,label=below:$0$] (3) at (1,0) {};
    \node[dnode,label=below:$0$] (4) at (2,0) {};
    \node[dnode,label=below:$0$] (5) at (3,0) {};
    \node[dnode,label=below:$0$] (6) at (4,0) {};
    \node[dnode,label=below:$0$] (7) at (5,0) {};

    \path (0) edge[dashed, sedge] (1)
          (1) edge[sedge] (3)
          (3) edge[sedge] (4)
          (4) edge[sedge] (5)
 		      edge[sedge] (2)
          (5) edge[sedge] (6)
          (6) edge[sedge] (7);          
\end{tikzpicture}

\vspace{5pt}
\begin{tikzpicture} 
    \draw (-2,0.5) node[anchor=east]  {$\sigma^{E_7}_{\{3\}}:$};

   \node[dnode,label=below:$0$] (0) at (-1,0) {};
    \node[dnode,label=below:$0$] (1) at (0,0) {};
    \node[dnode,label=above:$0$] (2) at (2,1) {};
    \node[dnode,label=below:$0$] (3) at (1,0) {};
    \node[dnode,label=below:$0$] (4) at (2,0) {};
    \node[dnode,label=below:$1$] (5) at (3,0) {};
    \node[dnode,label=below:$0$] (6) at (4,0) {};
    \node[dnode,label=below:$0$] (7) at (5,0) {};

    \path (0) edge[dashed, sedge] (1)
          (1) edge[sedge] (3)
          (3) edge[sedge] (4)
          (4) edge[sedge] (5)
 		      edge[sedge] (2)
          (5) edge[sedge] (6)
          (6) edge[sedge] (7);          
\end{tikzpicture}

Finally, for $\mfg = E_8$, we have missed the torsion automorphisms of order $2, 3, 4, 5, 8$. They can be recovered by weighted Dynkin diagrams of nilpotent elements of regular semi-simple type. We list them as follows:

\begin{tikzpicture} 
    \draw (-1,0.5) node[anchor=east]  {$\sigma^{E_8}_{\{2\}}:$};

    \node[dnode,label=below:$0$] (0) at (0,0) {};
    \node[dnode,label=below:$0$] (1) at (1,0) {};
    \node[dnode,label=below:$0$] (2) at (2,0) {};
    \node[dnode,label=below:$0$] (3) at (3,0) {};
    \node[dnode,label=below:$0$] (4) at (4,0) {};
    \node[dnode,label=below:$0$] (5) at (5,0) {};
    \node[dnode,label=below:$0$] (6) at (6,0) {};
    \node[dnode,label=below:$1$] (7) at (7,0) {};
    \node[dnode,label=above:$0$] (8) at (5,1) {};

    \path (0) edge[sedge,dashed] (1)
          (1) edge[sedge] (2)
          (2) edge[sedge] (3)
          (3) edge[sedge] (4)
          (4) edge[sedge] (5)                        
          (5) edge[sedge] (6)
                 edge[sedge] (8)
          (6) edge[sedge] (7);
          
\end{tikzpicture}

\vspace{5pt}
\begin{tikzpicture} 
    \draw (-1,0.5) node[anchor=east]  {$\sigma^{E_8}_{\{3\}}:$};

    \node[dnode,label=below:$0$] (0) at (0,0) {};
    \node[dnode,label=below:$0$] (1) at (1,0) {};
    \node[dnode,label=below:$0$] (2) at (2,0) {};
    \node[dnode,label=below:$0$] (3) at (3,0) {};
    \node[dnode,label=below:$0$] (4) at (4,0) {};
    \node[dnode,label=below:$0$] (5) at (5,0) {};
    \node[dnode,label=below:$0$] (6) at (6,0) {};
    \node[dnode,label=below:$0$] (7) at (7,0) {};
    \node[dnode,label=above:$1$] (8) at (5,1) {};

    \path (0) edge[sedge,dashed] (1)
          (1) edge[sedge] (2)
          (2) edge[sedge] (3)
          (3) edge[sedge] (4)
          (4) edge[sedge] (5)                        
          (5) edge[sedge] (6)
                 edge[sedge] (8)
          (6) edge[sedge] (7);
          
\end{tikzpicture}

\vspace{5pt}
\begin{tikzpicture} 
    \draw (-1,0.5) node[anchor=east]  {$\sigma^{E_8}_{\{4\}}:$};

    \node[dnode,label=below:$0$] (0) at (0,0) {};
    \node[dnode,label=below:$0$] (1) at (1,0) {};
    \node[dnode,label=below:$0$] (2) at (2,0) {};
    \node[dnode,label=below:$1$] (3) at (3,0) {};
    \node[dnode,label=below:$0$] (4) at (4,0) {};
    \node[dnode,label=below:$0$] (5) at (5,0) {};
    \node[dnode,label=below:$0$] (6) at (6,0) {};
    \node[dnode,label=below:$0$] (7) at (7,0) {};
    \node[dnode,label=above:$0$] (8) at (5,1) {};

    \path (0) edge[sedge,dashed] (1)
          (1) edge[sedge] (2)
          (2) edge[sedge] (3)
          (3) edge[sedge] (4)
          (4) edge[sedge] (5)                        
          (5) edge[sedge] (6)
                 edge[sedge] (8)
          (6) edge[sedge] (7);
          
\end{tikzpicture}

\vspace{5pt}
\begin{tikzpicture} 
    \draw (-1,0.5) node[anchor=east]  {$\sigma^{E_8}_{\{5\}}:$};

    \node[dnode,label=below:$0$] (0) at (0,0) {};
    \node[dnode,label=below:$0$] (1) at (1,0) {};
    \node[dnode,label=below:$0$] (2) at (2,0) {};
    \node[dnode,label=below:$0$] (3) at (3,0) {};
    \node[dnode,label=below:$1$] (4) at (4,0) {};
    \node[dnode,label=below:$0$] (5) at (5,0) {};
    \node[dnode,label=below:$0$] (6) at (6,0) {};
    \node[dnode,label=below:$0$] (7) at (7,0) {};
    \node[dnode,label=above:$0$] (8) at (5,1) {};

    \path (0) edge[sedge,dashed] (1)
          (1) edge[sedge] (2)
          (2) edge[sedge] (3)
          (3) edge[sedge] (4)
          (4) edge[sedge] (5)                        
          (5) edge[sedge] (6)
                 edge[sedge] (8)
          (6) edge[sedge] (7);
          
\end{tikzpicture}

\vspace{5pt}
\begin{tikzpicture} 
    \draw (-1,0.5) node[anchor=east]  {$\sigma^{E_8}_{\{8\}}:$};

    \node[dnode,label=below:$0$] (0) at (0,0) {};
    \node[dnode,label=below:$1$] (1) at (1,0) {};
    \node[dnode,label=below:$0$] (2) at (2,0) {};
    \node[dnode,label=below:$0$] (3) at (3,0) {};
    \node[dnode,label=below:$0$] (4) at (4,0) {};
    \node[dnode,label=below:$1$] (5) at (5,0) {};
    \node[dnode,label=below:$0$] (6) at (6,0) {};
    \node[dnode,label=below:$0$] (7) at (7,0) {};
    \node[dnode,label=above:$0$] (8) at (5,1) {};

    \path (0) edge[sedge,dashed] (1)
          (1) edge[sedge] (2)
          (2) edge[sedge] (3)
          (3) edge[sedge] (4)
          (4) edge[sedge] (5)                        
          (5) edge[sedge] (6)
                 edge[sedge] (8)
          (6) edge[sedge] (7);
          
\end{tikzpicture}

\newpage

  \bibliography{ADrefs} 
 \bibliographystyle{JHEP}
\end{document}